\documentclass[12pt]{article}
\usepackage{rotate}
\newlength{\pubnumber} 

\catcode`\@=11
\@addtoreset{equation}{section}

\def\section{\@startsection{section}{1}{\z@}{3.5ex plus 1ex minus .2ex}
 {2.3ex plus .2ex}{\large\bf}}
\def\subsection{\@startsection{subsection}{2}{\z@}{2.3ex plus .2ex}
 {2.3ex plus .2ex}{\bf}}

\catcode`@=11
\long\def\@caption#1[#2]#3{\par\addcontentsline{\csname
  ext@#1\endcsname}{#1}{\protect\numberline{\csname
  the#1\endcsname}{\ignorespaces #2}}\begingroup
    \small
    \@parboxrestore
    \@makecaption{\csname fnum@#1\endcsname}{\ignorespaces #3}\par
  \endgroup}
\catcode`@=12

\input epsf.tex

\begin{document}
\begin{titlepage}
\samepage{
\setcounter{page}{1}
\rightline{IASSNS-HEP-95/97}
\rightline{\tt hep-th/9602045}
\rightline{(to appear in {\it Physics Reports}\/)}
\rightline{September 1996}
\vfill
\begin{center}
 {\Large \bf String Theory and the Path to Unification: \\
       A Review of Recent Developments\\}
\vfill
 {\large Keith R. Dienes\footnote{
   E-mail address: dienes@sns.ias.edu}\\}
\vspace{.12in}
 {\it  School of Natural Sciences, Institute for Advanced Study\\
  Olden Lane, Princeton, N.J. 08540 USA\\}
\end{center}
\vfill
\begin{abstract}
  {\rm
  This is a pedagogical review article surveying the various approaches
  towards understanding gauge coupling unification within string theory.
  As is well known, one of the major problems confronting string
   phenomenology has been an apparent discrepancy between the scale
   of gauge coupling unification predicted within string theory, and the
   unification scale expected within the framework of the
   Minimal Supersymmetric Standard Model (MSSM).
   In this article, I provide an overview of the different
   approaches that have been taken in recent years towards reconciling
   these two scales, and outline some of the major recent developments
   in each.  These approaches include string GUT models;
   higher affine levels and non-standard hypercharge
   normalizations;
   heavy string threshold corrections;
   light supersymmetric thresholds;
   effects from intermediate-scale gauge and
   matter structure beyond the MSSM;  strings without supersymmetry;
   and strings at strong coupling. }
\end{abstract}
\smallskip}
\vfill
\end{titlepage}

\setcounter{footnote}{0}


\def\beq{\begin{equation}}
\def\eeq{\end{equation}}
\def\beqn{\begin{eqnarray}}
\def\eeqn{\end{eqnarray}}

\def\ie{{\it i.e.}}
\def\eg{{\it e.g.}}

\def\bone{{\bf 1}}

\def\Str{{{\rm Str}\,}}
\def\rep#1{{\bf{#1}}}
\def\KM{{Ka\v{c}-Moody}}
\def\bQ{{\bf Q}}
\def\bbox#1{\parbox[t]{0.85 in}{{#1}}}
\def\bbreak{\vfill\break}


\def\inbar{\,\vrule height1.5ex width.4pt depth0pt}

\def\IC{\relax\hbox{$\inbar\kern-.3em{\rm C}$}}
\def\IQ{\relax\hbox{$\inbar\kern-.3em{\rm Q}$}}
\def\IR{\relax{\rm I\kern-.18em R}}
 \font\cmss=cmss10 \font\cmsss=cmss10 at 7pt
\def\IZ{\relax\ifmmode\mathchoice
 {\hbox{\cmss Z\kern-.4em Z}}{\hbox{\cmss Z\kern-.4em Z}}
 {\lower.9pt\hbox{\cmsss Z\kern-.4em Z}}
 {\lower1.2pt\hbox{\cmsss Z\kern-.4em Z}}\else{\cmss Z\kern-.4em Z}\fi}

\def\NPB#1#2#3{{\it Nucl.\ Phys.}\/ {\bf B#1} (19#2) #3}
\def\PLB#1#2#3{{\it Phys.\ Lett.}\/ {\bf B#1} (19#2) #3}
\def\PRD#1#2#3{{\it Phys.\ Rev.}\/ {\bf D#1} (19#2) #3}
\def\PRL#1#2#3{{\it Phys.\ Rev.\ Lett.}\/ {\bf #1} (19#2) #3}
\def\PRT#1#2#3{{\it Phys.\ Rep.}\/ {\bf#1} (19#2) #3}
\def\CMP#1#2#3{{\it Commun.\ Math.\ Phys.}\/ {\bf#1} (19#2) #3}
\def\MODA#1#2#3{{\it Mod.\ Phys.\ Lett.}\/ {\bf A#1} (19#2) #3}
\def\IJMP#1#2#3{{\it Int.\ J.\ Mod.\ Phys.}\/ {\bf A#1} (19#2) #3}
\def\NUVC#1#2#3{{\it Nuovo Cimento}\/ {\bf #1A} (#2) #3}
\def\etal{{\it et al.\/}}

\hyphenation{su-per-sym-met-ric non-su-per-sym-met-ric}
\hyphenation{space-time-super-sym-met-ric}
\hyphenation{mod-u-lar mod-u-lar--in-var-i-ant}


\vfill\eject
\setcounter{page}{2}
\tableofcontents

\vfill\eject
\setcounter{footnote}{0}
\section{Introduction}

As is well-known,
string theories achieve remarkable success in answering
some of the most vexing problems of theoretical high-energy
physics.  With string theory, we now have for the first time
a consistent theoretical framework which is finite and which simultaneously
incorporates both quantum gravity and chiral supersymmetric gauge theories
in a natural fashion.
An important goal, therefore, is to determine the extent to
which this framework is capable of describing other more
phenomenological features of the low-energy world.

In this review article, I shall focus on one such feature:
the unification of gauge couplings.
There are various reasons why this is a particularly compelling
feature to study.  On the one hand, the unification of gauge
couplings --- like the appearance of gravity or of gauge symmetry
in the first place --- is a feature intrinsic to string theory,
one whose appearance has basic, model-independent origins.
On the other hand, viewing the situation from an
experimental perspective, the unification of the
gauge couplings is arguably the highest-energy
phenomenon that any extrapolation from low-energy data
can uncover;  in this sense it sits at what is believed to
be the frontier between
our low-energy $SU(3)\times SU(2)\times U(1)$
world, and whatever may lie beyond.
Thus, the unification of gauge couplings
provides a fertile meeting-ground
where string theory can be tested against the results
of low-energy experimentation.

At first glance, string theory
appears to fail this test:  it predicts, {\it a priori}, a
unification of gauge couplings at a scale $M_{\rm string}\approx
5\times 10^{17}$ GeV, approximately a factor of 20 higher
than the expected scale $M_{\rm MSSM}\approx 2 \times 10^{16}$ GeV
obtained through extrapolations from low-energy data
within the framework of the Minimal Supersymmetric Standard
Model (MSSM).  While this may seem
to be a small difference in an absolute sense (amounting to
only 10\% of the logarithms of these mass scales),
this discrepancy nevertheless translates into
predictions for the
low-energy gauge couplings that differ by many standard
deviations from their experimentally observed values.
This is therefore a major problem for string phenomenology.

Fortunately, there are various effects which
may modify these naive predictions, and thereby reconcile
these two unification scales.  These include:  the appearance of a possible
grand-unified (GUT) symmetry at the intermediate scale $M_{\rm MSSM}$
(which would then unify with gravity and any other ``hidden-sector''
gauge symmetries at $M_{\rm string}$);
the possibility that the MSSM gauge group is realized
in string theory through non-standard higher-level
affine gauge symmetries and/or exotic hypercharge normalizations
(which would alter the boundary conditions of the gauge couplings
at unification);
possible large ``heavy string threshold corrections'' (which would
effectively lower the predicted value of $M_{\rm string}$);
possible effects due to light SUSY thresholds (arising from
the breaking of supersymmetry at a relatively low energy scale);
and the appearance of extra matter beyond the MSSM (as often arises
in realistic string models).
There even exist unification scenarios based
on strings {\it without}\/ spacetime supersymmetry,
and on strings at strong coupling.
It is presently unknown, however,
which of these scenarios (or which combination of scenarios)
can successfully explain the apparent discrepancy between
$M_{\rm MSSM}$ and $M_{\rm string}$ in string theory.
In other words, it is not known
which ``path to unification'', if any,
string theory ultimately chooses.

In this article, I shall summarize the basic
status of each of these possibilities, and outline some
of the recent developments in each of these areas.
As we shall see, some of these paths are quite
feasible, and can actually reconcile string-scale unification
with low-energy data.  Others, by contrast,
are tied to more subtle issues in string theory,
and await further insight.

It is precisely for such reasons that this review has been written.
Given the recent experimental results confirming gauge coupling
unification within the MSSM, there is now considerable interest
among low-energy phenomenologists in the potential that gauge
coupling unification holds for uncovering and probing new physics at
very high energy scales.  It is therefore particularly important
at this time to survey the possibilities for new physics that are
suggested by string theory.  But there are also string-based reasons
why a current review should be particularly useful.  Over the
past decade, string model-building and string phenomenology have
matured to the point that specific phenomenological issues such as
gauge coupling unification can now be meaningfully and quantitatively
addressed.  Moreover, as we shall see, the past several years have
witnessed an explosion in the development of different string-based
unification scenarios, with ideas and results coming from many different
directions.  Indeed, each of these various ``paths to unification'' has
now been investigated in considerable detail, and the relevant issues that
are raised within each scenario have now been systematically explored.
This review should therefore serve not only to organize and summarize
the accomplishments achieved within each of these ``paths to unification'',
but also to point the way towards understanding how, through gauge coupling
unification, the predictions of string theory may eventually have a direct
bearing on low-energy physics.

This article is organized as follows.
In Sect.~2, I review the basic problem of gauge coupling
unification, and highlight some of the differences that
exist between gauge coupling unification in field theory and
in string theory.
In Sect.~3, I then provide an outline of the various approaches
that have been proposed for understanding gauge coupling
unification in string theory.
The seven sections which follow
(Sects.~4 through 10) then
discuss each of these approaches in turn, and survey
their relevant issues, problems, and current status.
In particular, Sect.~4 focuses on string GUT models;
Sect.~5 deals with non-standard affine levels and hypercharge normalizations;
Sect.~6 discusses heavy string threshold corrections;
Sect.~7 analyzes light SUSY thresholds and intermediate-scale gauge structure;
Sect.~8 considers extra matter beyond the MSSM;
Sect.~9 introduces gauge coupling unification via strings
without spacetime supersymmetry;
and Sect.~10 outlines a proposal based on strings at strong coupling.
Finally, in Sect.~11, I conclude with a brief summary
and suggestions for further research.

\vfill\eject
\noindent {\it Disclaimers}
\bigskip

This review article is aimed at discussing recent progress
in one specific area:  the unification of gauge couplings within
string theory.
As such, this review does not attempt to cover the vast literature of
field-theoretic unification models, nor (at the other end)
does it attempt to discuss general aspects of string
model-building or string phenomenology.
For reviews of the former subject, the reader
should consult Ref.~\cite{unifreview};
likewise,
for recent reviews of the latter subject, the reader
is urged to consult Refs.~\cite{stringreview,phenreview}.
Although
certain portions of this review are
based upon research
\cite{unif1,unif2,unif3,unif4}
that I have performed in joint collaborations,
I have nevertheless attempted to place
these results in context by surveying related recent works by
other authors as well.  My hope is therefore
that this article presents
a fairly complete survey of the issues surrounding
gauge coupling unification in string theory, including most lines
of development that have been
advanced through the present time (September 1996).
Finally,
since my goal has been to present a pedagogical and (hopefully)
non-technical introduction to recent progress in this field,
I have  avoided the detailed mathematics that is involved
in any particular approach.  Therefore, for further details ---
or for applications to related issues beyond the scope of
this review ---
the reader should consult the relevant references.


\setcounter{footnote}{0}
\section{Background}

\subsection{The problem of gauge coupling unification}

\begin{figure}
\centerline{
   \epsfxsize 3.3 truein \epsfbox {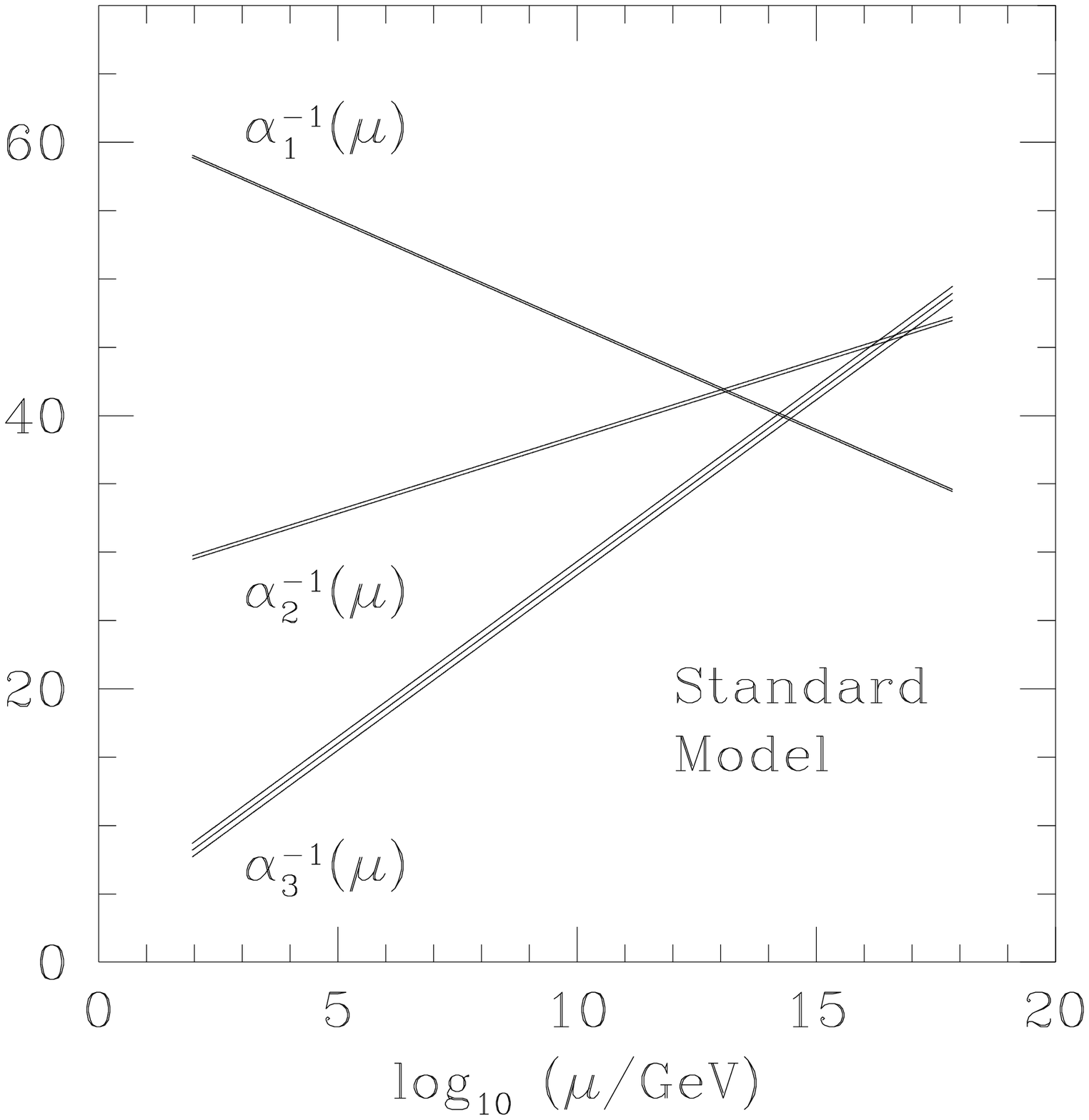}
    }
\caption{One-loop evolution of the gauge couplings within
    the (non-supersymmetric) Standard Model.
    Here $\alpha_1\equiv (5/3)\alpha_Y$,
    where $\alpha_Y$ is the hypercharge coupling
    in the conventional normalization.  The relative width of each line
    reflects current experimental uncertainties.}
\label{introfiga}
\vskip 0.25 truein
\centerline{
   \epsfxsize 3.3 truein \epsfbox {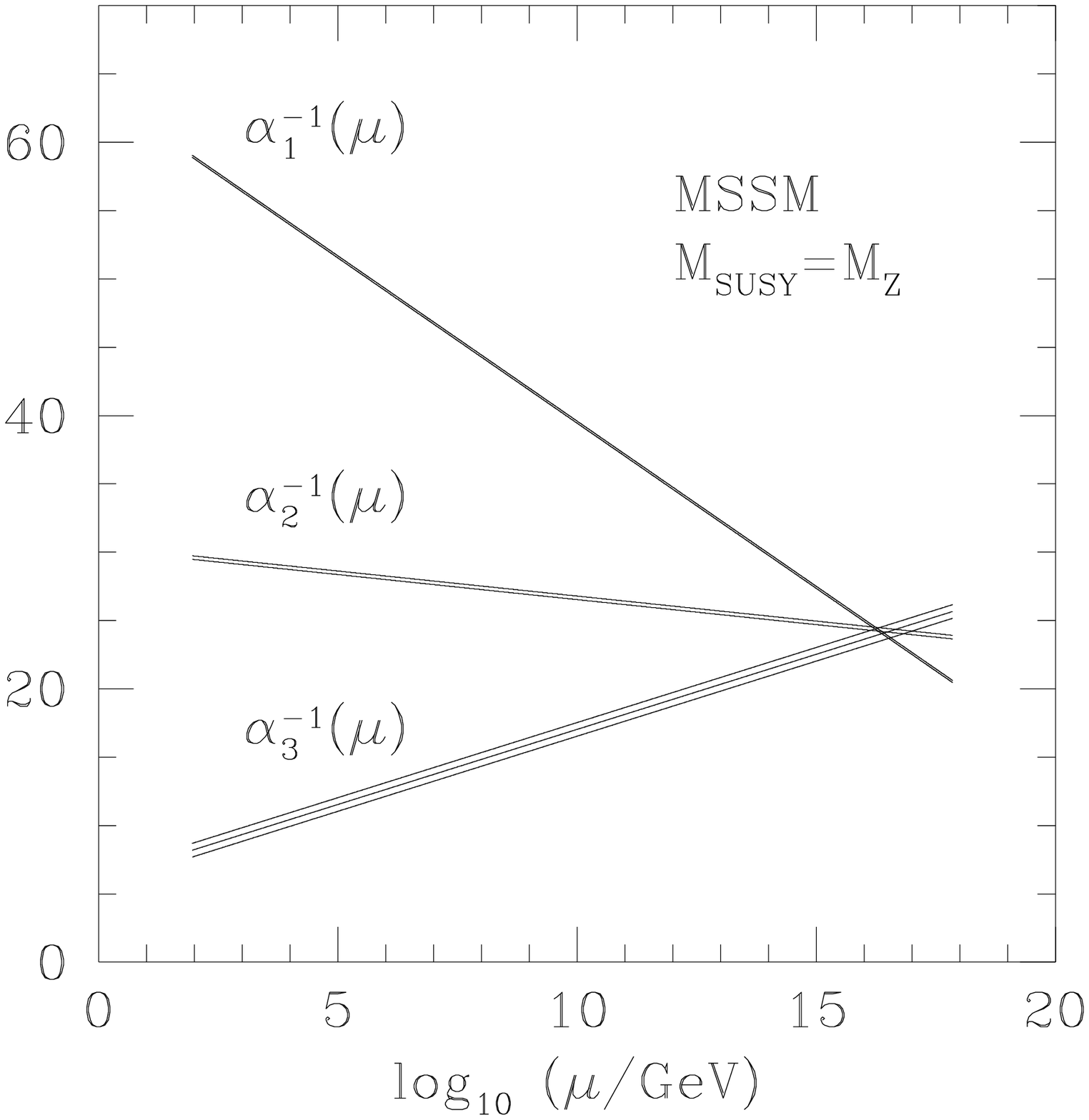}
    }
\caption{One-loop evolution of the gauge couplings within
   the Minimal Supersymmetric
   Standard Model (MSSM), assuming
   supersymmetric thresholds at the $Z$ scale.
   As in Fig.~\protect\ref{introfiga},
   $\alpha_1\equiv (5/3)\alpha_Y$, where $\alpha_Y$ is
   the hypercharge coupling in the conventional normalization.
   The relative width of each line reflects current experimental
   uncertainties.}
\label{introfigb}
\end{figure}

The Standard Model of particle physics is by now extremely
well-established, and accounts for virtually all presently
available experimental data.
Moreover, one particular extension of the Standard Model,
namely the Minimal Supersymmetric Standard Model (MSSM) \cite{SUSYreviews},
successfully incorporates the Standard Model within the
framework of a supersymmetric theory, thereby
improving the finiteness properties of the theory and
providing, for example, an elegant solution to
the technical gauge hierarchy problem.

This introduction of $N=1$
supersymmetry, however, also has another profound effect:
it brings about a unification of the gauge couplings,
as illustrated in Figs.~\ref{introfiga} and \ref{introfigb}.
This unification can be seen as follows.
At the $Z$ scale $M_Z\approx 91.16$ GeV,
the experimentally accepted
values for the hypercharge, electroweak, and strong gauge couplings
are respectively given
(within the $\overline{\rm MS}$ renormalization group scheme) as
\cite{exptcouplings}
\beqn
        \alpha_Y^{-1}(M_Z)|_{\overline{\rm MS}} &\equiv&  98.29 \pm 0.13
\nonumber\\
        \alpha_2^{-1}(M_Z)|_{\overline{\rm MS}} &\equiv&  29.61 \pm 0.13
\nonumber\\
        \alpha_3^{-1}(M_Z)|_{\overline{\rm MS}} &\equiv&  8.3 \pm 0.5 ~
\label{lowenergycouplingsa}
\eeqn
where $\alpha_i\equiv g_i^2/(4\pi)$ for $i=Y,2,3$.
In Eq.~(\ref{lowenergycouplingsa})
we have assumed the conventional hypercharge normalization
in which the Standard Model
right-handed singlet electron state $e_R$ has unit hypercharge.
We then extrapolate these couplings to higher energy scales
$\mu$ via the standard one-loop renormalization group equations (RGE's)
of the form
\beq
   \alpha_i^{-1}(\mu) ~=~ \alpha_i^{-1}(M_Z) ~-~
        {b_i\over 4\pi}\,\ln\,{\mu^2\over M_Z^2}~.
\label{mssmrge}
\eeq
Note that the one-loop beta-function coefficients $b_i$ that govern
this logarithmic running depend on the matter  content of the theory.
It is therefore here that the introduction of $N=1$ supersymmetry
plays a role ({\it i.e.}\/, by introducing superpartner states and
an extra Higgs doublet into the theory).
Specifically, one finds that these coefficients take the values
\beq
          (b_Y,b_2,b_3)~=~ \cases{
           ~(7,-3,-7) & within the Standard Model \cr
           ~(11,1,-3) & within the MSSM.\cr}
\eeq
Using the beta-function coefficients $b_i$ of the Standard Model and
extrapolating the low-energy couplings upwards according
to Eq.~(\ref{mssmrge}), one then finds that the three gauge couplings fail to
meet at any scale.
This is illustrated in Fig.~\ref{introfiga}.
By contrast, performing this extrapolation within the MSSM, one
discovers \cite{unifreview} an apparent unification of gauge couplings
of the form
\beq
           {5\over 3}\,\alpha_Y (M_{\rm MSSM}) ~=~
           \alpha_2 (M_{\rm MSSM}) ~=~
           \alpha_3 (M_{\rm MSSM}) ~\approx~ {1\over 25} ~
\label{MSSMunification}
\eeq
at the scale
\beq
    M_{\rm MSSM}~\approx~ 2\,\times\, 10^{16} ~ {\rm GeV}~.
\label{MSSMscale}
\eeq
This situation is shown in Fig.~\ref{introfigb}.
The fact that the gauge couplings unify within the MSSM is
usually interpreted as
evidence not only for $N=1$ supersymmetry,
but also for the existence of a single large grand-unified gauge
group $G_{\rm GUT}$
which breaks to $SU(3)\times SU(2)\times U(1)_Y$
at the MSSM scale $M_{\rm MSSM}$.
Indeed, with this interpretation, even the factor of $5/3$ appearing
in Eq.~(\ref{MSSMunification}) has a natural explanation, for it essentially
represents the group-theoretic factor by which the
conventional hypercharge generator
must be rescaled in order to be unified along with
the $SU(2)$ and $SU(3)$ generators within a single non-abelian
group $G_{\rm GUT}$ such as $SU(5)$, $SO(10)$, or $E_6$.

Thus, the popular field-theoretic scenario that is currently
envisioned is as follows.
At high energies far
above $M_{\rm MSSM}$, we have
$N=1$ supersymmetry and some grand-unified group $G_{\rm GUT}$,
with all matter falling into supersymmetric representations of this group.
Then, at the MSSM scale, this group is presumed to break
directly to $SU(3)\times SU(2)\times U(1)_Y$, and any extra
states that do not appear within the MSSM will have masses near $M_{\rm MSSM}$
and thus not affect the running of gauge couplings below this scale.
The MSSM itself is then presumed to govern physics all the way down to the
scale
$M_{\rm SUSY}$ at which SUSY-breaking occurs,
and then finally, below
$M_{\rm SUSY}$, we expect to see
merely the Standard-Model gauge group and spectrum.
The scale $M_{\rm SUSY}$ is set, of course, with two considerations
in mind:  it must be sufficiently high to explain
why the lightest superparticles have not yet been
observed, and it must be sufficiently low that the gauge hierarchy
is protected.
This in turn constrains various measures of SUSY-breaking, such
as the value of the mass supertrace ${\rm Str}(M^2)$.

On the face of it, this is a fairly compelling picture.
There are, however, a number of outstanding problems
that are not addressed within this scenario.  First, the
unification scale $M_{\rm MSSM}$ is quite close to the
Planck scale $M_{\rm Planck}=\sqrt{1/G_N} \approx 10^{19}$ GeV, yet
gravity is not incorporated into this picture.
Second, one would hope to explain the spectrum of the
Standard Model and the MSSM, in particular the values of
the many arbitrary free parameters which describe the fermion
masses and couplings.  Indeed, one might even seek
an explanation of more basic parameters such as
the number of generations or even the choice of gauge group.
Third, if we truly expect some sort of GUT theory above
$M_{\rm MSSM}$, we face the problem of the proton lifetime;
stabilizing the proton requires a successful doublet-triplet
splitting mechanism.
Finally, we may even ask why we should expect a GUT theory at
all.  After all, the appearance of a grand-unified theory is
essentially a theoretical prejudice, and is not required
in any way for the theoretical consistency of the model.
In other words, the unification of gauge couplings may just
be a happy accident.\footnote{
   Given that two non-parallel lines intersect in a point,
   it is arguably only a single coincidence that a
   third line intersects at the same point, and not a conspiracy
   between three separate couplings.  Of course,
   {\it a priori}\/, the unification scale thus obtained
     could have been lower than $M_Z$, or higher than
     $M_{\rm Planck}$.
   In a similar vein, we remark that the introduction of
   $N=1$ supersymmetry is not the only manner in which
   a unification of gauge couplings at $2\times 10^{16}$
   GeV can be achieved.  One alternate possibility starting
   from the {\it non-supersymmetric} Standard Model utilizes
   the introduction of extra multiplets at intermediate
   mass scales (see, {\it e.g.}, Ref.~\cite{framp});  another
   possibility (which leads to an even higher unification scale)
   will be discussed in Sect.~9.    }

\subsection{String theory vs.\ field theory}

String theory, however, has the potential to address all
of these shortcomings.  First, it naturally incorporates
quantum gravity, in the sense that a spin-two massless particle
(the graviton) always appears in the string spectrum.  Second, $N=1$
supersymmetric field theories with non-abelian gauge groups
and chiral matter
naturally appear as the low-energy limits of a certain phenomenologically
appealing class of string theories
(the {\it heterotic}\/ strings) \cite{heterotic}.
Third, such string theories may in principle provide a uniform framework for
understanding all of the features of low-energy phenomenology,
such as the appearance of three generations, the
fermion mass matrices, and even a doublet-triplet splitting
mechanism \cite{phensucc}.  Indeed, string theories ultimately contain
no free parameters!

But most importantly for the purposes of this article,
string theories also imply a natural unification of
the couplings.  Indeed, regardless of the particular string
model in question and independently of whether there exists any
unifying GUT gauge symmetry in the model, it turns out that
the gauge and gravitational couplings in heterotic string theory
 {\it always automatically unify}\/ at tree-level to
form one dimensionless coupling constant $g_{\rm string}$
\cite{Ginsparg}:
\beq
    8\pi {G_N\over \alpha'} ~=~ g_i^2\,k_i ~=~ g_{\rm string}^2 ~.
\label{unification}
\eeq
This unification relation holds because the gravitational
and gauge interactions all arise from the same underlying sectors
in the heterotic string, and can therefore be related to each other.
Here $G_N$ is the gravitational coupling (Newton's constant);
$\alpha'$ is the Regge slope (which sets the mass scale for
excitations of the string);
$g_i$ is the gauge coupling for each gauge group factor $G_i$;
and $k_i$, which appears as the corresponding normalization factor for
the gauge coupling $g_i$,
is the so-called  {\it affine level}\/
(also often called the {\it Ka\v{c}-Moody level}\/) at which the
group factor $G_i$ is realized.

These affine levels have a simple origin in string theory, and can
be understood as follows.
In (classical) heterotic string theory,
all gauge symmetries are ultimately realized in the form
of worldsheet affine Lie algebras with central
extensions \cite{KMreview}.
Explicitly, this means that if one computes the operator product
expansions (OPE's) between the worldsheet currents $J^a(z)$ corresponding
to any non-abelian group factor $G_i$ appearing in a given string model,
one obtains a result of the form
\beq
       J^a(z) J^b(w) ~\sim~ {if^{abc}\over z-w}\,J^c(w)
           ~+~ k \,{\vec \alpha_h^2\over 2}\, {\delta^{ab}\over (z-w)^2}~+~ ...
\label{OPE}
\eeq
Here $f^{abc}$ are the structure
constants of the Lie algebra, and  $\vec\alpha_h^2$ is the
squared length of the longest root $\vec\alpha_h$.
While the first term in Eq.~(\ref{OPE})
has the expected form of the usual Lie algebra,
the second term (the so-called ``central extension'')
appears as double-pole Schwinger contact term.
As indicated in Eq.~(\ref{OPE}),
the ``level'' $k$ is then defined as the coefficient of
this double-pole term,
and the gauge symmetry is said to have been ``realized at level $k$''.
(The specific definition of $k$ in the case of abelian groups
will be presented in Sect.~5.1.)
Current-algebra relations of the form in Eq.~(\ref{OPE}) are those
of so-called {\it affine Lie algebras}\/, which
are also often called \KM\ algebras in the physics literature.
Such algebras were first discovered by mathematicians in
Ref.~\cite{KacMoody}, and later independently by physicists in
Ref.~\cite{Halpern}.  The affine levels $k$ that concern us here
were first discovered in Ref.~\cite{Halpern}.
Note that the length $\vec\alpha_h^2$ is inserted into the definition
in Eq.~(\ref{OPE}) so that the level $k$ is invariant under trivial rescalings
of
the
currents $J^a(z)$.  For non-abelian gauge groups, it turns out that
the levels $k$ are restricted to be positive integers, while
for $U(1)$ gauge groups they can take arbitrary model-dependent values.

We see, then, that string theory appears to give us precisely
the features we want, with a prediction for the unification
of the couplings in Eq.~(\ref{unification}) that is strikingly
reminiscent of the ``observed'' MSSM unification in
Eq.~(\ref{MSSMunification}).
There are, however, some crucial differences between
the unification of gauge couplings in field theory and in string theory.
First, string theory, unlike field theory, is intrinsically a {\it finite}\/
theory;  thus when we talk of a {\it running}\/ of the gauge couplings
in string theory, we are implicitly calculating within the framework of
the low-energy {\it effective}\/ field theory which is derived from
only the massless ({\it i.e.}, observable) states of the full string spectrum.
Second, again in contrast to field theory,
in string theory all couplings are actually {\it dynamical variables}\/
whose values are fixed by the expectation values of certain {\it moduli}\/
fields.
For example, the string coupling $g_{\rm string}$ is related at tree-level
to the VEV of a certain modulus, the {\it dilaton}\/ $\phi$, via
a relation of the form $g_{\rm string}\sim e^{-\langle \phi \rangle}$
\cite{Witten}.
These moduli fields --- which are massless gauge-neutral
Lorentz-scalar fields --- essentially parametrize an entire space of
possible ground states (or ``vacua'') of the string,
and have an effective potential which is classically flat
and which remains flat to all orders in perturbation theory.
For this reason one does not know, {\it a priori}\/, the value of the
string coupling $g_{\rm string}$ at unification, much less the true
string ground state from which to perform our calculations in the first place.

A third distinguishing feature between field-theoretic and string-theoretic
unification concerns the presence of the affine levels $k_i$
in the unification relation (\ref{unification}).
It is clear from Eqs.~(\ref{unification}) and (\ref{OPE}) that these factors
essentially appear as normalizations for the gauge couplings $g_i$
(or equivalently for the gauge symmetry currents $J_a$), and indeed
such normalizations are familiar from ordinary field-theoretic
GUT scenarios such as those based on $SU(5)$ or $SO(10)$ embeddings
in which the hypercharge generator $Y$ must be rescaled by a factor $k_Y=5/3$
in order to be unified within the larger GUT symmetry group.
The new feature from string theory, however, is that such normalizations
$k_i$ now also appear for the {\it non-abelian}\/ gauge factors as well.
We shall see, however, that the most-easily constructed string models
have $k_i=1$ for the non-abelian gauge factors.

But once again,
for the purposes of this article, the most important difference
between gauge coupling unification in field theory and string theory
is the {\it scale}\/ of the unification.
As we have already discussed, in field theory this scale is determined
via an extrapolation of the measured low-energy couplings within the framework
of the MSSM,
ultimately yielding $M_{\rm MSSM}\equiv 2\times 10^{16}$ GeV.
This number, deduced on the basis of experimental measurement of the
low-energy couplings, appears without theoretical justification.
In string theory, by contrast, the scale of unification $M_{\rm string}$
is fixed by the intrinsic scale of the theory itself.
Since string theory is a theory of quantum gravity, its natural
scale $M_{\rm string}$ is ultimately related to the Planck scale
\beq
           M_{\rm Planck}~=~\sqrt{1/ G_N}
            ~\approx~ 1.22\times 10^{19} ~{\rm GeV}~,
\eeq
and is given by
\beq
    M_{\rm string} ~ = ~ g_{\rm string}\, M_{\rm Planck} ~\sim ~
\sqrt{1/\alpha'}
\eeq
where $g_{\rm string}$ is the string coupling.
For realistic string models,
the string coupling $g_{\rm string}$ should be
of order one at unification;  this is not only necessary
for rough agreement with experiment
at low energies, but also guarantees
that the effective four-dimensional field theory will be
weakly coupled \cite{scales}, making a perturbative analysis appropriate.
At tree-level, therefore, $M_{\rm string}$ is the scale at which the
unification
in Eq.~(\ref{unification}) is expected to take place.  One-loop string effects
have the potential to lower this scale somewhat,
however,
and indeed one finds \cite{Kaplunovsky}
that in the $\overline{\rm DR}$ renormalization scheme\footnote{
   The $\overline{\rm DR}$ scheme is the modified minimal
   subtraction scheme for dimensional reduction, wherein
   Dirac $\gamma$-matrix manipulations are performed in
   four dimensions.  This scheme therefore preserves
   spacetime supersymmetry in loop calculations.},
the scale of string unification is shifted down to
\beqn
     M_{\rm string} &=& {e^{(1-\gamma)/2} \,3^{-3/4}
       \over 4\pi} \,g_{\rm string}\, M_{\rm Planck}\nonumber\\
       &\approx& g_{\rm string}\,\times\,5.27\,\times\,10^{17}~~{\rm GeV}~
\label{Mstringg}
\eeqn
where $\gamma\approx 0.577$ is the Euler constant.
Thus, assuming that $g_{\rm string}\sim {\cal O}(1)$ at unification,
we have
\beq
     M_{\rm string}~\approx~ 5\,\times\,10^{17}~~{\rm GeV}~.
\label{Mstring}
\eeq
We thus see that a factor of approximately 20 or 25 separates the
predicted string unification scale from the MSSM unification scale.
Equivalently, the logarithms of these scales (which are arguably
the true measure of this discrepancy) differ by about 10\% of their
absolute values.

\begin{figure}[th]
\centerline{
   \epsfxsize 3.3 truein \epsfbox {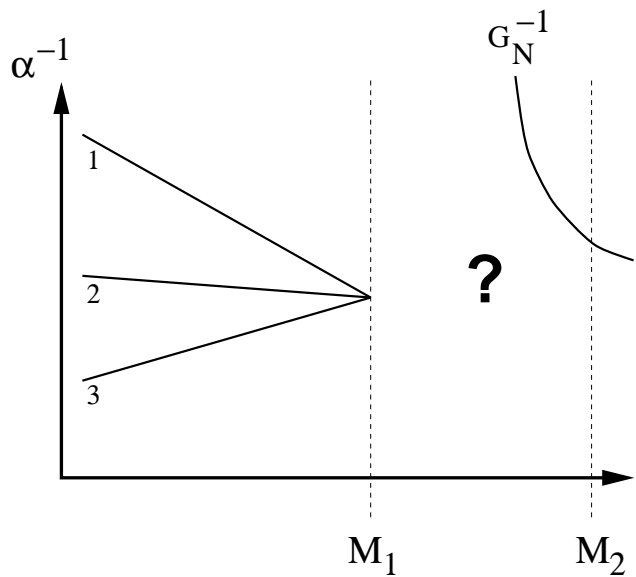}
    }
\caption{The fundamental problem of gauge coupling
   unification within string theory:  the scale
   $M_1\equiv M_{\rm MSSM}\approx 2\times 10^{16}$ GeV
   at which the gauge couplings are expected to unify within the MSSM is
   significantly below the scale $M_2\equiv M_{\rm string}\approx
   5\times 10^{17}$ GeV at which string theory predicts
   their unification with each other and with the
   gravitational coupling (Newton constant) $G_N$.}
\label{couplings1}
\end{figure}

This situation is sketched in Fig.~\ref{couplings1}, where we compare the
lower scale $M_1\equiv M_{\rm MSSM}$ at which the extrapolated gauge
couplings unify with the higher scale $M_2\equiv M_{\rm string}$ at
which string theory predicts their unification with each other and with
the gravitational coupling $G_N$.
There are several important comments to make regarding this figure.
First, since $G_N$ has mass dimension $-2$, its running
is dominated by {\it classical}\/ effects which are stronger than the purely
quantum-mechanical running experienced by the dimensionless gauge couplings
$\alpha_i$.  In order to see this explicitly, let us first recall that the
dimensionless
gauge couplings experience a scale dependence of the form
$\alpha_i^{-1}(\mu)\sim \mu^{f_i} \tilde\alpha_i^{-1}$ where $\tilde\alpha_i$
are {\it fixed}\/ numbers describing the strength of the gauge couplings,
and where the functions $f_i$ (which are
similar to anomalous dimensions) describe the quantum-mechanical
scale-dependence
of the gauge couplings.  It is, of course, the scale-dependent quantities
$\alpha_i^{-1}(\mu)$ that we have been considering all along, and we see that
if we relate $\alpha_i^{-1}(\mu)$ to $\alpha_i^{-1}(M_Z)$ by expanding $f_i$
to first order in the gauge couplings via $f_i=b_i \alpha_i(M_Z)/2\pi+...$,
we reproduce the one-loop renormalization group equations given in
Eq.~(\ref{mssmrge}). The situation for the gravitational coupling is similar.
We first define an analogous {\it dimensionless}\/ gravitational coupling
$G_N^{-1}(\mu)\sim \mu^{-2+f_G} \tilde G_N^{-1}$ where $\tilde G_N$ is the
usual fixed dimensionful gravitational coupling (Newton constant), and where
the
two terms in the exponent $-2+f_G$ respectively represent the classical and
quantum-mechanical contributions to the running.  It is appropriate to consider
this dimensionless gravitational coupling $G_N^{-1}(\mu)$ rather than $\tilde
G_N^{-1}$
since it is $G_N^{-1}(\mu)$ which represents the effective strength of the
gravitational
interaction at the scale $\mu$.  Since $f_G$ is proportional to $G_N$ and is
therefore
exceedingly small at energies below the Planck scale, we can disregard $f_G$
entirely and concentrate on the classical contribution.  This yields the
power-law
scale dependence $G_N^{-1}(\mu)/G_N^{-1}(M_Z)= (M_Z/\mu)^2$, which gives rise
to an
exponential curve when sketched relative to a logarithmic mass scale as in
Fig.~\ref{couplings1}.  In this context, it is also interesting
to note that any possible unification of gauge and gravitational couplings must
necessarily
take the form $G_N(\mu)=\alpha_i(\mu)$ (where we are neglecting overall
numerical factors),
since only couplings of similar mass dimensionalities can be equated.   Upon
comparison
with the string-theoretic tree-level unification prediction in
Eq.~(\ref{unification}), we then
immediately find that such a unification is possible only when
$\mu\sim 1/\sqrt{\alpha'}\sim M_{\rm string}$.
Thus, because string theory essentially relates a dimensionless gauge
coupling to a dimensionful gravitational coupling, it has the property that
the unification relation itself predicts the unification scale.
Of course,
the result $\mu\sim M_{\rm string}$ is merely the tree-level result,
while the one-loop corrected unification scale is given in
Eq.~(\ref{Mstringg}).
Finally, also note that for illustrative purposes we have greatly exaggerated
the difference between $M_{\rm MSSM}$ and $M_{\rm string}$ in this sketch.

\begin{figure}
\centerline{
   \epsfxsize 3.3 truein \epsfbox {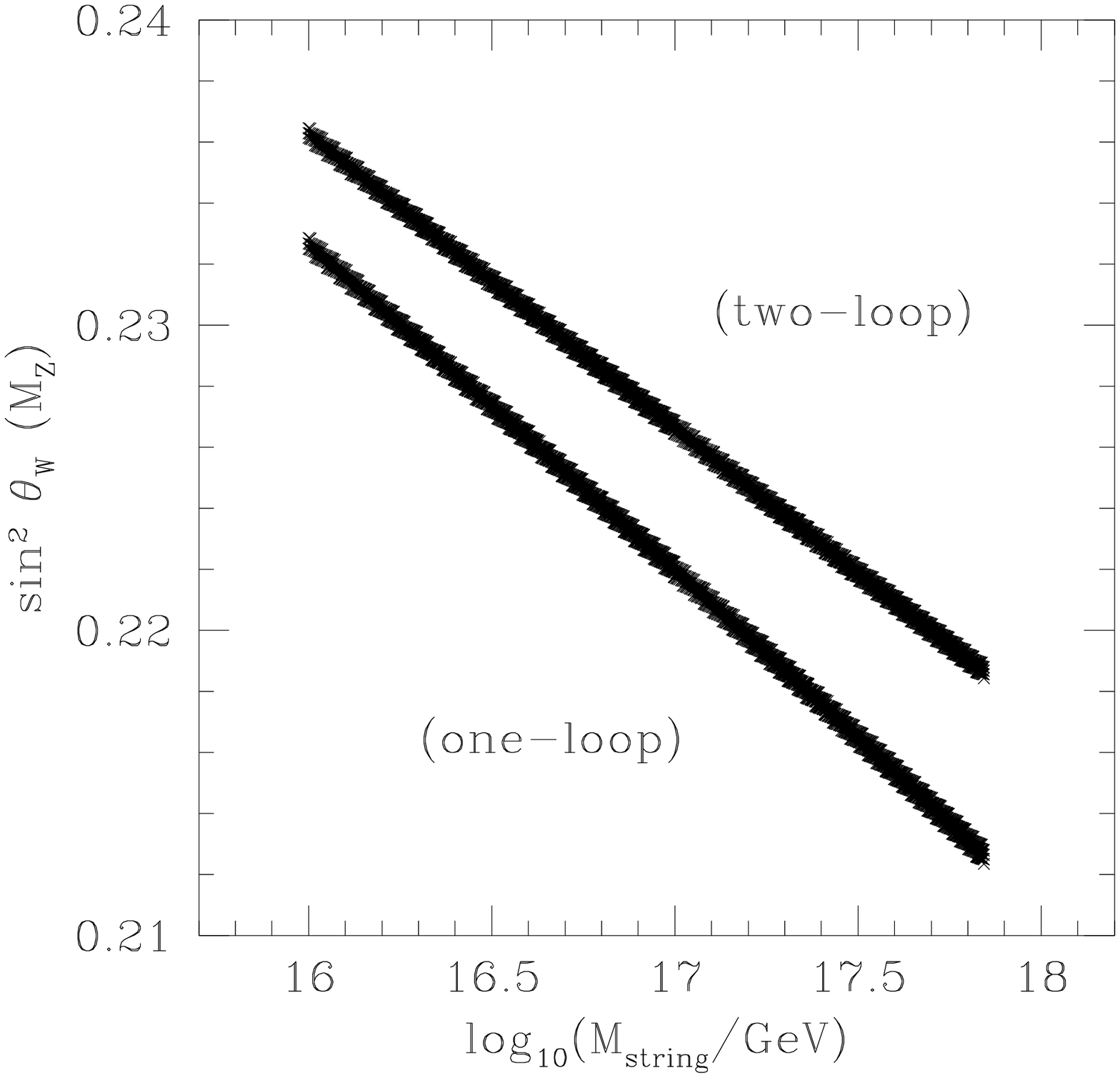}
    }
\caption{Dependence on the string scale:  the predicted value of
     the low-energy electroweak mixing angle $\sin^2\theta_W(M_Z)$,
    assuming unification at $M_{\rm string}$.
    Results for both one-loop and two-loop running are plotted.  }
\label{couplings_vs_Ma}
\vskip 0.25 truein
\centerline{
   \epsfxsize 3.3 truein \epsfbox {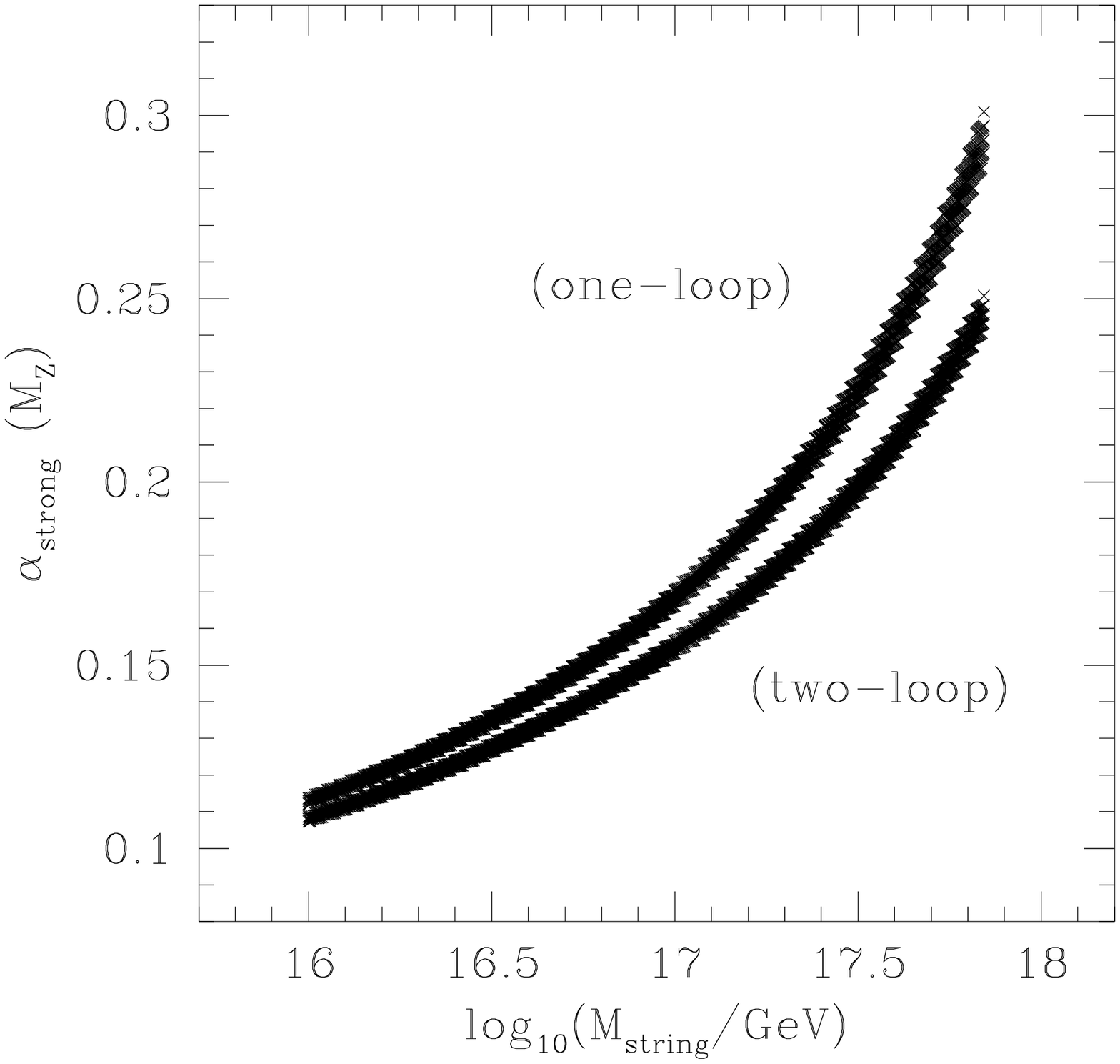}
    }
\caption{Dependence on the string scale:
    the predicted value of the low-energy coupling $\alpha_{\rm strong}(M_Z)$,
    assuming unification at $M_{\rm string}$.
    Results for both one-loop and two-loop running are plotted.  }
\label{couplings_vs_Mb}
\end{figure}

One may argue that this discrepancy between the two unification scales
is not a major problem, since it is
only a 10\% effect in terms of their logarithms.
More generally,
one may also argue that it is improper to worry
about a discrepancy between {\it scales}\/ of unification, since
such scales are dependent upon the particular renormalization scheme
employed, and hence have no physical significance.
Both of these observations are of course true,
but the gauge couplings themselves {\it are}\/ physical quantities, and
this discrepancy between $M_{\rm string}$ and $M_{\rm MSSM}$ implies that
the hypothesis of string-scale unification yields {\it incorrect}\/ values for
low-energy couplings.
In other words, if we take the predictions of string theory seriously
and assume that the gauge couplings
unify at $M_{\rm string}$ rather than at $M_{\rm MSSM}$, we find
that a straightforward extrapolation down to low energies
does not reproduce the correct values for these
couplings at the $Z$ scale.

In order to appreciate the seriousness of this problem,
recall that the experimentally measured values
for these couplings at the $Z$ scale
$M_Z\approx 91$ GeV in the $\overline{\rm MS}$
renormalization scheme were given in
Eq.~(\ref{lowenergycouplingsa}).
Because the value of the electromagnetic coupling
$\alpha_{\rm e.m.}(M_Z)=1/127.9$
is known with great precision and is trivially
related to the electroweak couplings $(\alpha_Y,\alpha_2)$
via the electroweak mixing angle
\beqn
    \alpha_Y^{-1} & =& \alpha_{\rm e.m.}^{-1} \, \cos^2\theta_W\nonumber\\
    \alpha_2^{-1} & =& \alpha_{\rm e.m.}^{-1} \, \sin^2\theta_W~,
\eeqn
it is traditional to take $\alpha_{\rm e.m.}(M_Z)$ as a fixed input
parameter and quote the values of $(g_Y,g_2)$
in terms of the single electroweak mixing angle $\sin^2\theta_W$.
Indeed, using Eq.~(\ref{lowenergycouplingsa}),
we then obtain the low-energy coupling parameters
\beqn
      \sin^2\theta_W(M_Z)\vert_{\overline{\rm MS}}&=&
             0.2315\pm0.001~\nonumber\\
      \alpha_{3}(M_Z)\vert_{\overline{\rm MS}} &=& 0.120\pm0.010~,
\label{lowenergycouplings}
\eeqn
and we shall see later that conversion from the $\overline{\rm MS}$ scheme to
the $\overline{\rm DR}$ scheme makes only a very small correction.
By contrast, in Figs.~\ref{couplings_vs_Ma} and \ref{couplings_vs_Mb}
we have plotted the values of the same low-energy coupling parameters
that one would obtain by assuming unification at different hypothetical values
of $M_{\rm string}$,
and then running down to low energies according to both one-loop and two-loop
analyses within the MSSM.
It is clear that for $M_{\rm string}$ taking the value indicated
in Eq.~(\ref{Mstring}), the predicted values of $\sin^2 \theta_W(M_Z)$
and $\alpha_{3}(M_Z)$ deviate
by many standard deviations from those that are measured.
Thus, what may have seemed to be a minor discrepancy between two
unification
scales becomes in fact a major problem for string phenomenology.


\setcounter{footnote}{0}
\section{Possible Paths to String-Scale Unification}

Faced with this situation, a number of possible ``paths'' towards
reconciliation have been proposed.  We shall here summarize the basic
features of each path,
and devote the following sections to
more detailed examinations of the issues involved in each.

\subsection{Overview of possible paths}

\begin{itemize}

\item \underbar{\sl String GUT models}\/:~
 Perhaps the most obvious path towards reconciling the scale
of string unification $M_{\rm string}$ with the apparent unification
scale $M_{\rm MSSM}$ is through the assumption of an intermediate-scale
unifying gauge group $G$, so that $SU(3)\times SU(2)\times U(1)_Y \subset G$.
In this scenario,
the gauge couplings would still unify at the intermediate scale $M_{\rm MSSM}$
to form the coupling $g_G$ of the intermediate group $G$,
but then $g_G$ would run up to the string scale $M_{\rm string}$ where
it would then unify, as required, with the gauge couplings of
any other (``hidden'') string gauge groups and with the gravitational coupling.
As far as string theory is concerned, this basic path to unification
has two possible sub-paths, depending on whether the GUT group $G$
is simple [as in $SU(5)$ or $SO(10)$], or non-simple
[as in the Pati-Salam unification scenario with $G=SO(6)\times SO(4)$,
or the flipped $SU(5)$ scenario with $G=SU(5)\times U(1)$].
We shall refer to the first path as the ``strict GUT'' path,
and the second as merely involving ``intermediate-scale gauge structure''.
As we shall see,
these two sub-paths have drastically different
stringy consequences and realizations.

\item \underbar{\sl Non-standard affine levels and
                 hypercharge normalizations}\/:~
A second possible path to unification retains the MSSM gauge
structure all the way up to the string scale,
and instead exploits the
fact that in string theory, the hypercharge normalization $k_Y$ need
not have the standard value $k_Y=5/3$ that it has in those GUT
scenarios which make use of $SU(5)$ or $SO(10)$ embeddings.  Indeed,
in string theory, the value that $k_Y$ may take is {\it a priori}\/
arbitrary.
Likewise, in all generality, the affine levels $(k_2,k_3)$ that describe the
non-abelian group factors $SU(2)$ and $SU(3)$ of the MSSM may also differ
from their ``usual'' value $k_2=k_3=1$, and therefore it is
possible that by building string models which realize
appropriately  chosen non-standard values of $(k_Y,k_2,k_3)$,
one can alter the running of the corresponding couplings in such a way
as to realize gauge coupling unification at $M_{\rm string}$ while
simultaneously
obtaining the proper values of the low-energy parameters
$\alpha_{\rm strong}$ and  $\sin^2\theta_W$.
Thus, this possibility would represent a purely string-theoretic effect.
As we shall discuss, however, it remains an open question whether
realistic string models with the required values of $(k_Y,k_2,k_3)$ can
be constructed.

\item \underbar{\sl Heavy string threshold corrections}\/:~
The next three possible paths to string-scale unification
all involve adding various ``correction terms'' to the renormalization
group equations (RGE's) of the MSSM.
For example, the next possible path we shall discuss involves the
so-called {\it heavy string threshold corrections}\/ which represent
the contributions from the infinite towers of massive ({\it i.e.},
Planck-scale)
string states that are otherwise neglected in an analysis of the purely
low-energy massless string spectrum.  Strictly speaking, such corrections
must be included in any string-theoretic analysis, and it is possible that
these corrections may be sufficiently large to reconcile string-scale
unification with the observed low-energy couplings.  Thus, like the
non-standard affine levels and hypercharge normalizations, this
too represents a purely stringy effect that would not arise
in ordinary field-theoretic scenarios.

\item \underbar{\sl Light SUSY thresholds}\/:~
A fourth possible path to unification involves
the corrections due to light SUSY-breaking thresholds near the
electroweak scale.
While the plot in Fig.~\ref{introfigb} assumes that the
superpartners of the MSSM states all have equal masses
at $M_Z$, it is expected that realistic SUSY-breaking
mechanisms will yield a somewhat different sparticle
spectroscopy.  The light SUSY-breaking thresholds
are the corrections that would be needed in order to
account for this, and are typically analyzed in purely
field-theoretic terms.

\item \underbar{\sl Extra non-MSSM matter}\/:~
A fifth possible path towards reconciling the string unification
scale with the MSSM unification scale involves the corrections due to
possible extra exotic matter beyond the MSSM.
Although introducing such additional matter is
completely {\it ad hoc}\/ from the field-theoretic point of view,
such matter appears naturally in many realistic string models, and
is in fact {\it required}\/ for their self-consistency.
Such matter also has the potential to significantly alter the
naive string predictions of the low-energy couplings, and
its effects must therefore be included.

\item \underbar{\sl Strings without supersymmetry}\/:~
Another  possible path towards unification, one which
is highly unconventional and which lies outside the paradigm of the MSSM,
involves strings {\it without}\/ supersymmetry --- {\it i.e.}, strings
which are non-supersymmetric at the Planck scale.  Such string models therefore
seek to reproduce the Standard Model, rather than the MSSM,  at low energies.
As a result of the freedom to adjust the hypercharge normalization
that exists within the string framework, it turns out that gauge coupling
unification can still be achieved in such models,
even without invoking supersymmetry.  Moreover,
the scale of gauge coupling unification in this scenario
surprisingly turns out to be
somewhat closer to the string scale
than it is within the MSSM.

\item \underbar{\sl Strings at strong coupling}\/:~
Finally, there also exists another possible path to unification
which --- unlike those above ---
is intrinsically non-perturbative, and which makes use
of some special features of the strong-coupling behavior
of strings in ten dimensions.

\end{itemize}

\begin{figure}
\centerline{ \epsfxsize 3.5 truein \epsfbox {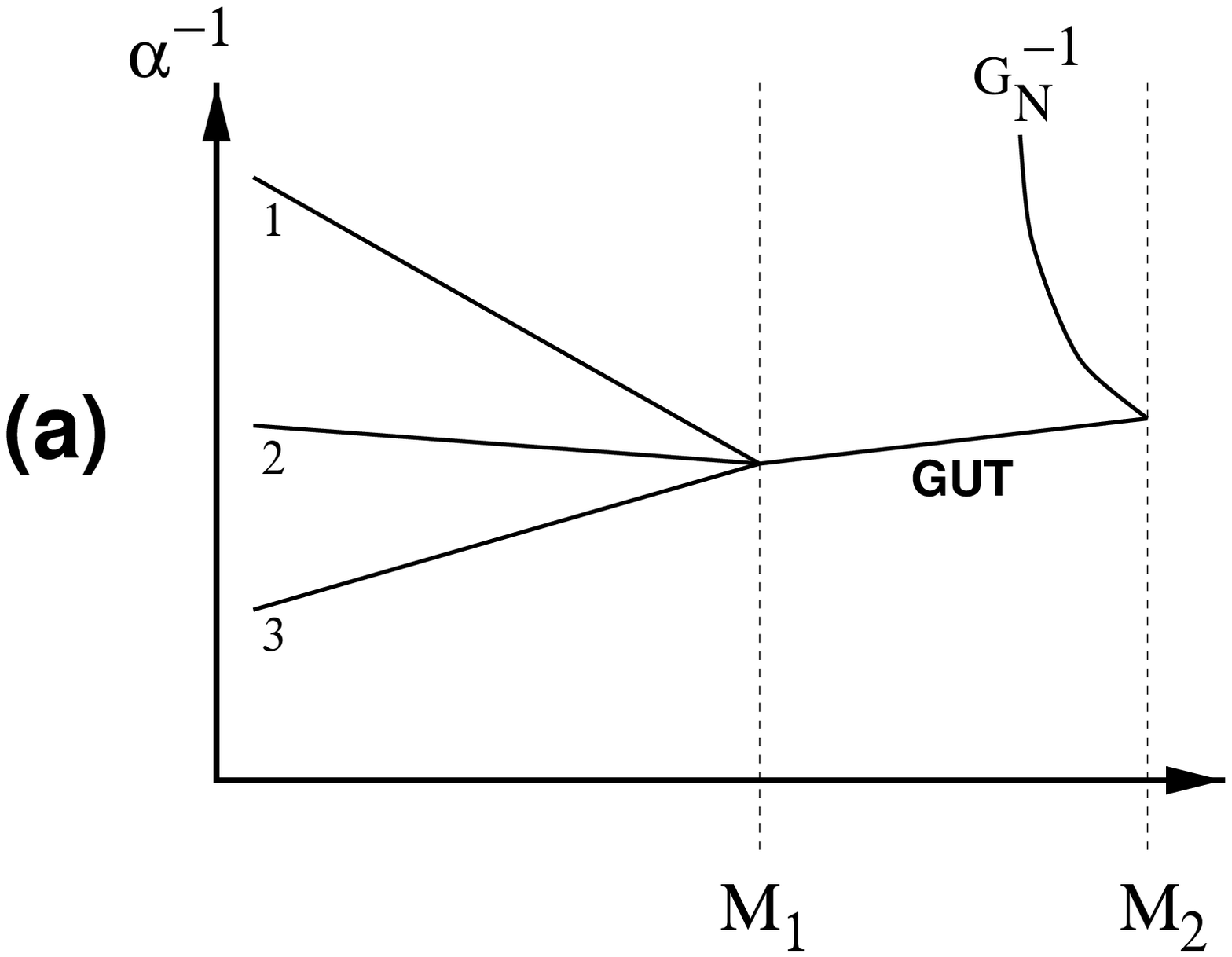}
              \hskip 0.10 truein
             \epsfxsize 3.5 truein \epsfbox {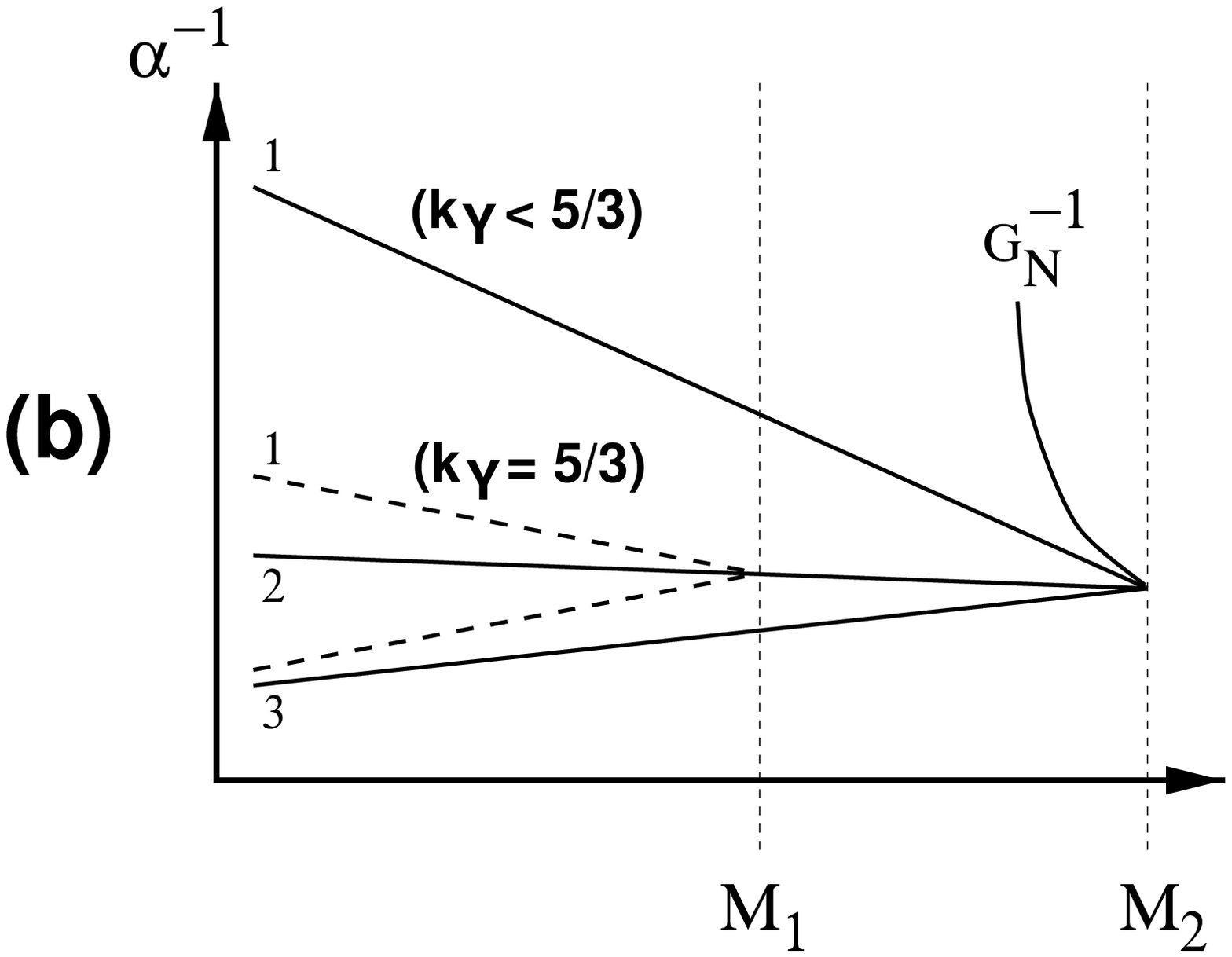} }
\centerline{ \epsfxsize 3.5 truein \epsfbox {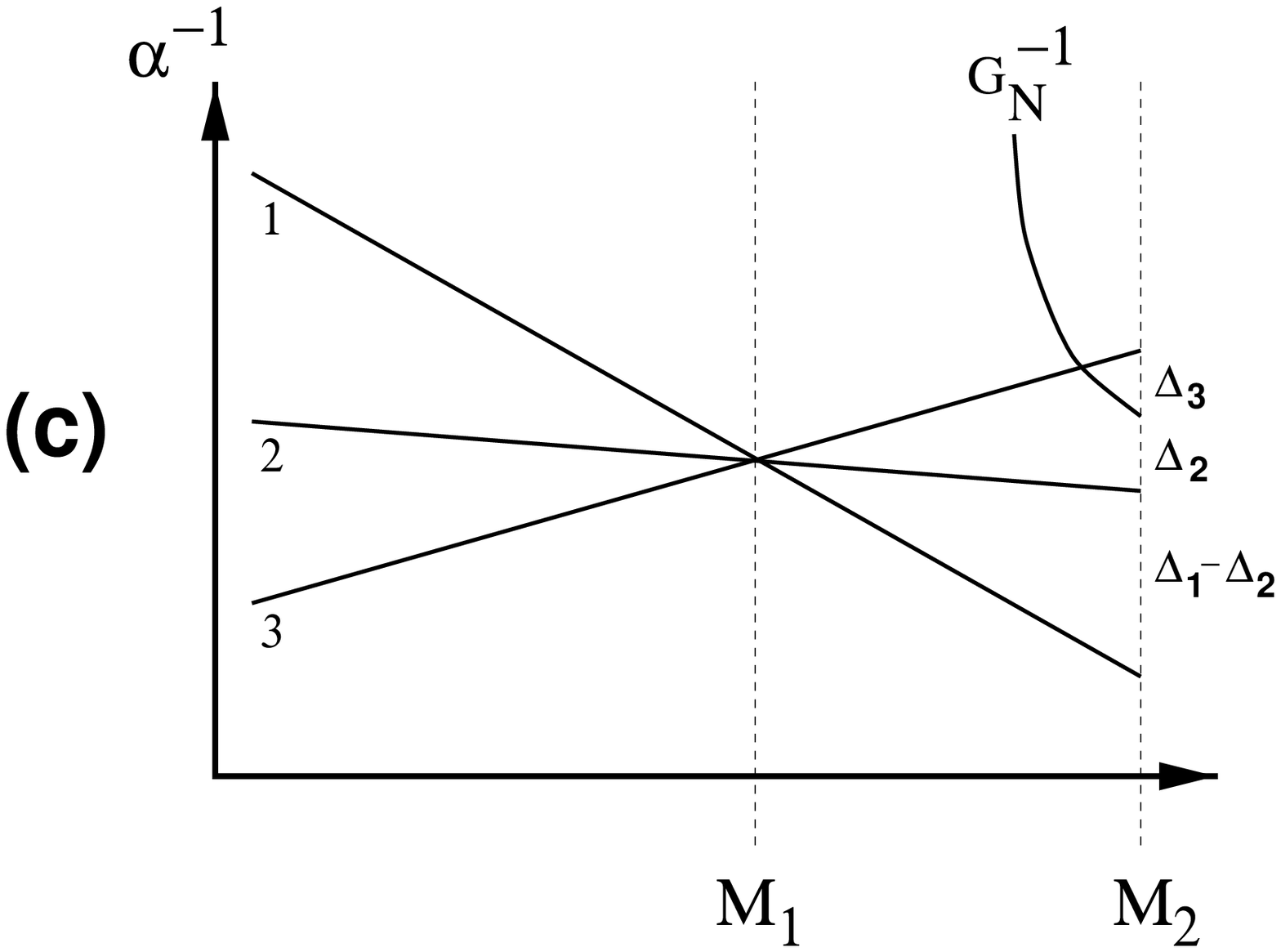}
              \hskip 0.10 truein
             \epsfxsize 3.5 truein \epsfbox {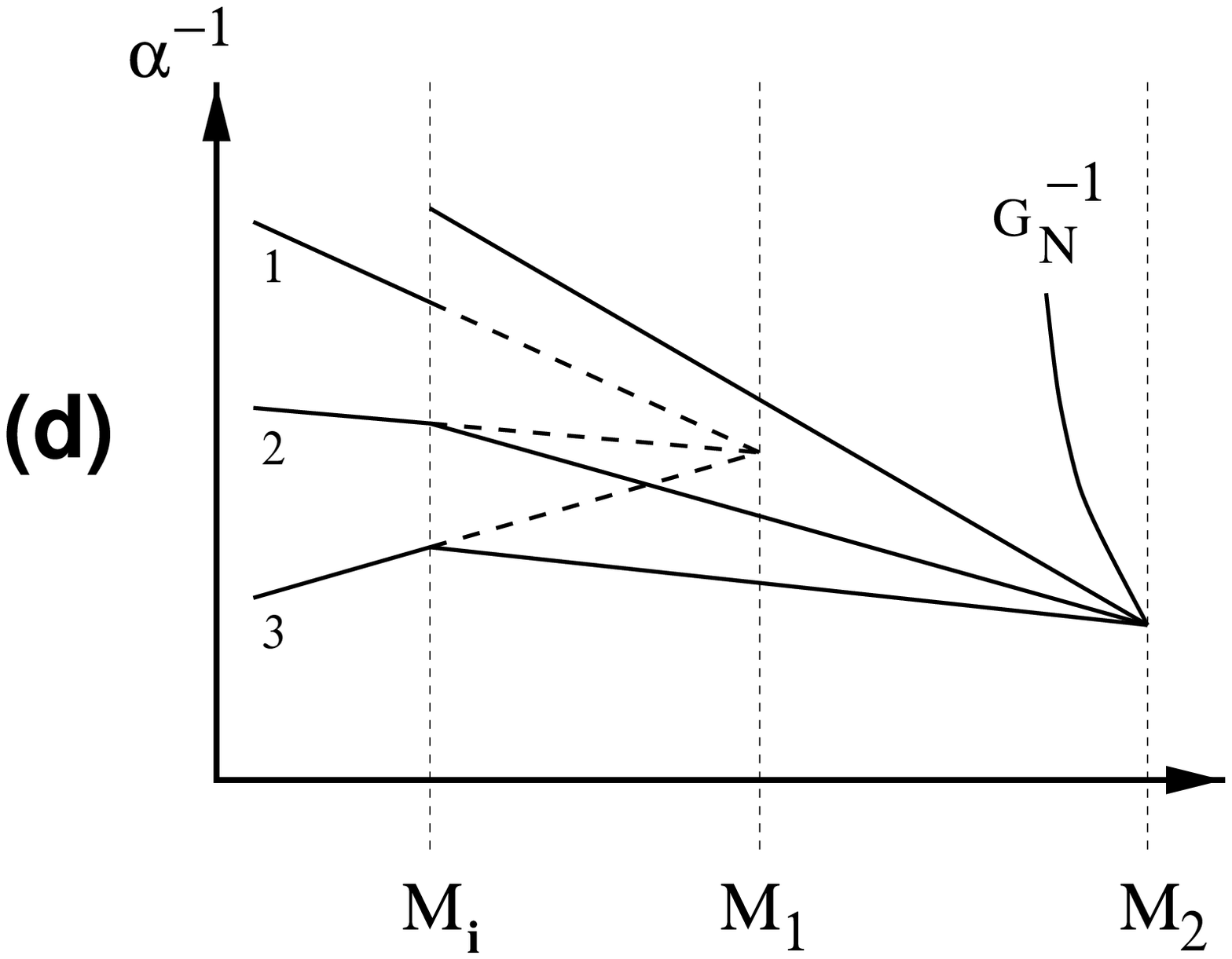} }
\centerline{ \epsfxsize 3.5 truein \epsfbox {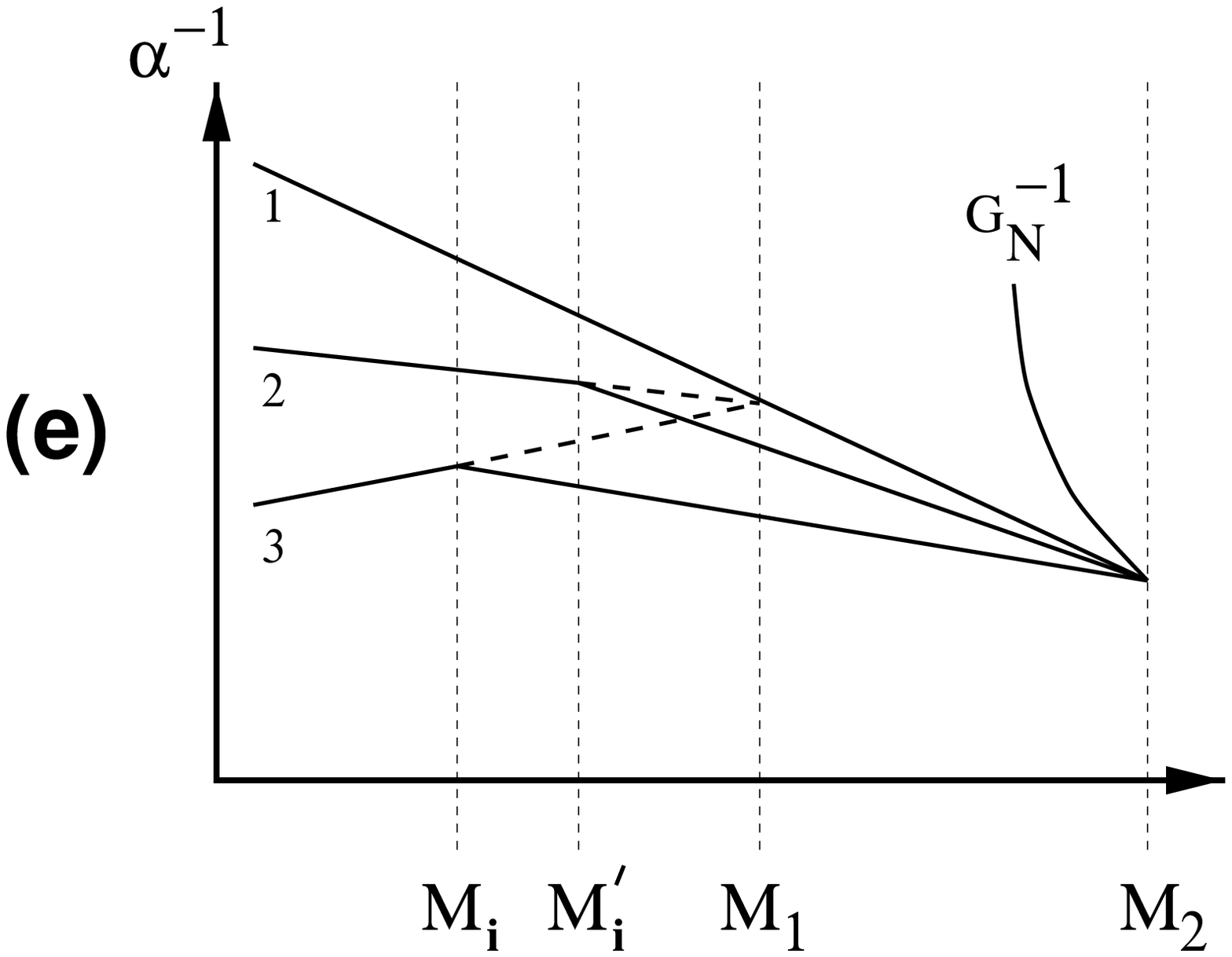}
              \hskip 0.10 truein
             \epsfxsize 3.5 truein \epsfbox {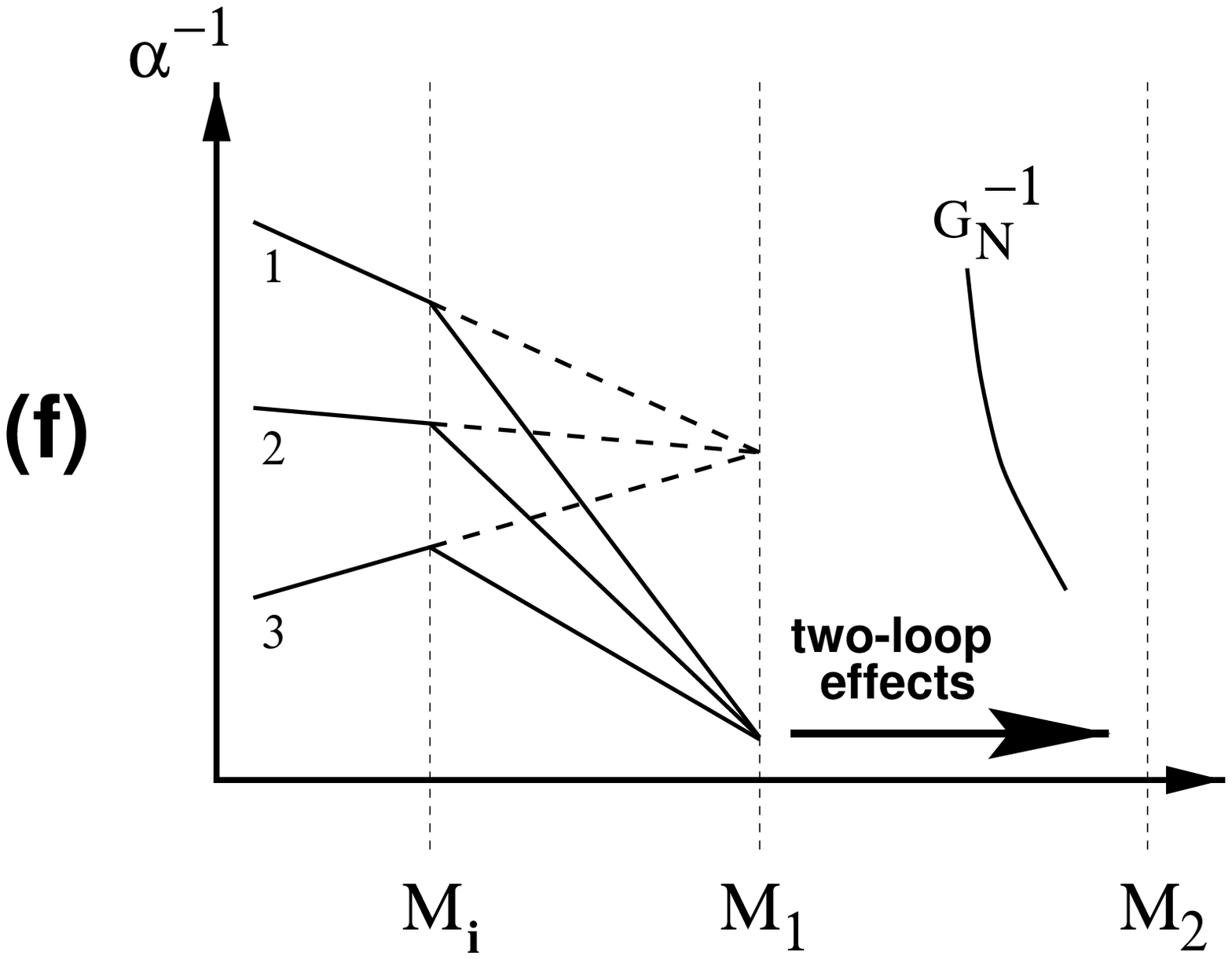} }
\caption{Various paths to unification, as discussed in the text.
   Each path provides a different solution to the fundamental
    problem posed in Fig.~\protect\ref{couplings1}.}
\label{paths}
\end{figure}

The above ``paths to unification''
are clearly very different from each other, and
thus imply different resolutions to the fundamental
problem posed in Fig.~\ref{couplings1}.
In Fig.~\ref{paths},
we have sketched some of these different resolutions;
we remind the reader that once again
these sketches are greatly exaggerated and are meant only to illustrate
the basic scenarios.
First, in Fig.~\ref{paths}(a),
we illustrate the standard GUT resolution in which
the three low-energy couplings emerge at $M_1\equiv M_{\rm MSSM}$
from a common GUT coupling which in turn unifies with the gravitational
coupling at $M_2\equiv M_{\rm string}$.
In Fig.~\ref{paths}(b),
by contrast, we illustrate the approach based on modifications
of the levels $k_i$.  [Recall that in all of these sketches we are
actually plotting not $\alpha_i^{-1}$ but $(k_i \alpha_i)^{-1}$.]
It is already clear from this sketch
that successful string-scale unification generally requires decreasing $k_Y$
and increasing $k_3$;  for convenience we have held $k_2$ constant
in this plot.
Next, in Fig.~\ref{paths}(c),
we sketch the scenario based on heavy string
threshold corrections.  In this scenario, the exact tree-level
unification relation (\ref{unification}) at $M_{\rm string}$ is corrected by
fixed thresholds $\Delta_i$ which must have specific
sizes and signs if they are to consistent with the observed
gauge coupling unification at $M_{\rm MSSM}$.
For example, it is already clear that we would require $\Delta_1-\Delta_2$
to be negative and $\Delta_3-\Delta_2$ to be positive.

The remaining sketches show the influence of
possible non-trivial physics at intermediate scales $M_i$
between $M_Z$ and $M_{\rm MSSM}$.
In Fig.~\ref{paths}(d),
for example, we illustrate a scenario based on
possible intermediate-scale gauge structure [such
as, {\it e.g.}\/, $SO(6)\times SO(4)$].
In such scenarios, the low-energy gauge couplings emerge only
at the intermediate scale $M_i$ at which the larger gauge
symmetry is broken, and the $U(1)$ gauge coupling generally
experiences a discontinuity.
Likewise, in Figs.~\ref{paths}(e) and \ref{paths}(f),
we show the effects of extra {\it matter}\/ beyond the MSSM.
In Fig.~\ref{paths}(e), we illustrate
the effects of potential extra color triplets and electroweak doublets
appearing at intermediate scales $M_i$ and $M'_i$ respectively;
for simplicity we have assumed that these extra states have
vanishing hypercharge.
It is clear that the effect of such extra matter states is essentially that of
a corrective lens which ``refocuses'' the running of the gauge couplings so
that they meet at $M_{\rm string}$ rather than at $M_{\rm MSSM}$.
(Light SUSY thresholds also have a similar effect.)
In Fig.~\ref{paths}(f),
by contrast, we show the effect of extra matter in
 {\it complete}\/ GUT multiplets.  While such multiplets do not
affect the unification {\it scale}\/ to one-loop order, we have sketched
a scenario in which they raise the unified
 {\it coupling}\/ to such an extent that two-loop effects may
become significant and then refocus the unification scale up to the
string scale.

It is also possible to sketch the two remaining ``paths to unification''
that we have mentioned above.  In particular, the scenario based upon strings
 {\it without}\/ supersymmetry will be illustrated in Sect.~9 (see
Fig.~\ref{nonsusystringfig} for a precise plot), while the scenario based
upon string non-perturbative effects is much more subtle, and
essentially weakens the string unification predictions in such
a way that the unification scale for the gauge couplings need no longer
coincide with the string scale derived from the gravitational
coupling.  Thus, in this scenario, the original mismatch
illustrated in Fig.~\ref{couplings1} would continue to apply, but would
no longer be viewed as problematic.

\subsection{Choosing between the paths:  General remarks}

Given these many potential ``paths to unification'', we are then left with one
over-riding question:  {\it Which path to string-scale gauge coupling
unification
does string theory actually take?}
Or equivalently, we may ask:  {\it To what extent can realistic string models
be constructed which exploit each of these possibilities?}

These questions are far more subtle
in string theory than they would be in ordinary field theory.
Indeed, in field theory, it would not seem to be too difficult to construct
models which exploit the GUT mechanism, or which introduce extra matter beyond
the MSSM.
In string theory, however, the situation is far more complicated because
these different paths are often related to each other in deep ways
that are not always immediately apparent.  For example, as we shall
see, any consistent realization of a ``strict GUT'' string
model requires that the GUT gauge symmetry $G$ be realized with an affine
level $k_G>1$.
Likewise, the presence of non-standard hypercharges $k_Y$ in string theory
can be shown to imply the existence of certain classes of exotic
non-MSSM matter with fractional electric charge.
Thus, the different ``paths to unification'' as we have outlined
them are not independent of each other, and we expect that
various combinations of all of these effects will play a role in
different types of string models.

It is important to understand why such
such unexpected connections arise in string theory,
and why such seemingly disparate features such as
affine levels, extra non-MSSM matter, and GUT gauge groups
are all ultimately tied together.
These connections essentially
occur because the fundamental object in string theory is the {\it string}\/
itself.  Each of the different worldsheet modes of
excitation of the string corresponds to a
different particle in spacetime.  Thus, in string theory, {\it four-dimensional
spacetime physics is ultimately the consequence of two-dimensional worldsheet
physics}.   This means that the four-dimensional particle spectra, gauge
symmetries,
couplings, {\it etc.}, that we obtain in a given string model
are all ultimately determined and constrained by worldsheet
symmetries.

There are numerous well-known examples of this interplay
between worldsheet and spacetime physics, examples in which a
given worldsheet symmetry has profound effects in spacetime.
For example,
worldsheet conformal invariance
sets the spacetime Hagedorn temperature of the theory
(with ensuing consequences in string thermodynamics), and
  establishes a critical spacetime dimension which is, in general,
 greater than four.  It is this which necessitates a compactification
to four dimensions, leading to an infinite moduli
space of possible phenomenologically distinct string
ground states.
Likewise, worldsheet supersymmetry
also has profound effects in spacetime:
it introduces spacetime fermions into the string spectrum, lowers the
critical spacetime dimension from 26 to 10, and sets an upper limit
of 22 on the rank of the corresponding gauge group in a classical heterotic
string.
Even more profound are the effects of worldsheet modular invariance
(or one-loop conformal anomaly cancellation):
in certain settings this removes the tachyon from the spacetime string
spectrum,
 introduces spacetime supersymmetry,
 and guarantees the ultraviolet finiteness of string one- and multi-loop
amplitudes.
 Indeed, modular invariance is also responsible for the appearance
 of so-called {\it GSO projections}\/ which remove certain states from
all mass levels of the string spectrum, and likewise   requires
the introduction of corresponding
 new sectors which {\it add}\/ new states to the string spectrum.
All of this happens simultaneously in a tightly constrained manner,
and one finds that
it is generally difficult to alter the properties of
one sector of a given string
model without seriously disturbing the features of another sector.

An important goal for string phenomenologists, then, is the development
of a {\it dictionary}\/, or table of relations, between worldsheet physics and
spacetime physics.  In this way we hope to  ultimately learn
what ``patterns'' of spacetime physics are allowed or consistent
with an underlying string theory.
More specifically, as far as unification is concerned,
we wish to determine which of the possible paths to unification are
mutually consistent and can be realized in actual realistic
string models.

In the rest of this article, therefore,
we shall outline the current status and
recent developments for each
of these possible paths to unification.
In other words, we shall be exploring the extent
to which the various unification mechanisms we have outlined
can be made consistent with the worldsheet symmetries
(such as conformal invariance, modular invariance,
and worldsheet supersymmetry) that underlie string theory.


\setcounter{footnote}{0}
\section{Path \#1:  String GUT models}

In this approach, we ask the question:  Can one realize the ``strict GUT''
scenario consistently in string theory?  In
other words, can we build a consistent and phenomenologically realistic
string model which realizes, say, a unified $SU(5)$ or $SO(10)$ gauge symmetry
at the string scale,
along with the appropriate matter content necessary for yielding three complete
MSSM representations, the appropriate electroweak Higgs representations,
as well as the GUT Higgs needed to break the GUT gauge symmetry group down to
$SU(3)\times SU(2)\times U(1)_Y$ at some lower scale?
As we shall see, obtaining the required gauge group is fairly easy.
By contrast, obtaining the required matter representations
turns out to be much more difficult.

Why is this so hard?  The short answer is that there naively seems
to be a clash between the three properties that we demand of our string theory:
unitarity of the underlying worldsheet conformal field theory,
existence of the required GUT Higgs in the massless spacetime spectrum,
and the existence of only three chiral generations.
The first two properties together imply that we need to realize our  GUT
symmetry group $G$ with an affine level $k_G\geq  2$.  This in turn has
historically
rendered the construction of corresponding three-generation string GUT
models difficult \cite{stringguts,aldaone,aldatwo,shynew,KTtwo,KTthree}
(though not impossible).
In this section, we shall briefly sketch the basic arguments and current
status of the string GUT approach.

\subsection{Why higher levels are needed}

We begin by discussing why grand-unified gauge groups
in string theory must be realized at higher affine levels.

Recall that in heterotic string theory, all gauge symmetries
are ultimately realized in the form of worldsheet affine Lie
algebras, with currents $J^a(z)$ satisfying operator product
expansions of the form given in Eq.~(\ref{OPE}).
However, given a gauge group $G$, the corresponding
level $k_G$ is not arbitrary, for there are two constraints
that must be satisfied.  First, if $G$ is non-abelian
(which is the case that we will be discussing here), we must
have $k_G\in \IZ^+$.  This restriction permits the algebra to have
unitary representations, as required for a consistent string model.
Second, in order to satisfy the heterotic-string conformal anomaly
constraints, we must also have $c(G_{k_G})\leq 22$
where $c(G_{k_G})$ is the contribution to the conformal anomaly
arising from the gauge group factor $G_{k_G}$.
This central charge contribution is defined as
\beq
        c(G_{k_G}) ~=~ {k_G~ {\rm dim}\/\, G \over k_G + \tilde h_G }~
\label{centcharge}
\eeq
where $\tilde h_G$ is the dual Coxeter number of $G$.
If this constraint is not satisfied,
then worldsheet conformal invariance cannot be maintained
at the quantum-mechanical level, and the string model will again
be inconsistent.

Given a level $k_G$ satisfying these two constraints, there are
then two further constraints that govern which corresponding
representations of $G$ can potentially appear in the massless
({\it i.e.}\/, observable) string spectrum.
Specifically, one finds \cite{KMreview} that the only
unitary representations $R$ of $G$ which can potentially be massless
are those for which
\beq
            0 ~\leq \sum_{i=1}^{{\rm rank}(G)} a_i^{(R)}\,m_i ~\leq~ k_G~
\label{unitarity}
\eeq
and
\beq
          h_R ~\equiv~
        {C_G^{(R)}/\vec \alpha_h^2 \over k_G+\tilde h_G} ~\leq~1~.
\label{masslessness}
\eeq
Here $a_i^{(R)}$ are the Dynkin labels of the highest weight of the
representation $R$;
$m_i$ are the so-called ``comarks'' (or ``dual Coxeter labels'')
corresponding to each simple root
$\vec \alpha_i$ of $G$;
$C_G^{(R)}$ is the eigenvalue of the quadratic Casimir acting on
the representation $R$;
and $\vec \alpha_h$ is the longest root of $G$.
The condition in Eq.~(\ref{unitarity}) guarantees that the representation
$R$ will be unitary, while the condition in Eq.~(\ref{masslessness})
reflects the requirement that the representation should be potentially
massless.
This latter requirement arises because the conformal dimension $h_R$
of a given state describes its spacetime mass (in Planck-mass units), and
is related to the number of underlying string excitations needed to produce
it.  Since the vacuum energy of the gauge sector of the heterotic
string is $-1$, only those states with $h_R\leq 1$ have the potential
to appear in the massless spectrum.  Of course, states with $h_R <1$
must carry additional quantum numbers beyond those of $G$ in order
to satisfy the full masslessness condition $h^{(\rm total)}=1$.

\begin{table}
\begin{rotate}{
 {\footnotesize
\begin{tabular}{c|c|c|c|c|c|c|c}
   ~& $SU(2)$ & $SU(3)$ & $SU(4)$ & $SU(5)$ & $SU(6)$ & $SO(10)$ & $E_6$ \\
   ~&  ~ & ~ & ~ & ($k_{\rm max}=55$) & ($k_{\rm max}=10$) & ($k_{\rm max}=7$)
& ($k_{\rm max}=4$) \\
\hline
\hline
$k=1$ &
           \bbox{
           $c=1:$\bbreak
           ~(\rep{2},~1/4)
           }
           &
           \bbox{
           $c=2:$\bbreak
           ~(\rep{3},~1/3)
           }
           &
           \bbox{
            $c=3:$\bbreak
             ~(\rep{4},~3/8)\bbreak
             ~(\rep{6},~1/2)
             }
           &
        \bbox{
         $c=4:$\bbreak
         ~(\rep{5},~2/5)\bbreak
           ~(\rep{10},~3/5)
           }
        &
        \bbox{
         $c=5:$\bbreak
         ~(\rep{6},~5/12)\bbreak
         ~(\rep{15},~2/3)\bbreak
           ~(\rep{20},~3/4)
           }
        &
        \bbox{
         $c=5:$\bbreak
           ~(\rep{10},~1/2)\bbreak
           ~(\rep{16},~5/8)
           }
        &
        \bbox{
         $c=6:$\bbreak
           ~(\rep{27},~2/3)
           }
        \\
\hline
$k=2$ &
           \bbox{
            $c=3/2:$\bbreak
            ~(\rep{2},~3/16)\bbreak
              ~(\rep{3},~1/2)
              }
           &
           \bbox{
            $c=16/5:$\bbreak
            ~(\rep{3},~4/15)\bbreak
              ~(\rep{6},~2/3)\bbreak
              ~(\rep{8},~3/5)
              }
           &
           \bbox{
            $c=5:$\bbreak
              ~(\rep{4},~5/16)\bbreak
              ~(\rep{6},~5/12)\bbreak
              ~(\rep{10},~3/4)\bbreak
              ~(\rep{15},~2/3)\bbreak
              ~(\rep{20},~13/16)\bbreak
              ~(\rep{20$^\prime$},~1)
              }
           &
        \bbox{
         $c=48/7:$\bbreak
           ~(\rep{5},~12/35)\bbreak
           ~(\rep{10},~18/35)\bbreak
           ~(\rep{15},~4/5)\bbreak
           ~(\rep{24},~5/7)\bbreak
           ~(\rep{40},~33/35)\bbreak
           ~(\rep{45},~32/35)
           }
         &
        \bbox{
         $c=35/4:$\bbreak
           ~(\rep{6},~35/96)\bbreak
           ~(\rep{15},~7/12)\bbreak
           ~(\rep{20},~21/32)\bbreak
           ~(\rep{21},~5/6)\bbreak
           ~(\rep{35},~3/4)\bbreak
           ~(\rep{84},~95/96)
           }
         &
        \bbox{
         $c=9:$\bbreak
           ~(\rep{10},~9/20)\bbreak
           ~(\rep{16},~9/16)\bbreak
           ~(\rep{45},~4/5)\bbreak
           ~(\rep{54},~1)
           }
         &
        \bbox{
         $c=78/7:$\bbreak
           ~(\rep{27},~13/21)\bbreak
           ~(\rep{78},~6/7)
           } \\
\hline
$k=3$ &
           \bbox{
            $c=9/5:$\bbreak
              ~(\rep{2},~3/20)\bbreak
              ~(\rep{3},~2/5)\bbreak
              ~(\rep{4},~3/4)
              }
            &
           \bbox{
            $c=4:$\bbreak
              ~(\rep{3},~2/9)\bbreak
              ~(\rep{6},~5/9)\bbreak
              ~(\rep{8},~1/2)\bbreak
              ~(\rep{10},~1)\bbreak
              ~(\rep{15},~8/9)
              }
             &
           \bbox{
            $c=45/7:$\bbreak
              ~(\rep{4},~15/56)\bbreak
              ~(\rep{6},~5/14)\bbreak
              ~(\rep{10},~9/14)\bbreak
              ~(\rep{15},~4/7)\bbreak
              ~(\rep{20},~39/56)\bbreak
              ~(\rep{20$^\prime$},~6/7)\bbreak
              ~(\rep{36},~55/56)
              }
             &
        \bbox{
         $c=9:$\bbreak
           ~(\rep{5},~3/10)\bbreak
           ~(\rep{10},~9/20)\bbreak
           ~(\rep{15},~7/10)\bbreak
           ~(\rep{24},~5/8)\bbreak
           ~(\rep{40},~33/40)\bbreak
           ~(\rep{45},~4/5)\bbreak
           ~(\rep{75},~1)
           }
          &
        \bbox{
         $c=35/3:$\bbreak
           ~(\rep{6},~35/108)\bbreak
           ~(\rep{15},~14/27)\bbreak
           ~(\rep{20},~7/12)\bbreak
           ~(\rep{21},~20/27)\bbreak
           ~(\rep{35},~2/3)\bbreak
           ~(\rep{70},~11/12)\bbreak
           ~(\rep{84},~95/108)\bbreak
           ~(\rep{105},~26/27)
           }
          &
        \bbox{
         $c=135/11:$\bbreak
           ~(\rep{10},~9/22)\bbreak
           ~(\rep{16},~45/88)\bbreak
           ~(\rep{45},~8/11)\bbreak
           ~(\rep{54},~10/11)\bbreak
           ~(\rep{120},~21/22)\bbreak
           ~(\rep{144},~85/88)
           }
          &
        \bbox{
         $c=78/5:$\bbreak
           ~(\rep{27},~26/45)\bbreak
           ~(\rep{78},~4/5)
           }
         \\
\hline
$k=4$ &
           \bbox{
            $c=2:$\bbreak
              ~(\rep{2},~1/8)\bbreak
              ~(\rep{3},~1/3)\bbreak
              ~(\rep{4},~5/8)\bbreak
              ~(\rep{5},~1)
              }
            &
           \bbox{
            $c=32/7:$\bbreak
              ~(\rep{3},~4/21)\bbreak
              ~(\rep{6},~10/21)\bbreak
              ~(\rep{8},~3/7)\bbreak
              ~(\rep{10},~6/7)\bbreak
              ~(\rep{15},~16/21)
              }
            &
           \bbox{
            $c=15/2:$\bbreak
              ~(\rep{4},~15/64)\bbreak
              ~(\rep{6},~5/16)\bbreak
              ~(\rep{10},~9/16)\bbreak
              ~(\rep{15},~1/2)\bbreak
              ~(\rep{20},~39/64)\bbreak
              ~(\rep{20$^\prime$},~3/4)\bbreak
              ~(\rep{20$^{\prime\prime}$},~63/64)\bbreak
              ~(\rep{36},~55/64)\bbreak
              ~(\rep{45},~1)\bbreak
              ~(\rep{64},~15/16)
              }
            &
        \bbox{
         $c=32/3:$\bbreak
           ~(\rep{5},~4/15)\bbreak
           ~(\rep{10},~2/5)\bbreak
           ~(\rep{15},~28/45)\bbreak
           ~(\rep{24},~5/9)\bbreak
           ~(\rep{40},~11/15)\bbreak
           ~(\rep{45},~32/45)\bbreak
           ~(\rep{50},~14/15)\bbreak
           ~(\rep{70},~14/15)\bbreak
           ~(\rep{75},~8/9)
           }
         &
        \bbox{
         $c=14:$\bbreak
           ~(\rep{6},~7/24)\bbreak
           ~(\rep{15},~7/15)\bbreak
           ~(\rep{20},~21/40)\bbreak
           ~(\rep{21},~2/3)\bbreak
           ~(\rep{35},~3/5)\bbreak
           ~(\rep{70},~33/40)\bbreak
           ~(\rep{84},~19/24)\bbreak
           ~(\rep{105},~13/15)\bbreak
           ~(\rep{120},~119/120)\bbreak
           ~(\rep{189},~1)
           }
         &
        \bbox{
         $c=15:$\bbreak
           ~(\rep{10},~3/8)\bbreak
           ~(\rep{16},~15/32)\bbreak
           ~(\rep{45},~2/3)\bbreak
           ~(\rep{54},~5/6)\bbreak
           ~(\rep{120},~7/8)\bbreak
           ~(\rep{144},~85/96)\bbreak
           ~(\rep{210},~1)
           }
         &
        \bbox{
         $c=39/2:$\bbreak
           ~(\rep{27},~13/24)\bbreak
           ~(\rep{78},~3/4)
           }
         \\
\hline
\end{tabular}
 }
 }
\end{rotate}
\caption{
 Unitary, potentially massless representations
 for various gauge groups and levels.
  Each representation $R$ is listed as $({\bf n},h_R)$
  where {\bf n} is its dimension and $h_R$ its conformal
  dimension.  For each
  group, the maximum string-allowed level $k_{\rm max}$ is
  determined by requiring that its central charge
  not exceed 22.}
\label{allowedrepstable}
\end{table}

The representations that survive these constraints are listed in
Table~\ref{allowedrepstable}
for the groups $SU(2)$, $SU(3)$, $SU(4)$, $SU(5)$, $SU(6)$, $SO(10)$,
and $E_6$.
Each entry is listed in the form $(\rep{n},h_R)$ where
$\rep{n}$ is the dimensionality of the representation
and $h_R$ is its conformal dimension.
Note that in this table, we have not listed
singlet representations [for which the corresponding
entry is (\rep{1},0) for all groups and levels].
We have also omitted
the complex-conjugate representations.
It is immediately apparent from this table that while the
fundamental representations of each group always appear for
all levels $k\geq 1$,
the {\it adjoint}\/ representations do not appear until levels
$k\geq  2$.  This is a crucial observation, since the
GUT Higgs field that is typically required in order to break a grand-unified
group such as $SU(5)$ or $SO(10)$ down to the MSSM gauge group
$SU(3)\times SU(2)\times U(1)_Y$ must transform in the adjoint
of the unified group.
Thus, we see that the dual requirements
of a unitary worldsheet theory and a potentially massless adjoint
Higgs representation force any
string GUT group $G$ to be realized at a level $k_G\geq 2$.

\subsection{Why higher levels are harder}

This, it turns out, is a profound requirement,
and completely alters the methods needed to construct such models.
We shall now explain why subtleties arise, and how they
are ultimately resolved.

\subsubsection{The subtleties}

In order to fully appreciate the subtleties that enter
the construction of string models with higher-level
gauge symmetries, let us first recall
the simplest string constructions ---
 those based on free worldsheet bosons or fermions.
These constructions are called {\it free-field}\/ constructions,
and encompass all of the string constructions (such as the
free-fermionic, lattice, or orbifold constructions)
which have formed the basis of string GUT model-building
attempts in recent years.
In a four-dimensional heterotic string,
the conformal anomaly
on the left-moving side
can be saturated by having $22$ internal bosons
$\Phi^I$ ($I=1,...,22$), or equivalently $22$ complex
fermions $\psi^I$.
If we treat these bosons or fermions
indistinguishably, this generates an internal symmetry group
$SO(44)$, and we can obtain other internal symmetry
groups by distinguishing  between these different worldsheet fields
({\it e.g.}, by giving different toroidal boundary conditions to
different fermions $\psi^I$).
 Such internal symmetry groups are then interpreted
as the gauge symmetry groups of the effective low-energy theory.

In general, the spacetime gauge bosons of such symmetry groups
fall into two classes:  those of the form
$\psi^\mu |0\rangle_R \otimes
            i\partial\phi^I  |0\rangle_L$
give rise to the 22 {\it Cartan}\/ generators of the gauge symmetry,
while those of the form
$\psi^\mu |0\rangle_R \otimes e^{i \alpha \phi^I} e^{i\beta \phi^J}
|0\rangle_L$
with $\alpha^2+\beta^2=2$
give rise to the {\it non-Cartan}\/ generators.
Equivalently, in the language of complex fermions, both classes of gauge bosons
take the simple form
 $\psi^\mu |0\rangle_R \otimes \overline{\psi}^I \psi^J |0\rangle_L$:
if $I= J$ (so that one left-moving fermionic excitation is the
antiparticle of the other), we obtain the Cartan gauge-boson states,
whereas if $I\not =J$  we obtain the non-Cartan states.
Taken together, the Cartan and non-Cartan states fill out the adjoint
representation of some Lie group.
The important point to notice here, however, is the fact
that in the fermionic formulation, {\it two}\/ fermionic excitations
are required on the left-moving side (or equivalently, we must have
$\alpha^2+\beta^2=2$ in the bosonic formulation).  Indeed, not only is this
required in order to produce the two-index tensor
representation that contains the adjoint representation
(as is particularly evident in the fermionic construction),
but precisely this many excitations are
also necessary in order for the resulting gauge-boson state
to be {\it massless}.
This is evident from the fact that
the conformal dimension of the state
$ e^{i \alpha \phi^I} e^{i\beta \phi^J} |0\rangle_L$
is given by $h=(\alpha^2+\beta^2)/2$.
We must have $h=1$ for a massless gauge boson.

The next step is to consider the corresponding charge lattice,
which can be constructed as follows.
Each of the left-moving worldsheet bosons $\phi^I$ has, associated
with it, a left-moving current $J_I\equiv i\partial \phi_I$ (or, in a fermionic
formulation, a current $J_I\equiv \overline{\psi}_I \psi_I$).
The eigenvalue $Q_I$
of this current when acting on a given state yields the charge
of that state.
Thus, we see that the complete left-moving charge of a given state
is a 22-dimensional vector ${\bf Q}$, and the charges of the above
gauge-boson states together comprise the root system
of a rank-22 gauge group (which can be simple or non-simple).
However, the properties of this gauge group are highly constrained.
For example, the fact that we require two fundamental excitations in order
to produce the gauge-boson state (or equivalently that we must
have $\alpha^2+\beta^2=2$ in a bosonic formulation) implies that
each non-zero root must have (length)$^2=2$.
Indeed, $h={\bf Q}^2/2$.
Thus, we see that we can obtain only {\it simply-laced}\/ gauge groups
in such constructions!  Moreover,
it turns out that in such constructions,
the GSO projections only have the power to project a
given {\it non}\/-Cartan root into or out of the spectrum.
Thus, while we are free to potentially alter
the particular gauge group in question via GSO projections, we cannot go beyond
the set of rank-22 simply-laced gauge groups.

As the final step,
let us now consider the affine level at which such groups are ultimately
realized.
Indeed, this is another property of the gauge group that
cannot be altered in such constructions.
As we saw in Eq.~(\ref{OPE}),
the affine level $k$ is defined through the OPE's of the currents $J_a$.
However, with fixed normalizations
for the currents $J_a$ and structure constants $f^{abc}$,
we see from Eq.~(\ref{OPE}) that
\beq
    k_G\cdot |\vec \alpha_h|^2 ~  =  ~{\rm constant} ~=~2~.
\label{lengthlevel}
\eeq
Thus, with roots of (length)$^2=2$, we see that our gauge symmetries
are realized at level $k_G=1$.
Indeed, if we wish to realize our gauge group at a higher level
({\it e.g.}, $k_G=2$), then we must somehow devise a special mechanism
for obtaining roots of smaller length ({\it e.g.}, length $=1$).
However, as discussed above, this would naively appear to
conflict with the masslessness requirement.
Thus, at first glance, it
would seem to be impossible to realize higher-level gauge
symmetries within free-field string models.

\subsubsection{The resolution}

Fortunately, however, there do exist various
methods which are capable of yielding higher-level gauge symmetries
within free-field string constructions.
Indeed, although these methods
are fairly complicated,
they share certain simple underlying features.
We shall now describe, in the language of the above discussion,
the underlying mechanism which enables such symmetries to be
realized.  The approach we shall follow
is discussed more fully in Ref.~\cite{unif4}.

As we have seen, the fundamental problem that we face is
that we need to realize our gauge-boson string states
as massless states, but with smaller corresponding charge vectors (roots).
To do this, let us for the moment
imagine that we could somehow {\it project}\/ these roots onto
a certain hyperplane in the 22-dimensional charge space,
and consider only the surviving components of these roots.
Clearly, thanks to this projection, the ``effective length''
of our roots would be shortened.  Indeed, if we cleverly
choose the orientation of this hyperplane of projection,
we can imagine that our shortened, projected roots could
either
\begin{itemize}
\item combine with other longer, unprojected roots ({\it i.e.},
     roots which originally lay in the projection hyperplane)
      to fill out the root system of a {\it non}\/-simply laced
      gauge group;  or
\item combine with other similarly shortened roots to
    fill out the root system of a {\it higher-level}\/ gauge symmetry.
\end{itemize}
Thus, such a projection would be exactly what is required.

The question then arises, however:  how can we achieve or interpret
such a hyperplane projection in charge space?  Clearly, such a
projection would imply that one or more dimensions of the charge lattice
should no longer be ``counted'' towards building the gauge group, or
equivalently that one or more of the gauge quantum numbers should be lost.
Indeed, such a projection would entail a {\it loss of rank}\/
(commonly called ``rank-cutting''), which corresponds to a loss
of Cartan generators.
Thus, we see that we can achieve the required projection if and only
if we can somehow construct a special GSO projection which,
unlike those described above, is capable of projecting out
a {\it Cartan root}\/!

What kinds of string constructions can give rise to such
unusual GSO projections?
Clearly, as we indicated, such GSO projections cannot arise
in the simplest free-field constructions based on free bosons or complex
fermions.
Instead, we require highly ``twisted'' orbifolds (typically
asymmetric, non-abelian orbifolds
\cite{orbifoldconstr,asymmorbifolds,oldorbmodels,KT}),
or equivalently constructions based on so-called ``necessarily
real fermions'' \cite{freefermions,realfreefermions}.  The
technology for constructing such string theories is
continually being developed and refined
\cite{stringguts,aldaone,aldatwo,shynew,KT}.
The basic point, however, is that such ``dimensional truncations''
of the charge lattice are the common underlying feature in
all free-field constructions of higher-level string models.
Moreover, we also see
from this point of view that such higher-level gauge symmetries
go hand-in-hand with rank cutting and the appearance of
non-simply laced gauge groups.

\begin{figure}[ht]
\centerline{\epsfxsize 3.5 truein \epsfbox {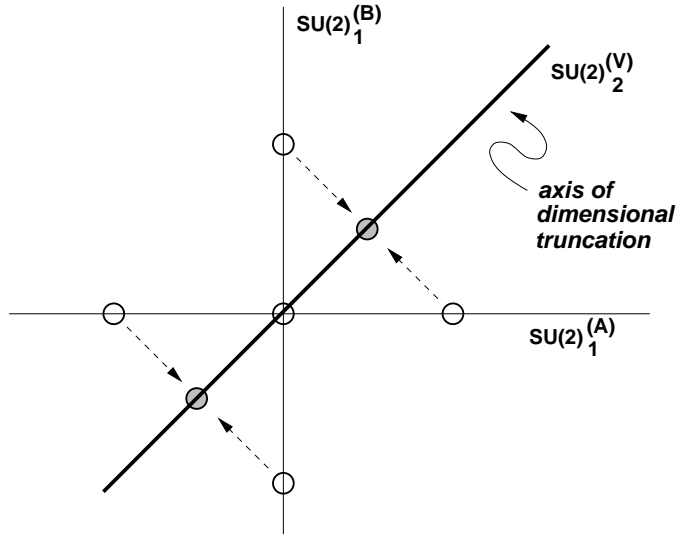}}
\caption{ The root system of $SU(2)_1^{(A)}\times SU(2)_1^{(B)}$ (denoted
by open circles), and its
dimensional truncation onto the diagonal subgroup $SU(2)_2^{(V)}$
(with new non-zero roots denoted by shaded circles).}
\label{su2su2}
\end{figure}

As an example of how such ``dimensional truncations'' work,
let us consider the well-known
method of achieving a level-two affine Lie algebra which consists of
tensoring together two copies of any group $G$ at level one,
and then modding out by the symmetry interchanging the
two group factors.  This leaves behind the diagonal
subgroup $G$ at level two.
This construction is well-known in the case $G=E_8$, where it serves
as the underlying mechanism responsible
for the non-supersymmetric (level-two) single-$E_8$ string model in
ten dimensions \cite{Eeightmodel}.
For simplicity, let us analyze this construction
for the case $G=SU(2)$.
If we start with an $SU(2)_1^{(A)}\times SU(2)_1^{(B)}$ gauge symmetry,
as illustrated in Fig.~\ref{su2su2}, then modding
out by their interchange symmetry [{\it i.e.},
constructing the diagonal subgroup $SU(2)^{(V)}$] corresponds
to projecting or dimensionally truncating the roots
onto the diagonal axis corresponding to
the diagonal Cartan generator $J_z^{(V)}=J_z^{(A)}+J_z^{(B)}$.
As we can see from Fig.~\ref{su2su2},
this reproduces the $SU(2)$ root system, but scaled
so that roots which formerly had length $\sqrt{2}$ now
have length $1$.
Thus we have realized $SU(2)$ at level two as the diagonal
survivor of this dimensional-truncation procedure.
Indeed, it is clear that this diagram generalizes
to any groups $G_1^{(A)}\times G_1^{(B)}$, since the roots
of the two level-one root systems are always orthogonal to each other,
and hence project onto the diagonal axes with
a reduction in length by a factor $\cos\, 45^\circ=1/\sqrt{2}$.
Moreover, we see that this also generalizes to {\it any}\/ number of identical
group factors tensored together,
$G_1^{(1)}\times G_1^{(2)}\times...G_1^{(n)}$,
leaving behind the completely diagonal subgroup $G$ at level $n$.
More sophisticated examples of such dimensional truncations, along
with descriptions of their basic properties and uses, can
be found in Ref.~\cite{unif4}.

\subsection{Current status and recent developments}

Thus far we have
described, in a model-independent way, the basic ingredients
that go into the realization of higher-level gauge symmetries
in string theory.  However, the construction of an actual satisfactory
string GUT {\it model}\/ is a far more difficult affair.
In particular, while the methods for realizing the special
GSO projections that effect rank-cutting are now well-understood,
it is still necessary to reconcile these special projections
with the ``ordinary'' GSO projections that yield, for example,
(1)  three and only three massless chiral generations;
(2)  no extra exotic chiral matter; and
(3)  the proper Higgs content and couplings
     ({\it e.g.}, in order to avoid proton decay).
Indeed, in string GUT models (just as in ordinary field-theoretic
GUT models), the problem of proton decay is a vital issue.

Until recently, it appeared that such a reconciliation might
not be possible.  In particular, despite various
early attempts \cite{stringguts,aldaone},
no three-generation models with higher-level GUT symmetries had
been constructed, and it might have seemed that
the requirement of three generations might, by itself, be in fundamental
conflict with the requirement of higher-level gauge groups.
However, it has recently been shown that this is not the case
for a variety of GUT gauge groups and levels.
A summary of those groups and levels for which consistent
three-generation string GUT models have been constructed
appears in Table~\ref{tableGUTmodels}.
In particular,
level-two $SU(5)$ models were obtained in
Ref.~\cite{aldatwo} using a symmetric orbifold construction
and in Ref.~\cite{shynew}
using the free-fermionic construction \cite{freefermions,realfreefermions}.
By contrast, level-three $SU(5)$, $SO(10)$, and $E_6$ models
were obtained in Refs.~\cite{KTtwo,KTthree}
using an {\it asymmetric}\/ orbifold construction developed in Ref.~\cite{KT}.
All of these models realize their higher-level gauge symmetries
via the diagonal embeddings discussed above, and contain extra
matter (beyond the three massless generations and adjoint
Higgs representations) in their massless spectra.

\begin{table}[htb]
\centerline{
\begin{tabular}{r||c|c|c}
     ~ & ~~$k=2$~~ & ~~$k=3$~~ & ~~$k=4$~~ \\
\hline
\hline
$SU(5)$ & $\surd $ & $\surd \surd$ &  ? \\
$SO(10)$ & ? & $\surd \surd$ &  ? \\
$E_6$ & ? & $\surd \surd$ &  ? \\
\end{tabular}
  }
\caption{Progress in the construction of three-generation
  higher-level string GUT models.
  A single check ($\surd$) indicates the construction
  of three-generation string models with
  extra chiral matter in the massless spectrum, whereas
  a double check ($\surd\surd$) signifies
  the construction of three-generation string models with
  extra non-chiral (vector-like) matter in the massless spectrum.
  No three-generation models have been constructed
  without additional chiral or vector-like matter.
  Question marks indicate that no three-generation
  models have yet been constructed.}
\label{tableGUTmodels}
\end{table}

The existence of these three-generation models clearly demonstrates
that such string GUT models are possible.
Unfortunately, it is not yet clear whether such models are
entirely ``realistic'', for the other problems listed above
tend to remain unresolved for the string GUT models.
For example, the three-generation level-two $SU(5)$
models constructed in Refs.~\cite{aldatwo,shynew} all contain extra
chiral $\rep{15}$ representations of $SU(5)$, which would give rise to extra
color sextets after GUT symmetry breaking.
Indeed, because these states are chiral, they remain light and are thus
phenomenologically unacceptable.
Likewise, the three-generation level-three models, although  promising,
have yet to be examined in detail, and in particular
it remains to examine their superpotentials and Higgs couplings
to see if realistic low-energy phenomenologies are possible and
if any undesirable extra vector-like matter can be made superheavy.
However, the pace of recent progress
in both the free-fermionic and orbifold
constructions --- along with the recent construction of three-generation
higher-level string models --- suggests that
the problems faced in this approach are now simply
those of constructing {\it realistic}\/ models, and
are not fundamental.
Further advances in model construction are therefore likely.

There have also been recent developments in understanding general
properties of such higher-level string GUT models.
All of the above string models that have been constructed use
the so-called ``diagonal construction''
in which the level-$k$ GUT gauge group $G$
is realized as the diagonal subgroup within a
$k$-fold tensor-product
of group $G$ at level one:
\beq
         G_k ~\subset~  G ~\times~ G ~\times~ ... ~\times ~G~.
\label{diagonalembedding}
\eeq
Such diagonal embeddings were described in Sect.~4.2.2,
and were illustrated in Fig.~\ref{su2su2} for the case of $SU(2)_2$.
Unfortunately, however,
the diagonal construction is a very expensive and inefficient way
of realizing such higher-level gauge groups, for it requires
that one first build a string model with the larger gauge group
$G\times G \times...\times G$, and then subsequently break to the diagonal
subgroup.  Indeed,
because of its larger rank,
this tensor-product group typically occupies many more dimensions
of the charge lattice than would be required
for only the higher-level subgroup;
it also requires greater central charge than the
higher-level subgroup itself would require.
Thus, such diagonal realizations of higher-level
gauge symmetries come with extra hidden ``costs'' beyond
those due to the higher-level gauge symmetries themselves,
and imply that such string models will have smaller hidden-sector gauge
symmetries
than would otherwise be possible.  This in turn means that
one has less flexibility for controlling or arranging
many other desired phenomenological features of a given string model.
It would therefore be useful to have a general method of surveying
whether the diagonal embeddings in Eq.~(\ref{diagonalembedding}) are
the only embeddings that can be employed for
realizing higher-level GUT gauge symmetries in free-field string constructions,
or whether other more efficient embeddings are possible.

\begin{table}[t]
\centerline{
\begin{tabular}{rcl|c|c|c}
 $G'_{k} $ & $\subset$ &  $G$ &  $\Delta c$ & $\Delta r$ & $\xi$ \\
\hline
\hline
  $SU(5)_2$  & &  Diagonal only & $8/7$ & 4 &  \\
  $SU(5)_3$  & & $SU(10)_1$  & $0$ & 5 & $+3$ \\
  $SU(5)_4$  & &  $SU(15)_1$  & $10/3$ & 10 & $+2$  \\
  $SU(5)_4$  & &  $SU(16)_1$  & $13/3$ & 11 & $+1$ \\
\hline
  $ SU(6)_2 $  & &  Diagonal only & $5/4$ & 5 &  \\
  $ SU(6)_3 $  & &  Diagonal only & $10/3$ & 10 &  \\
  $ SU(6)_4 $ & & $SU(15)_1 $  & $0$   & 9 & $+6$ \\
\hline
  $SO(10)_2$  & &   $SU(10)_1$  & $0$  & 4 & $+1$ \\
  $ SO(10)_3 $  & &  Diagonal only & $30/11$ & 10 &  \\
  $SO(10)_4$  & &  $SU(16)_1$  & $0$ & 10 & $+5$ \\
  $SO(10)_4$ & & $[SU(10)_1]^2 $  & $3$  & 13 & $+2$ \\
  $SO(10)_4$ & & $SU(10)_1 \times [SO(10)_1]^2$ & $4$ & 14 & $+1$ \\
  $ SO(10)_{k>4} $  & &  Impossible &  & & \\
\hline
  $ (E_6)_{2} $  & &  Diagonal only & $6/7$ & 6 & \\
  $ (E_6)_{3} $  & &  Diagonal only & $12/5$ & 12 & \\
  $ (E_6)_{k> 3} $  & &  Impossible &  & & \\
\hline
\end{tabular}
    }
\caption{The complete list of all possible non-diagonal
          free-field string embeddings that can give rise
          to $SU(5)_k$, $SU(6)_k$, $SO(10)_k$, and $(E_6)_k$
          at levels $2\leq k \leq 4$.  Here $\Delta r$ and $\Delta c$
          are the extra rank and central charge required for each embedding,
          and $\xi$ is the ``savings'' in these quantities relative
          to the corresponding diagonal embedding.}
\label{GUTembeddings}
\end{table}

Such a general survey has recently been completed \cite{unif4}.
By studying the dimensional truncations that are necessary
in order to achieve higher-level gauge symmetries, it has
been shown \cite{unif4} that each such truncation corresponds uniquely
to a so-called ``irregular'' embedding in group theory.
Given this information, it has now been possible to classify
all possible ways of realizing higher-level gauge symmetries
in free-field string theories \cite{unif4}, and a complete list
of all embeddings for the GUT groups
$SU(5)$, $SU(6)$, $SO(10)$, and $E_6$ for levels $k=2,3,4$
is given in Table~\ref{GUTembeddings}.
For each such embedding $G'_k\subset G$, we have also listed
the extra ranks and central charges that are required for its realization,
\beqn
         \Delta r &=&  {\rm rank}\, G ~-~ {\rm rank}\, G'\nonumber\\
         \Delta c &=&  c(G) ~-~ c(G'_{k})~,
\eeqn
where the central charges are computed according to Eq.~(\ref{centcharge}).
Note that in many cases, the only possible higher-level embeddings
are in fact the diagonal embeddings.
However, it is evident from Table~\ref{GUTembeddings}
that there also exist many new classes of embeddings
which might form the basis for new types
of string GUT model constructions.
In particular, some of these new embeddings
require less rank and central charge for their realization
than the diagonal embeddings, and are therefore significantly
more efficient.
For such embeddings, we have defined the quantity $\xi$ which appears
in Table~\ref{GUTembeddings} in such a way as to measure this improved
efficiency in central charge and rank relative to the diagonal embedding:
\beq
     \xi ~\equiv~  k~c(G'_{1}) ~-~ c(G) ~=~
                 k\,({\rm rank}\,G') ~-~ {\rm rank}\,G ~.
\eeq
Thus, embeddings with $\xi>0$ are
always more efficient than their diagonal counterparts.

This complete classification of higher-level GUT embeddings
has also made it possible to obtain new restrictions on
the possibilities for potentially viable string GUT models.
Certain phenomenological characteristics of
string GUT models are, of course, immediately obvious from
Table~\ref{allowedrepstable} directly.  For example,
since the ${\bf 54}$ representation of $SO(10)$ at level two
already saturates the $h_R=1$ masslessness constraint,
this representation cannot carry any additional quantum numbers
under any gauge symmetries beyond $SO(10)$ in such models.
This in turn restricts its allowed couplings \cite{aldatwo}, essentially
ruling out couplings of the forms
$X\cdot {\bf 54}\cdot {\bf 54}$ or $X\cdot {\bf 54}\cdot {\bf 54'}$
where $X$ is any chiral field which transforms as a singlet under $SO(10)$.
Likewise, again following similar sorts of arguments,
it has been shown \cite{aldatwo} that all
superpotential terms in string GUT
models must have dimension $\geq 4$.  In other words, explicit
mass terms (which would have dimension two) are ruled out.
Other phenomenological consequences of the results in
Table~\ref{allowedrepstable}, especially as they restrict the possibilities
for hidden-sector phenomenology, can be found in Ref.~\cite{ELN}.
Indeed, such string GUT ``selection rules'' are completely general,
and apply to all string constructions.

However, given the recent classification of GUT embeddings in
Table~\ref{GUTembeddings},
it is now possible to go beyond these sorts of constraints in the
case of string models using free-field constructions.
For example, it has been shown in Ref.~\cite{unif4} that $SO(10)$
can never be realized at levels exceeding four in free-field string
constructions;
this result clearly goes beyond the simple central-charge constraint (which
would
have permitted realizations up to $k=7$), and thereby implies, for
instance, that it
is impossible to realize the useful {\bf 126} representation within
such $SO(10)$ string GUT models.  A more detailed study \cite{KDso10paper}
shows, in fact,
that in free-field heterotic $SO(10)$ string models using diagonal
embeddings, all representations larger than the {\bf 16}\/
must always transform as singlets under all gauge symmetries beyond
$SO(10)$;  moreover, such constructions can never give rise to
the ${\bf 120}$ or ${\bf 144}$ representations
of $SO(10)$.  These results hold regardless of the affine level
at which $SO(10)$ is realized ({\it i.e.}, despite the general
results of Table~\ref{allowedrepstable}),
and essentially reflect the additional costs
that are involved when realizing higher-level gauge symmetries through
diagonal embeddings in free-field string constructions
\cite{unif4,KDso10paper}.
Indeed, the only known exception to these rules takes place in
a phenomenologically unrealistic {\it non-chiral}\/ string model
(see, {\it e.g.}\/, Appendix B of Ref.~\cite{KTthree}).
Likewise, a similar analysis \cite{KDso10paper} for $E_6$ shows
that $E_6$ can never be realized beyond level three
in free-field constructions (even though the central-charge constraint
would have permitted level four);  moreover,
one finds that the adjoint ${\bf 78}$ representation must
always transform as a singlet under all gauge symmetries beyond $E_6$.
Taken together, then, these results thus severely limit the types
of phenomenologically realistic field-theoretic $SO(10)$ and $E_6$
models that can be obtained using such string constructions.
These issues are discussed more fully in Ref.~\cite{KDso10paper}.

Of course, there are various ways around these difficulties.
One method is to employ the {\it non-diagonal}\/ embeddings
listed in Table~\ref{GUTembeddings};  at present, these
new embeddings have not been explored in the literature,
but it has been shown that they may have improved phenomenological
prospects \cite{unif4,KDso10paper}.
Another method is to adopt a string construction
which is not based on free worldsheet fields.
Such constructions include, for example, those based on tensor products
of Wess-Zumino-Witten models \cite{WZW}, such as Gepner or
Kazama-Suzuki models.
These constructions are quite difficult to implement in practice, however,
and thus have not been investigated for phenomenological purposes.

\subsection{String GUT models without higher levels?}

The complicated issues that arise when
constructing higher-level string GUT models should not
be taken to indicate that the GUT idea cannot be made to work
in a simple manner in string theory.
Indeed, even if no completely realistic higher-level string GUT
models are ultimately constructed,
there exists an entirely different approach which,
although technically not a ``string GUT'', is equally
compelling.
These are the so-called $G\times G$ models,
which have been examined in both
string-theoretic \cite{aldaone,aldatwo,GxGmodels,Finnell}
and field-theoretic \cite{GxGfield}
contexts.
The basic idea is as follows.
We have seen in Fig.~\ref{su2su2}
that one can realize a gauge symmetry group $G$
at level $k_G=2$ by starting with
a gauge symmetry $G\times G$ realized at level one,
and then modding out by the $\IZ_2$ symmetry that interchanges the
two group factors.
Indeed, this is just the diagonal construction.
In the string GUT models that we have been discussing up to this point,
this final step (the $\IZ_2$ modding) is done
 {\it within the string construction itself}\/, so that
at the Planck scale, the string gauge symmetry group is $G$ at
level $k=2$.
However, an alternative possibility is simply
to construct a string model with the level-one gauge
symmetry $G\times G$, and
then do the final breaking to the diagonal subgroup $G$ in the
 {\it effective field theory}\/ at some lower scale below
the Planck scale.
It turns out that
all that is required for this breaking is a Higgs
in the {\it fundamental}\/ representation of $G\times G$, and after the
breaking one effectively obtains the adjoint Higgs
required for further breaking to the MSSM gauge group:
\beqn
     SU(5):&~~~~~ & (\rep{5},\overline{\rep{5}}) ~\to~ \rep{24} ~+~
\rep{1}~\nonumber\\
     SU(6):&~~~~~ & (\rep{6},\overline{\rep{6}}) ~\to~ \rep{35} ~+~
\rep{1}~\nonumber\\
     SO(10):&~~~~~ & (\rep{10},\rep{10}) ~\to~ \rep{45} ~+~
\rep{54}~+~\rep{1}~.
\eeqn
Thus, simply by utilizing the fundamental representations available at level
one
at the string scale,
one can nevertheless produce effective adjoint representations at lower scales.
String models utilizing this mechanism
have been investigated by
several authors \cite{aldaone,aldatwo,GxGmodels,Finnell},
and have met with roughly the same level of success as achieved in
the level-two constructions.  In particular,
three-generation $G\times G$ string models have
been constructed for the case of $G=SU(5)$,
and in fact one such model \cite{Finnell}  apparently
contains no extra chiral matter in its massless spectrum.
Note that while this mechanism works in principle for
$SU(5)$, $SU(6)$, and $SO(10)$, it cannot be used for $E_6$ since
the necessary initial representation $(\rep{27},\overline{\rep{27}})$
would have $h>1$ and thus could not appear in the massless spectrum.

We see, then, that the $G\times G$ models
clearly represent an alternative scenario which,
although not a strict ``string GUT'', could still resolve
the gauge coupling unification problem in the GUT manner,
with a unification of MSSM couplings at $M_{\rm MSSM}$, followed
by the running of a single GUT coupling between $M_{\rm MSSM}$
and $M_{\rm string}$.
An interesting fact which has recently been pointed out \cite{bachas},
however, is that such $G\times G$ models
typically give rise to
moduli which transform in the adjoint of the final gauge group $G$.
As shown in Refs.~\cite{bachas,FGM} for the case of
color octet moduli and electroweak triplet moduli with vanishing hypercharge,
such extra adjoint states have the potential to alter the running of
the gauge couplings in such a way as to fix the original
discrepancy between $M_{\rm MSSM}$ and $M_{\rm string}$
which motivated the consideration of the GUT in the first
place.
Of course, it is not known {\it a priori}\/ whether these states
will actually have the masses required to resolve this
discrepancy.
In any case, the phenomenological problems
that afflict the higher-level string models using
diagonal embeddings become even more severe for these level-one
$G\times G$ models, and thus must still be addressed.

Finally, we mention that the subtleties with higher levels
can also be avoided by employing a GUT scenario that does not
require an adjoint Higgs for breaking to the MSSM gauge group.
An example of such a GUT scenario is flipped $SU(5)$
\cite{flippedsu5fieldtheory},
in which the required gauge symmetry is $G =SU(5)\times U(1)$ and the
matter embedding within $G$ is ``flipped'' relative to that
of ordinary $SU(5)$.
Another example is the Pati-Salam unification scenario
\cite{patisalam}, for which  $G= SO(6)\times SO(4)$.
These field-theoretic unification scenarios have been realized
within consistent three-generation string models \cite{flipped,alrmodel},
and will be discussed at various points in Sects.~6--8.
Furthermore, even within the level-one ``strict GUT'' scenarios,
it may be possible to overcome the absence of
adjoint Higgs representations
by utilizing the string states with fractional
electric charge that often appear in such string models \cite{Wen,Schellekens}.
As recently suggested in Ref.~\cite{DLT},
such states can in principle serve as ``preons'' which may
bind together under the influence of hidden-sector
gauge interactions in order to form {\it effective}\/ adjoint
Higgs representations.  We shall discuss the appearance of fractionally charged
string states in Sect.~5.4.

Thus, we see that although the string GUT approach to understanding
gauge coupling unification is quite subtle and complicated,
there has been substantial progress in recent years both in understanding
the general properties of such theories as well as in the construction
of realistic three-generation string GUT models.


\setcounter{footnote}{0}
\section{Path \#2:  Unification via Non-Standard Levels $(k_Y,k_2,k_3)$}

The second possible path to string-scale unification {\it preserves}\/ the
MSSM gauge structure between the string scale and the $Z$ scale, and
instead attempts to reconcile the discrepancy between the MSSM unification
scale and the string scale by adjusting the values of the string
affine levels $(k_Y,k_2,k_3)$.
The MSSM unification scale, of course,
is determined under the assumption that
$(k_Y,k_2,k_3)$ have their usual MSSM values $(5/3,1,1)$ respectively.
However, in string theory the possibility
exists for the MSSM gauge structure to be realized with different
values of $(k_Y,k_2,k_3)$, thereby altering,
as in Eq.~(\ref{unification}), the predicted boundary
conditions for gauge couplings at unification.
The possibility of adjusting these parameters (especially
$k_Y$) was first proposed in Ref.~\cite{ibanez},
and has been considered more recently
in Refs.~\cite{prevattempts,unif3,shynew,allanach}.

\subsection{What is $k_Y$?}

Before discussing whether such realizations are possible,
we first discuss the definition of $k_Y$.
As we explained in Sect.~4, $k_2$ and $k_3$ are
the affine levels of the $SU(2)$ and $SU(3)$ factors of
the MSSM gauge group, and these levels are defined
through the relation (\ref{OPE}).
The hypercharge group factor $U(1)_Y$, by contrast,
is {\it abelian}\/,
and thus there are no corresponding structure constants $f^{abc}$
or non-zero roots $\vec\alpha_h$ through which
to fix the magnitudes of the terms
in the hypercharge current-current OPE.
In other words, there is no way in which
a unique normalization for the hypercharge current $J_Y$
can be chosen, and consequently there is no intrinsic definition
of $k_Y$ which follows uniquely from the algebra of hypercharge currents.
Thus, the definition of $k_Y$ requires a {\it convention}.

In order to fix a convention, we examine something physical, namely
scattering amplitudes.  The following argument originates in
Ref.~\cite{Ginsparg},
and details for the case of abelian groups can be found in
Refs.~\cite{FIQS,unif3}.
For a non-abelian group, the gauge bosons experience not only couplings
to gravity (through vertices of the form $WWg$ where $W$ is a non-abelian
gauge boson and $g$ is a graviton),
but also trilinear self-couplings (through vertices of the form $WWW$).
It then turns out that $k_G$ is essentially the ratio of these
vertices.  Indeed, the $WWg$ vertex factor is derived from the double
pole term in Eq.~(\ref{OPE}), whereas the $WWW$ vertex factor is derived
from the  single pole term in Eq.~(\ref{OPE}).
For an {\it abelian}\/ group, however, we have only vertices of
the form $YYg$.  We nevertheless {\it define}\/ a normalization
$k_Y$
for the $U(1)_Y$ current $J_Y$ in such a way that it has the same coupling
to gravity as a non-abelian current.
As shown in Ref.~\cite{Ginsparg}, this is tantamount to requiring that $J_Y$
have a normalization giving rise to the OPE
\beq
         J_Y(z)\,J_Y(w) ~=~ {1 \over (z-w)^2} ~+~ {\rm regular}~.
\label{unitnorm}
\eeq
Given this normalization, we then
compute the ratio of the corresponding coupling $g_Y$
with the gravitational coupling, and find
\beq
        8\pi \, {G_N\over \alpha'}~=~ 2\, g_Y^2~.
\eeq
Thus, in analogy with Eq.~(\ref{unification}), we define $k_Y=2$ for an
abelian current normalized according to Eq.~(\ref{unitnorm}).

Given these results, it is then straightforward to determine the value of
$k_Y$ corresponding to {\it any}\/ $U(1)$ current:  we simply determine
the coefficient of the double-pole term in the
corresponding current-current OPE,
\beq
         J_Y(z)\,J_Y(w) ~=~ {k_Y/2 \over (z-w)^2} ~+~ {\rm regular}~.
\label{kYdef}
\eeq
Defining $k_Y$ in this way then insures that the
tree-level coupling constant relation
(\ref{unification}) holds for abelian group factors as well:
\beq
        8\pi \, {G_N\over \alpha'}~=~ k_Y\, g_Y^2~.
\eeq
Thus, in this manner, an invariant meaning can be given to
the ``affine level'' $k_Y$ of any abelian group factor $U(1)_Y$.

In practice, in a given string model,
the $U(1)$ hypercharge current $J_Y$ will typically be realized as a linear
combination of the elementary $U(1)$ currents $J_i$ ($i=1,...,22$)
which comprise the Cartan subalgebra of the gauge group
of the model:
\beq
               J_Y ~=~ \sum_i \, a_i \, J_i~.
\label{Yembedding}
\eeq
Here $a_i$ are coefficients which represent the
 {\it embedding}\/ of the hypercharge current $J_Y$ into the
charge lattice of the model.
Since each of these individual Cartan currents is normalized
according to Eq.~(\ref{unitnorm}), we then find that
\beq
        J_Y(z)\,J_Y(w)~=~  {\sum_i {a_i}^2 \over (z-w)^2}~+~ {\rm regular}~.
\eeq
Thus, for the current (\ref{Yembedding}), we have
\beq
           k_Y ~=~ 2\, \sum_i \, {a_i}^2~.
\label{kYa}
\eeq
The value of $k_Y$ is therefore completely determined
by the particular hypercharge {\it embedding}\/ $\lbrace a_i\rbrace$,
and is a model-dependent quantity which depends on the way in which
the hypercharge current is realized in a particular string construction.

\subsection{Phenomenologically preferred values of $(k_Y,k_2,k_3)$}

Given these definitions of the levels for the abelian and non-abelian
group factors,
we must now determine the phenomenologically preferred
values of $(k_Y,k_2,k_3)$ which would reconcile string unification
with the experimentally observed values \cite{exptcouplings}
of the low-energy couplings given in Eq.~(\ref{lowenergycouplings}).

The analysis is straightforward, and details can
be found in Ref.~\cite{unif3}.
We begin with the one-loop renormalization group equations (RGE's)
for the gauge couplings in the effective low-energy theory:
\beq
     {{16\pi^2}\over{g_i^2(M_Z)}}~=~k_i\,{{16\pi^2}\over{g_{\rm string}^2}}~+~
     b_i\,\ln{ M^2_{\rm string}\over{M_Z}^2}~+~\Delta_i^{\rm (total)}~.
\label{onelooprunning}
\eeq
Here $b_i$ are the one-loop beta-function coefficients,
and we keep the affine levels $(k_Y,k_2,k_3)$ arbitrary.
In Eq.~(\ref{onelooprunning}), the quantities $\Delta_i^{\rm (total)}$
represent the combined corrections from various string-theoretic
and field-theoretic effects such as
heavy string threshold corrections, light SUSY thresholds, intermediate
gauge structure, and extra string-predicted matter beyond the
MSSM.
These corrections are all highly model-dependent, as they are
strongly influenced by the particular string model or compactification
under study.
For the purposes of the present analysis, we shall ignore all
of these corrections, for our goal in this section is to determine the
extent to which a suitable foundation for string-scale gauge coupling
unification can be established by choosing appropriate values
for $(k_Y,k_2,k_3)$, {\it without}\/ resorting to large corrections from these
other sources.  (The effects of these other sources will
be discussed in the following sections.)
There are, however, some important corrections that are
model-{\it independent}\/:
these include the corrections that arise from
two-loop effects, the effects of minimal Yukawa couplings,
and the effects of renormalization-group scheme conversion
(from the $\overline{\rm DR}$-scheme used in calculating $M_{\rm string}$
to the $\overline{\rm MS}$-scheme used in extracting the
values of the low-energy gauge couplings from experiment).
It turns out that these latter effects are quite sizable,
and must be included.

\begin{figure}
\centerline{
   \epsfxsize 3.2 truein \epsfbox {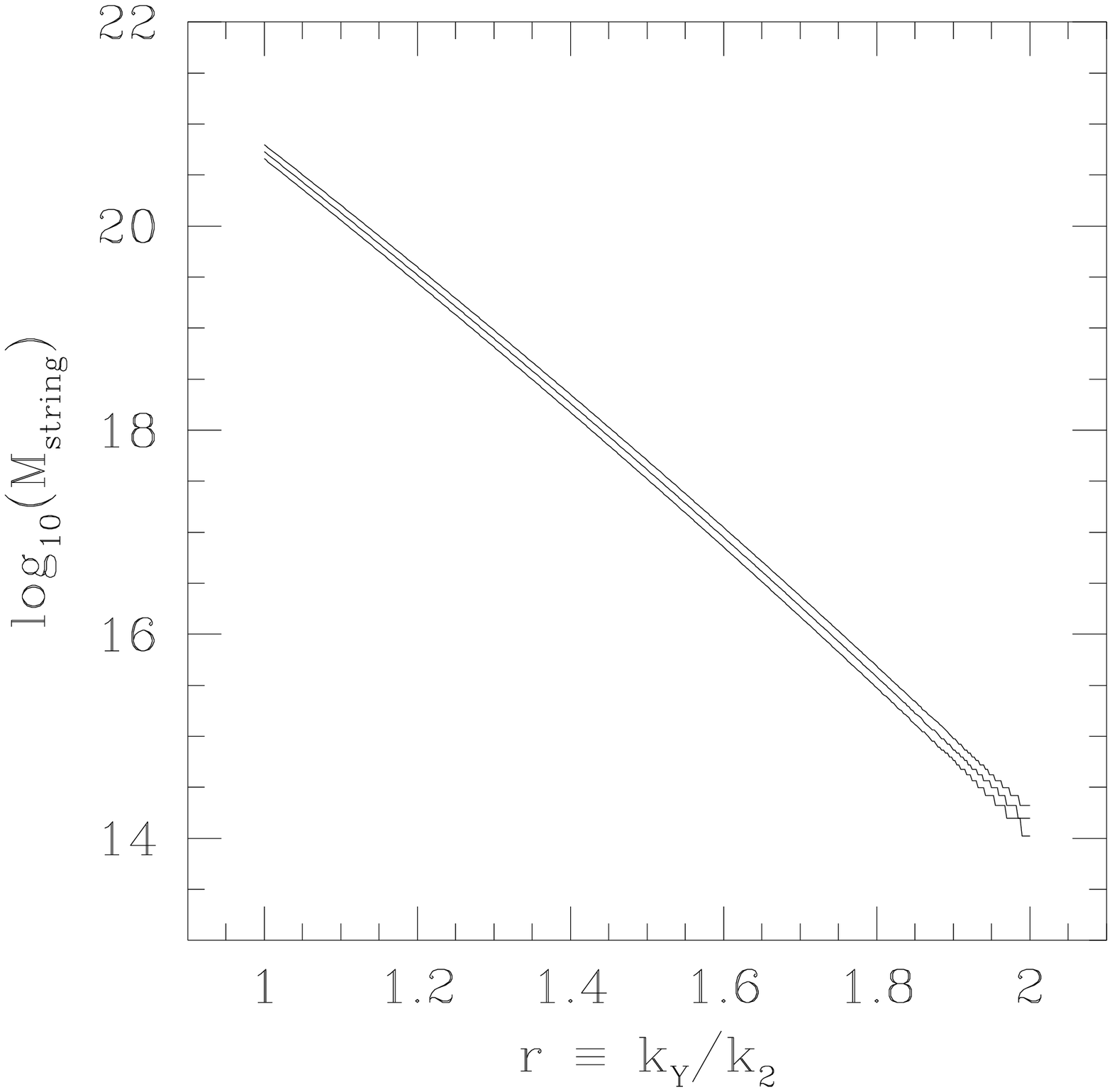}
    }
\caption{Dependence of the unification scale
  $M_{\rm string}$
    on the chosen value of $r\equiv k_Y/k_2$.
   The different curves
  correspond to different values of $\sin^2\theta_W(M_Z)$ within
  its experimental limits, with
 the lower curve arising for greater values.  }
\label{plotMrpa}
\vskip 0.25 truein
\centerline{
   \epsfxsize 3.2 truein \epsfbox {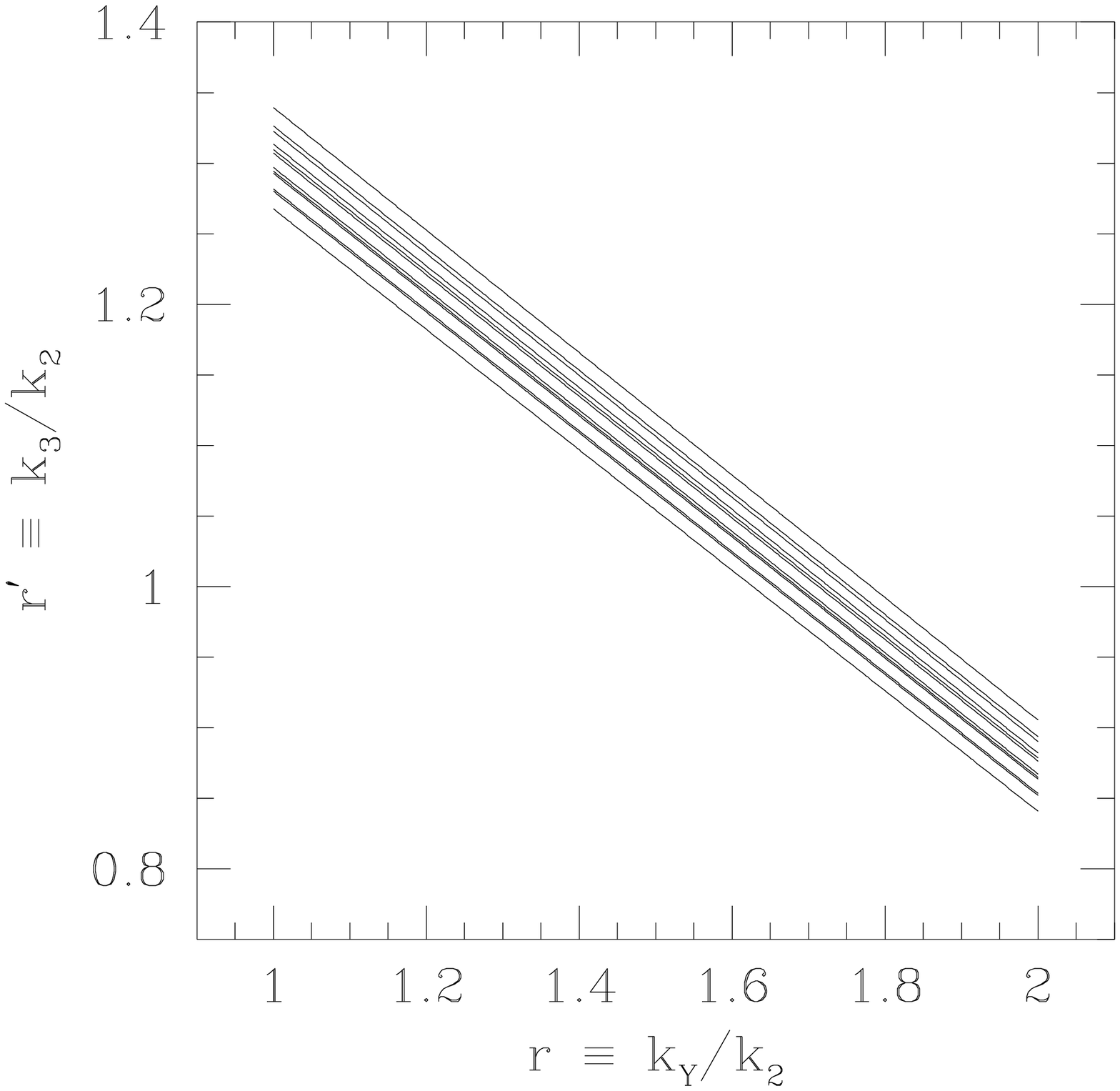}
    }
\caption{Parameter space of allowed values of $r'\equiv k_3/k_2$
  and $r\equiv k_Y/k_2$ that are consistent with the experimentally
   observed low-energy couplings.
 The different curves correspond to different values of the
 couplings within their experimental uncertainties,
  with the higher curves arising for smaller values of
 $\sin^2\theta_W(M_Z)$ and $\alpha_{\rm strong}(M_Z)$, and the
 lower curves arising for greater values.
 Different points on any single curve in this plot
  correspond to different unification scales.  }
\label{plotMrpb}
\end{figure}

Given the one-loop RGE's (\ref{onelooprunning}), we
then eliminate the direct dependence on the unknown string coupling
$g_{\rm string}$ by solving these equations simultaneously
in order to determine the dependence of the
low-energy couplings on the affine levels $(k_Y,k_2,k_3)$.
If we define the ratios of the levels
\beq
           r ~\equiv~ {k_Y\over k_2}
               ~~~~~~~ {\rm and}     ~~~~~~~
           r' ~\equiv~ {k_3\over k_2}~,
\eeq
we then find that the low-energy couplings
$\sin^2\theta_W$ and $\alpha_{\rm strong}$
at the $Z$ scale are given in terms of these ratios as
\beqn
   \sin^2\theta_W(M_Z)  &=&
    {1\over {1+r}}\, \left\lbrack
      1~-~ \left(b_Y - {r}\, b_2\right)\,{a\over 2\pi}\,
      \ln {M_{\rm string}\over M_Z}
       \right\rbrack  ~+~ \Delta^{(\rm sin)} \nonumber\\
     \alpha_{\rm strong}^{-1}(M_Z) &=&
     {r'\over 1+r}\,\left\lbrack
    {1\over a} ~-~ \left( b_Y + b_2 - {1+r\over r'}\,b_3\right)
    \,{1\over 2\pi}\,
  \,\ln {M_{\rm string}\over M_Z}\right\rbrack ~+~
        \Delta^{(\alpha)}~.~~~~~~
\label{RGEs}
\eeqn
Here $a\equiv \alpha_{\rm e.m.}(M_Z)$ is the electromagnetic coupling
at the $Z$ scale, and the correction terms in these equations are
\beqn
     \Delta^{\rm (sin)} &=& -\,{1\over 1+r}\,
    {a\over 4\pi}\,(\Delta_Y-r\,\Delta_2)\nonumber\\
     \Delta^{(\alpha)} &=& -\,{r'\over 1+r}\,
     {1\over 4\pi}\,\left(\Delta_Y+\Delta_2- {1+r\over r'}\,\Delta_3\right)~
\label{sinalphacorrections}
\eeqn
where $(\Delta_Y,\Delta_2,\Delta_3)$ are the combined two-loop, Yukawa,
and scheme-conversion corrections to the individual
couplings $(g_Y,g_2,g_3)$ in Eq.~(\ref{onelooprunning}).
It can be shown \cite{unif3} that these corrections $\Delta_{Y,2,3}$
are approximately independent of the affine levels $k_i$.
Thus, from Eqs.~(\ref{RGEs}) and (\ref{sinalphacorrections}),
we see that the magnitude of $\sin^2\theta_W$ depends on only
the single ratio $r$,
while the magnitude of $\alpha_{\rm strong}$ depends
on only the single ratio $r'/(1+r)$.
These observations imply that
the value of $k_Y/k_2$ can be determined purely by the value of
the coupling $\sin^2\theta_W$ at the $Z$ scale, while
the value of $k_3/k_2$ can then be adjusted in order to maintain
an acceptable value for $\alpha_{\rm strong}$ at the $Z$ scale.

It is straightforward to evaluate these constraints on $(r,r')$.
For the MSSM with
three generations and two Higgs representations,
we have $(b_Y,b_2,b_3)=(11,1,-3)$;
we likewise take as fixed input parameters
\cite{exptinputs} the $Z$ mass $M_Z \equiv 91.161\pm 0.031$ GeV and
the electromagnetic coupling at the $Z$-scale
$a^{-1}\equiv\alpha_{\rm e.m.}(M_Z)^{-1}= 127.9\pm 0.1$.
Given the experimentally measured low-energy couplings
listed in Eq.~(\ref{lowenergycouplings})
and the calculated values of the combined two-loop, Yukawa, and
scheme-conversion
corrections \cite{unif3}
\beq
            \Delta_Y ~\approx~ 11.6     ~,~~~~~~
            \Delta_2 ~\approx~ 13.0    ~,~~~~~~
            \Delta_3 ~\approx~  7.0      ~,
\label{totalcorrections}
\eeq
we can now determine which values of $(r,r')$ yield a successful unification.
The results \cite{unif3} are shown in Figs.~\ref{plotMrpa} and \ref{plotMrpb}.
In Fig.~\ref{plotMrpa},
we show the dependence of
the scale of unification on the choice of the ratio $r$;
as expected, we find that $r=5/3$ leads to a unification scale approximately at
$M_{\rm MSSM}\approx 2\times 10^{16}$ GeV, while unification
at the desired string scale $M_{\rm string}\approx 5\times 10^{17}$ GeV
occurs only for smaller values of $r$, typically $r\approx 1.5$.
Note that this curve relies on only the low-energy input from
$a\equiv \alpha_{\rm e.m.}(M_Z)$ and $\sin^2\theta_W(M_Z)$.
In Fig.~\ref{plotMrpb}, by contrast,
we see how the value
of $r'$ must then be correspondingly adjusted
in order to maintain an experimentally acceptable value for
$\alpha_{\rm strong}(M_Z)$.
Thus, Fig.~\ref{plotMrpb}
summarizes those combinations of $(r,r')$
which are consistent with the phenomenologically acceptable
values for each of the three low-energy couplings.
It is clear from this figure that decreases in $r$ must generally
be accompanied by increases in $r'$ in order to obtain
acceptable low-energy couplings;
moreover, only the approximate region $1.5\leq r\leq 1.8$ is capable
of yielding $r'\approx 1$.
This is an important constraint, for the non-abelian
affine levels $k_2$ and $k_3$
are restricted to be {\it integers},
and thus arbitrary non-integer values of $r'$ could generally
be approximated only with extremely high levels $k_2,k_3\gg 1$.

The analysis thus far has only constrained the values of
the {\it ratios}\/ of the levels $(k_Y,k_2,k_3)$, for
the renormalization group equations (\ref{RGEs}) give us no constraints
concerning the {\it absolute sizes}\/ of the affine levels.
Fortunately, however, there is one additional constraint which must be imposed
in order to reflect the intrinsically stringy nature of the unification.
In field theory, there are ordinarily two independent parameters
associated with unification:  the value of the coupling at the
unification scale, and the unification scale itself.  In string theory,
by contrast, these two parameters are tied together via Eq.~(\ref{Mstringg}).
Thus, in string theory it is actually not sufficient to determine the
ratios $k_Y/k_2$ and $k_3/k_2$
by merely demanding that they agree with low-energy data.
Rather, we must also demand that if our low-energy couplings are run up
to the unification point in a manner corresponding to certain values
of the levels $(k_Y,k_2,k_3)$,
then the value of the predicted coupling $g_{\rm string}$ at the unification
scale must be in agreement with that unification scale.
If this final constraint is not met, we have not achieved
a truly ``stringy'' unification.

This constraint may be imposed as follows \cite{unif3}.  We can
choose one of our three couplings at low energy (the
electromagnetic coupling, for example), and allow it to run
upwards in energy until its value as a function of the mass scale
coincides with the mass scale itself as in Eq.~(\ref{Mstringg}).
This then yields the following transcendental
self-consistency equation \cite{unif3}
which constrains the allowed values of the coupling (or equivalently,
the unification scale)
as a function of the affine levels that govern the running:
\beq
  {1\over \alpha_{\rm str}} ~=~ {1\over k_Y+k_2}\,\left\lbrack {1\over a}
     ~-~ {b_Y+b_2\over 2\pi} \,\ln\,{\sqrt{4\pi\alpha_{\rm str}}\,
  (5.27\times 10^{17}\,{\rm GeV}) \over M_Z\,({\rm GeV}) }\right\rbrack~+~
   \Delta^{(\rm trans)}~.~~~~~
\label{transcendental}
\eeq
Here $\alpha_{\rm str}\equiv g_{\rm string}^2/(4\pi)$, and the
combined two-loop, Yukawa, and scheme-conversion correction
$\Delta^{(\rm trans)}$ is given in terms of the individual corresponding
$\Delta_i$ in Eq.~(\ref{onelooprunning}) by
\beq
        \Delta^{(\rm trans)}  ~=~ -\,{\Delta_Y+\Delta_2\over k_Y + k_2}~.
\label{twoloopcorrectiontrans}
\eeq
The important thing to notice about this self-consistency constraint
is that it depends on the {\it absolute values}\/ of the affine levels,
and not only on their ratios.  Thus, by combining this constraint with
our previous experimental constraints, we can fix not only the
ratios $k_Y/k_2$ and $k_3/k_2$ of the affine levels,
but also their absolute values.

\begin{figure}
\centerline{
   \epsfxsize 3.3 truein \epsfbox {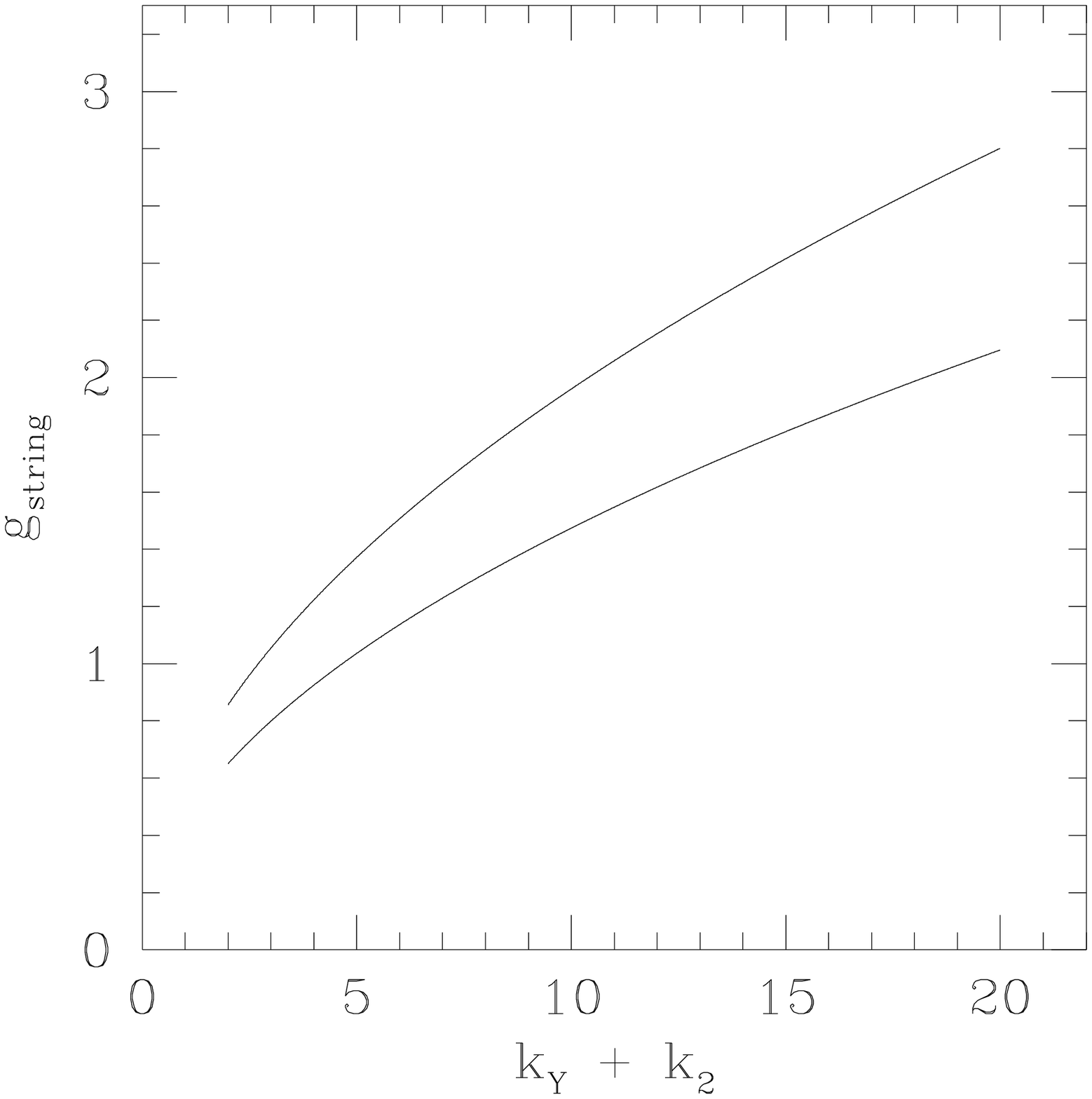}
    }
\caption{
  Numerical solutions
  to the transcendental equation (\protect\ref{transcendental})
  for different values of $k_Y+k_2$,
   with the two-loop, Yukawa, and scheme-conversion
  corrections included (upper curve), and omitted (lower curve).  }
\label{plotgk2a}
\vskip 0.25 truein
\centerline{
   \epsfxsize 3.3 truein \epsfbox {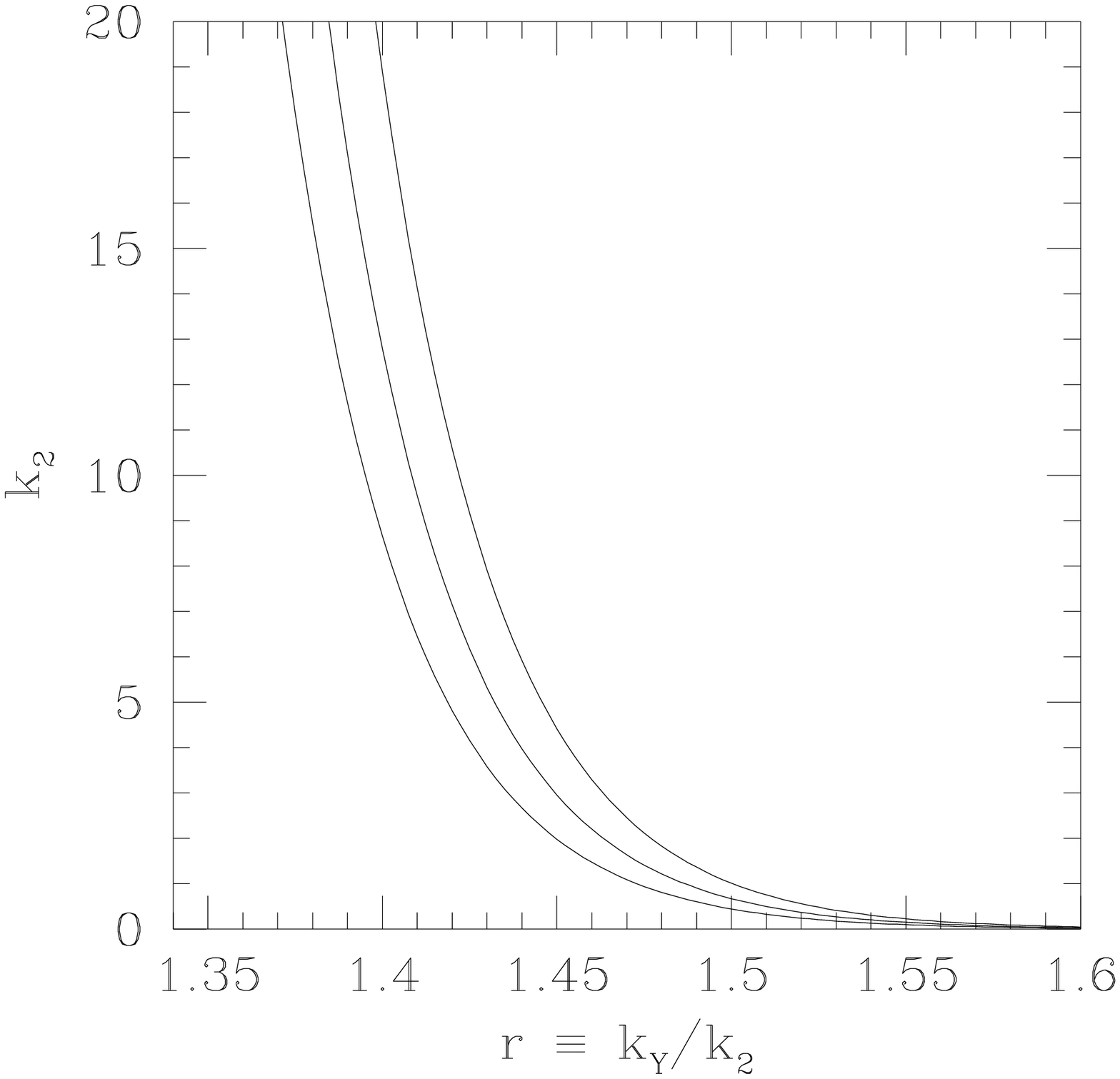}
    }
\caption{
  Dependence of
  $k_2$ on $r\equiv k_Y/k_2$.  In this plot,
  the different curves correspond
   to different values of $\sin^2\theta_W(M_Z)$, with the
   lower/left curves arising for greater values.}
\label{plotgk2b}
\end{figure}

These final results are shown in Figs.~\ref{plotgk2a} and \ref{plotgk2b}.
In Fig.~\ref{plotgk2a},
we display
the numerical solutions to
the transcendental self-consistency equation (\ref{transcendental}),
plotting $g_{\rm string}\equiv\sqrt{4\pi\alpha_{\rm str}}$
as a function of $k_Y+k_2$, both with and without
the two-loop, Yukawa, and scheme-conversion corrections.
As we can see, the effect of the corrections (\ref{twoloopcorrectiontrans})
is quite significant.
For $k_Y+k_2=5/3+1=8/3$, we find, as expected, that $0.7\leq g_{\rm string}
\leq 1.0$,
while for higher values of $k_Y+k_2$, the required string coupling increases
significantly.
Finally, combining this result with
Fig.~\ref{plotMrpb},
we obtain Fig.~\ref{plotgk2b}:
this strong dependence between $r\equiv k_Y/k_2$ and $k_2$
arises because the choice of a particular value of
$r\equiv k_Y/k_2$ not only sets a particular unification scale, but with that
choice of scale comes a certain scale for the string coupling via
Eq.~(\ref{Mstring}), and this in turn requires, via Eq.~(\ref{transcendental}),
a certain absolute magnitude for the affine levels.
This combined dependence is clearly quite dramatic.
In particular, we see from this figure that values of $r$ in the range
$1.45\leq r\leq 1.5$
are consistent with relatively small affine levels $k_{2,3}=1,2$, but
that for smaller values of $r$, the required affine levels increase
dramatically.
Of course, the precise placement of the curves in this plot
depends quite strongly
on the exact value of $\sin^2\theta_W(M_Z)$ (as shown), as well as on
the various threshold corrections we have been neglecting.
Indeed, although we expect these curves to remain essentially independent
of these corrections above $r\approx 1.45$, we see that the region $r<1.45$
depends {\it exponentially}\/ on the precise corrections.
Thus, the lower-$r$ regions of these curves are best
interpreted
as qualitative only.  Nevertheless, the general shapes of these curves,
evidently requiring extremely large changes in the absolute values of
$k_2$ and $k_3$ for seemingly modest shifts in the
ratio $r\equiv k_Y/k_2$,
are striking and should provide extremely strong constraints on
realistic string model-building.

It is clear, then,
that only certain tightly-constrained values
$(k_Y,k_2,k_3)$ are phenomenologically allowed if we are to achieve
string-scale unification with only the MSSM spectrum, and {\it without}\/
resorting to large corrections from heavy string thresholds or
extra non-MSSM matter.  Indeed, from
Figs.~\ref{plotMrpb} and \ref{plotgk2b},
we see that the best  phenomenologically allowed regions
of $(k_Y,k_2,k_3)$ exist in a narrow band stretching through
different values of $r$ in the range $1.4\leq r\leq 1.5$.
For example,
at the lower end of this band ($r\approx 1.4$) we have
values such as $(k_Y,k_2,k_3)\approx(21,15,17)$,
whereas at intermediate regions of the band
($r\approx 1.42$) we have values such as $(k_Y,k_2,k_3)\approx (14.2,10,11)$,
and at higher regions ($r\approx 1.5$) we have
$(k_Y,k_2,k_3)\approx (1.5,1,1)$ or $(3,2,2)$.

Of course, strictly speaking, the {\it smaller}\/ values of
$r\approx 1.4$ are phenomenologically preferred, since they
require the larger absolute values of $(k_2,k_3)$ which in turn
enable us to more precisely approximate the required corresponding
values of $r'$
shown in Fig.~\ref{plotMrpb}.
However, for  the purposes of {\it string model-building},
it turns out that smaller values for the non-abelian affine levels,
such as $k_2,k_3=1 ~{\rm or}~2$, are strongly preferred on practical grounds.
Indeed, not only does it grow increasingly difficult to build string models
with higher values of $(k_2,k_3)$, but unwanted $SU(3)$ and $SU(2)$
representations of increasing dimensionality begin to appear in the
massless spectrum as these non-abelian levels are increased.
We have seen both of these features in  Sect.~4.
Thus, for the purposes of string model-building,
the preferred regions
of parameter space are actually in the {\it higher}\/-$r$ region,
with $r\approx 1.5$ and with $k_2=k_3=1$ or $2$.
One would then hope that relatively {\it small}\/ corrections from other
sources such heavy string thresholds
would increase the effective value
of $r'$ from $r'=1$ to $r'\approx 1.05-1.1$.
Furthermore, in this higher-$r$ region, such corrections are not
expected to have a major effect on the validity of the
curves in Fig~\ref{plotgk2b}.

Thus, in this ``path to unification'',
one attempts to realize string models or embeddings with
affine levels in the range
\beq
       {\rm preferred~values}:~~~~~~~\cases{
        k_2~=~k_3~=~1 ~~{\rm or}~~ 2&~\cr
        r~\equiv~ k_Y/k_2~\approx~ 1.45 - 1.5~&~.\cr}
\label{preferredvalues}
\eeq

\subsection{Hypercharge embeddings with $k_Y/k_2<5/3$ ?}

Given the preferred affine levels in Eq.~(\ref{preferredvalues}),
the next question is to ask whether such levels
(or combinations of levels)
can be realized in string theory.
What are the constraints that govern which combinations
of $(k_Y,k_2,k_3)$ are mutually realizable in self-consistent
realistic string models?

Certain constraints are rather simple to formulate.
For example, as we have already mentioned,
worldsheet symmetries alone (namely, unitarity of the
worldsheet conformal field theory) dictate
that for non-abelian groups $G$, the corresponding affine levels
$k_G$ are restricted to be integers.  Thus $k_2,k_3\in \IZ$.
By contrast, $k_Y$ is the normalization of the {\it abelian}\/
group factor $U(1)_Y$, and thus
the value of $k_Y$ is arbitrary
(as far as the worldsheet theory is concerned).
Nevertheless, there are also spacetime phenomenological constraints
that also come into play.
For example, it is straightforward to show \cite{Schellekens}
that any phenomenologically consistent string
model containing the MSSM  right-handed electron singlet state $e_R$
must have $k_Y\geq 1$.
This constraint essentially arises because
the right-handed electron is a color- and electroweak-singlet in the MSSM.
This implies that the sole contribution to the conformal dimension of
this state from the Standard Model
gauge group factors is solely that from its weak hypercharge $Y_{e_R}$.
This state must also be massless, however.  From
Eq.~(\ref{masslessness}), we then find
\beq
                 {{\left(Y_{e_R}\right)}^2\over k_Y} ~\leq~1~,
\eeq
and since $Y_{e_R}=1$ in the appropriate normalizations,
we find $k_Y\geq 1$.
Indeed, given all of the MSSM representations, it turns out that this is
the strongest constraint that can be obtained in this manner.

The question then arises as to whether other phenomenological constraints can
also be imposed on the value of $k_Y$.  A general expectation
might be that $k_Y=5/3$ is the minimal value allowed, simply because
the standard $SO(10)$ embedding is an
extremely economical way to embed all three generations with universal
weak hypercharge assignment.  Indeed, all of the realistic
level-one string models constructed to date have $k_Y\geq 5/3$, and
most trivial extensions of the $SO(10)$ embedding tend
to {\it increase}\/ the value of $k_Y$.
Indeed, various initial searches for string models realizing $k_Y<5/3$
had been unsuccessful \cite{prevattempts}.
However, the possibility remains that there might
exist special, isolated string models
or constructions which manage to circumvent this bound, and populate the
range $1\leq k_Y<5/3$.
Thus, it is an important question to determine what values of $k_Y$ are
allowed in string theory, and to determine whether the desired
values $k_Y\leq 5/3$ (or, more generally, $k_Y/k_2\leq 5/3$)
are even possible.

As we discussed at the beginning of this section, the
``level'' $k_Y$ corresponding to any abelian group factor is
defined through the relation (\ref{kYdef}).   Given a particular
hypercharge embedding as in Eq.~(\ref{Yembedding}),
the value of $k_Y$ is determined by Eq.~(\ref{kYa}).
Of course, the particular hypercharge embedding that arises
in a given string model is fixed in turn
by the embeddings of the corresponding MSSM representations.
Specifically, as we discussed in Sect.~4,
corresponding to each MSSM state $s\in \lbrace
Q_L,u_R,d_R,L_L,e_R,H^+,H^-\rbrace$
there exists a charge vector $\bQ^{(s)}$,
and we must carefully choose the coefficients $a_i$ in such a way that
the quantity $\sum_i a_i Q_i^{(s)} $ reproduces the correct MSSM hypercharge
assignment $Y^{(s)}$ for each state.
This is a strong constraint, and it is not always possible to find
such solutions in a given string model.

As a first step, therefore, we seek to determine whether there exist
consistent MSSM embeddings that permit hypercharge solutions with $k_Y<5/3$.
If so, we then seek to know whether such embeddings can be
realized in consistent and realistic string models.

\subsubsection{``Minimal'' hypercharge embeddings}

We first concentrate on recent results concerning
the so-called ``minimal'' embeddings,
since most realistic string models constructed to date employ such
embeddings.
By a ``minimal'' embedding, we refer to a hypercharge embedding in which
the coefficients $a_i$ vanish for all of the lattice directions
except those corresponding to the non-abelian MSSM gauge group
factors $SU(3)_C\times SU(2)_W$.
In most standard level-one string constructions, the color $SU(3)_C$
group factor is embedded within the first three lattice components ($i=1,2,3$),
and the electroweak $SU(2)_W$ group factor
is embedded within the next two components ($i=4,5$).
Thus, in such cases, minimal hypercharge embeddings
are those for which $a_i=0$ for $i\geq 6$.

Due to their tightly constrained nature, such minimal embeddings
are particularly straightforward to analyze, and one can systematically
attempt to construct MSSM embeddings ({\it i.e.}, charge vectors $\bQ$
for each of the MSSM representations plus Higgs representations)
in such a way that they permit a consistent hypercharge
assignment with $k_Y<5/3$.
Of course, these MSSM embeddings are subject to a
variety of constraints \cite{unif3}.
First, each MSSM charge vector must have the correct form corresponding
to its representation under $SU(3)_C\times SU(2)_L$.  Second, the
length of each such charge vector is constrained by the appropriate
masslessness
condition.  Third, each representation must have a charge vector
which allows it to be realized as a {\it chiral}\/ state;  indeed,
some charge vectors can be shown to correspond to non-chiral states
only, with their chiral conjugates always present in the model.
Another potential constraint that one may choose to impose is that
these MSSM charge vectors should be consistent with having
the required mass terms arise at the cubic level of the superpotential.
Finally, one may also impose a {\it moding}\/ constraint on the charge
vectors;  such a constraint generally reflects a particular set
of allowed boundary conditions in a fermionic string construction,
or equivalently a certain $\IZ_N$ twist
in a bosonic formulation.

By analyzing all possible minimal embeddings, various results have
recently been obtained.
First, it has been shown that for a large class of realistic
string models, one must always have \cite{unif3}
\beq
       k_Y ~= ~5/3\, ,~11/3\,,~ {\rm or}~ 14/3~.
\eeq
This class includes all level-one string models
with minimal embeddings in
which the MSSM matter arises in sectors with $\IZ/2$ modings.
Thus, for such models, we must always have $k_Y\geq 5/3$.

By contrast, it has also been shown that it is
possible to circumvent this result by considering minimal
embeddings beyond this class
({\it i.e.}, embeddings with higher affine levels for the non-abelian
MSSM group factors and/or higher modings for the MSSM matter).
Indeed, although {\it almost all}\/
of the consistent hypercharge embeddings beyond this class have
$k_Y/k_2 \geq 5/3$, it has been found \cite{unif3} that
there do exist several isolated embeddings which have $k_Y<5/3$.
For example, given the level-one $SU(3)_C\times SU(2)_L$ embedding
which is typically employed in the so-called ``NAHE'' models \cite{NAHE},
the following MSSM matter embedding
\beqn
      Q_L:~~&&~~~~~\bQ=(-1/2,-1/2,1/2,0,1)\nonumber\\
      u_R:~~&&~~~~~\bQ=(1/4,1/4,-3/4,-3/4,-3/4)\nonumber\\
      d_R:~~&&~~~~~\bQ=(3/4,3/4,-1/4,-1/4,-1/4)\nonumber\\
      L_L:~~&&~~~~~\bQ=(-1/4,-1/4,-1/4,-3/4,1/4)\nonumber\\
      e_R:~~&&~~~~~\bQ=(1/2,1/2,1/2,1/2,1/2)\nonumber\\
    H^+:~~&&~~~~~\bQ=(1/4,1/4,1/4,3/4,-1/4)\nonumber\\
    H^-:~~&&~~~~~\bQ=(-1/4,-1/4,-1/4,-3/4,1/4)~
\eeqn
has a consistent hypercharge solution
\beq
  Y~=~{5\over 12}\,(Q_1+Q_2+Q_3)~+~ {3\over 8}\,(Q_4+Q_5)~.
\eeq
This embedding, which satisfies all of the above-mentioned
constraints including the cubic-level mass-term constraint,
corresponds to $k_Y=77/48$, which is less than $5/3$.
Clearly, this embedding makes use of $\IZ/4$ modings.
Likewise, there also exist $k_Y/k_2<5/3$ embeddings which
rely on having $k_2>1$ \cite{unif3}.
Unfortunately, it is not yet known whether such
embeddings can be realized in actual string models.
Thus, it is not known whether such embeddings are consistent
with other desirable phenomenological properties such
as $N=1$ spacetime SUSY, three and only three chiral generations,
anomaly cancellation, and no unacceptable exotic matter.

\subsubsection{``Extended'' hypercharge embeddings}

An alternative recent approach is to examine the so-called ``extended''
hypercharge embeddings \cite{274model,shynew} in which
the hypercharge current involves Cartan generators beyond those
employed in realizing the non-abelian MSSM group factors.
Because the structure of such embeddings is significantly less
constrained than those of the minimal embeddings,
they cannot be analyzed in a similar constructive manner.
Nevertheless, by adequately sampling the space of string
models ({\it e.g.}, by computer), it is possible to
find models that employ such extended hypercharge
embeddings \cite{shynew}.
Moreover, it has recently been found that some
of these models exhibit hypercharge embeddings with $k_Y<5/3$
along with $N=1$ spacetime SUSY, three complete chiral
generations, and $U(1)$ anomaly cancellation.
This, then, confirms the existence of string models with
$k_Y<5/3$.
In fact, several of these models also exhibit {\it continuously variable}\/
values of $k_Y$;  this surprising flexibility results from
the freedom to continuously deform the particle identification
in these models,
ultimately providing a one-parameter family of possible charge
assignments and corresponding values of $k_Y$ within a single
model.
Unfortunately, all of these models turn out to contain
fractionally charged states that could survive in their light spectra.
Whether such states may be removed
without destroying these features or increasing $k_Y$
remains an open question.

\subsection{Avoiding fractionally charged states}

The appearance of color-neutral string states
carrying fractional electric charge is
a well-known problem in string theory \cite{Wen}, and is one of the problems
that
must be faced in any attempt to build a realistic string model.
Such states represent  a problem because of the strong
constraints that stem from the failure
of numerous experimental searches
for such fractionally charged particles \cite{fractionalexpt}.
Unfortunately, the problem of color-neutral fractionally charged
states in string theory becomes even more severe when attempting
to build string models with non-standard values of $(k_Y,k_2,k_3)$,
for there exists a deep relation \cite{Schellekens} between the
values of these levels and the types of fractionally charged states
that can appear in the corresponding string model.
We shall now discuss this relation and some recent developments
concerning its implications for gauge coupling unification.

The basic idea behind this relation can be found in Ref.~\cite{Schellekens},
which in  turn rests upon ideas developed in Refs.~\cite{schellyank,keni}.
Let us begin by supposing that our string model is based upon a rational
worldsheet conformal field theory, and is
to contain only those states which (after forming color-neutral
bound states) carry integer electric charge.
Given this requirement, it is possible to construct a certain
current $\hat J$ in the worldsheet conformal field theory which satisfies
two properties.  First, $\hat J$ must be a {\it simple current}\/, meaning
that it must have a one-term fusion rule with all
other primary fields $\phi$ in the theory:
\beq
        \hat J ~\times~ \phi ~=~ \phi'~.
\label{uniquefusion}
\eeq
Second, $\hat J$ must be {\it local}\/ with respect to all other primary
fields  $\phi$ which appear in the theory, so that the $\hat J\phi$ OPE
takes the form
\beq
     \hat J(z) \phi(w) ~\sim~ { \phi'(w) \over (z-w)^{\alpha}}
                  ~+~ ...~
\label{opelocal}
\eeq
where the exponent $\alpha \equiv h(\hat J) + h(\phi) - h(\phi')$
is always an integer.
Since this exponent can ultimately be related to the electric charge of the
state created by $\phi$, it is the assumption of charge integrality
for all of the string states which allows us to
construct a current $\hat J$ for which Eq.~(\ref{opelocal}) is
always satisfied.
Given any current $\hat J$ meeting these
two conditions, it can then be shown that such a current must be a field
which is also present in the theory.
However, if this current is to be present in the theory,
modular invariance then requires that its conformal dimension
be an integer.  Since the conformal dimension of $\hat J$ depends on
the affine levels $(k_Y,k_2,k_3)$, we thereby obtain a constraint on these
levels.
In other words, we thereby
determine which combinations of affine levels are consistent
with modular invariance and integrally charged states.

The result one finds \cite{Schellekens} by carrying out this
procedure is relatively simple:  string models can be modular invariant
and free of color-neutral fractionally charged states
if and only if
\beq
     {k_3\over 3} ~+~ {k_2 \over 4} ~+~ {k_Y\over 4} ~\equiv~ 0
                ~~~~~~({\rm mod}~1)~.
\label{simplelevelcond}
\eeq
It is easy to see, then, that for $k_2=k_3=1$, the minimum
allowed value of $k_Y$ is indeed $k_Y=5/3$.
Thus, for level-one string models with only integer-charged states,
we have
\beq
                      k_Y ~\geq~ 5/3~.
\eeq
We also obtain the same result for level-two models.
Recall from Eq.~(\ref{preferredvalues})
that gauge coupling unification requires $k_2=k_3$.
Taking $k_2=k_3=2$ in Eq.~(\ref{simplelevelcond}) shows that
$k_Y\geq 10/3$, so that once again $k_Y/k_2 \geq 5/3$.

In general, though, string models {\it do}\/ contain color-neutral
fractionally charged states.
In fact, as also shown in Ref.~\cite{Schellekens},
it is impossible to have a level-one $SU(3)\times SU(2)\times U(1)_Y$
string model with $k_Y=5/3$ {\it without}\/ having such fractionally charged
states in the corresponding spectrum, for any GSO projections that would
remove all of the fractionally charged states
in such cases would also simultaneously promote
the $SU(3)\times SU(2)\times U(1)_Y$ gauge symmetry to level-one $SU(5)$.
How then can we reconcile our desire to
have string models with $k_Y<5/3$ while simultaneously avoiding
fractionally charged states?

One possibility is that such states can appear provided that
they are {\it confined}\/ under the influence of an additional confining
``hypercolor'' gauge interaction coming from a hidden sector beyond
the MSSM.  Indeed, heterotic string models typically have large (rank $\leq
18$)
hidden-sector gauge groups $G$, and thus it is conceivable that although
fractionally charged states may appear if $k_Y<5/3$, such states may fall into
the particular representations of $G$ that permit them to be confined
by $G$ into bound states with integral electric charge.
However, it is first necessary to generalize the
condition (\ref{simplelevelcond})
to such cases, and to determine if there actually exist such
confining scenarios which
permit solutions with $k_Y\leq 5/3$.
While various generalizations
of Eq.~(\ref{simplelevelcond})
have been obtained \cite{ELN,prevschell} for
the specific binding scenarios and hypercolor groups
available in certain string models with $k_Y\geq 5/3$,
we here require a more general analysis which is capable of
surveying all values of $k_Y$ and all possible binding groups $G$.

Such an analysis has recently been performed \cite{unif3},
and the results are as follows.
For $k_2=k_3=1$, it is found that there are only sixteen
possible
choices of simple hypercolor groups $G$ and corresponding
values of $k_Y<5/3$
for which all fractionally charged states can potentially be confined.
Those with values of $k_Y$ in the phenomenologically
interesting range $1.4\leq k_Y\leq 1.5$ are as follows:
\beq
\begin{tabular}{c|c}
  ~~~~$G$ ~~~~&   ~~~$k_Y$~~~      \\
\hline
     $ SU(10)_1$  & $ 22/15 $\\
     $ SU(17)_1$ & $     73/51 $\\
     $ SU(19)_1$ & $   83/57 $\\
\end{tabular}
\label{results1}
\eeq
The subscript on the hypercolor group $G$ indicates its level.
For $k_2=k_3=2$, by contrast, there are more solutions.  Those
permitting values of $k_Y$ in the range
$1.4\leq k_Y/k_2\leq 1.5$ are listed below:
\beq
\begin{tabular}{c|c|| c | c}
  ~~~~$G$ ~~~~&   ~~~~$k_Y/k_2$~~~~~       &
  ~~~~$G$ ~~~~&   ~~~$k_Y/k_2$~~~~       \\
\hline
 $ SU(4)_3    $ &   17/12     &
 $ SU(12)_1    $ &    17/12     \\
 $ SU(6)_5    $ &     3/2     &
 $ SU(17)_1    $ &    73/51     \\
 $ SU(9)_1    $ &     13/9     &
 $ SO(18)_1    $ &   17/12     \\
 $ SU(10)_2    $ &   22/15     &
 $ SO(34)_1    $ &    17/12     \\
 $ SU(11)_1    $ &   49/33     &
 $ ~  $ & $ ~$    \\
\end{tabular}
\label{results2}
\eeq
Thus, for $k_2=k_3=2$, we see that
relatively small hypercolor
groups can now confine the fractionally
charged states that arise
for $1.4\leq k_Y/k_2\leq 1.5$.  In particular, $SU(4)_3$, $SU(6)_5$,
and $SU(9)_1$
are the most likely candidates for realization in
consistent higher-level string models.

There are also other possible ways of avoiding
fractionally charged states.
Since the charge-integrality constraints we have discussed
apply to {\it all}\/ string states,
such constraints might be evaded if we require that only
the {\it massless}\/ states be integrally charged.
Another possibility would be to impose the even weaker
requirement that only those massless states which
are {\it chiral}\/ must be integrally
charged;  after all, vector-like states which are fractionally charged and
massless at tree level can acquire potentially large masses at higher loops
({\it e.g.}, via the shift of the moduli which is generally
required in order to break pseudo-anomalous $U(1)$ gauge symmetries and restore
spacetime supersymmetry \cite{DSWshift}).
Indeed,
such vacuum shifts along flat directions have the potential
to make various fractionally charged states superheavy
\cite{Huet,Alonmassestimate}, and the mechanism by which this occurs
will be discussed in more detail in Sect.~8.2.
Such scenarios, however, are highly model-dependent, and thus
cannot be readily incorporated into this sort of general
analysis.
They therefore must be analyzed in a model-by-model fashion.


\setcounter{footnote}{0}
\section{Path \#3:  Heavy String Threshold Corrections}

We now turn to the paths that involve adding ``correction terms''
to the MSSM RGE's.  These corrections include those that arise
from heavy string thresholds,
light SUSY thresholds, possible intermediate-scale gauge structure, and
possible intermediate-scale matter beyond the MSSM.
In all of these scenarios, therefore, we essentially preserve the MSSM gauge
structure with the standard MSSM levels $(k_Y,k_2,k_3)=(5/3,1,1)$,
and instead seek to determine whether the correction terms
$\Delta_i$ that such effects generate
can be sufficient to resolve the discrepancy
between the naive MSSM unification scale and the
unification scale expected within string theory.
These correction terms $\Delta_i$ influence the
running of the gauge couplings as indicated in Eq.~(\ref{onelooprunning}).
In this section we shall concentrate on the corrections
that arise from the heavy string thresholds,
and defer our discussion of the remaining correction terms
to Sects.~7 and 8.

\subsection{Calculating heavy string threshold corrections}

The so-called ``heavy string threshold corrections'' are the one-loop
corrections from the infinite towers of massive string states that
are otherwise
ignored in studies of the low-energy spectrum.
Thus, these corrections represent a purely ``stringy'' effect.
Although these states have Planck-scale masses,
there are infinitely many of them, and thus
the combined contributions $\tilde \Delta_i$ from such heavy
string modes have the potential to be quite sizable.
Since these corrections depend on the precise distributions
of states at all mass levels of the string, they depend crucially
on the particular choice of string vacuum.  Thus, these corrections
are generally moduli-dependent.

\begin{figure}[th]
\centerline{
   \epsfxsize 5.0 truein \epsfbox {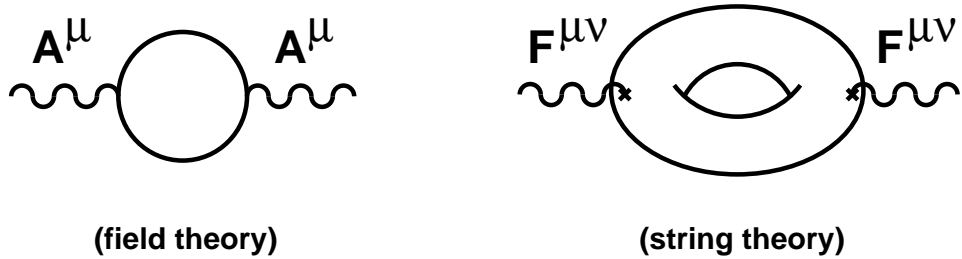}
    }
\caption{Calculation of the running of gauge couplings
   in field theory and string theory.
   In field theory, the background-field method involves evaluating
    the one-loop vacuum polarization diagram shown at left.
   In string theory, this polarization diagram generalizes to become the
    torus diagram shown at right.
    However, the complete calculation in string theory
    (including gravitational back-reactions)
    requires the calculation of the full correlation
    function $\langle F^a_{\mu\nu} F_a^{\mu\nu}\rangle$
    evaluated on the torus in the presence of a suitable regulator.}
\label{torusdiagram}
\end{figure}

In general, for a given string vacuum and gauge group $G$,
the corresponding threshold corrections
$\tilde \Delta_G$ are determined to one-loop order
by calculating the
one-loop renormalization of the gauge couplings
in the presence of a background gauge field.  The
threshold correction $\tilde\Delta_G$ is then deduced by comparing
the coefficients of
the gauge field-strength terms ${1\over 4} F^a_{\mu\nu} F_a^{\mu\nu}$
in the bare and one-loop effective Lagrangians.
As illustrated in Fig.~\ref{torusdiagram},
this renormalization calculation is essentially analogous to the
standard field-theoretic one-loop vacuum polarization calculation,
except that in string theory the full calculation involves evaluating
the two-dimensional (worldsheet) correlation function
$\langle F^a_{\mu\nu} F_a^{\mu\nu}\rangle$ on the torus.
This procedure generalizes the field-theoretic vacuum polarization
calculation in such a way that the full gravitational back-reactions
of string theory are automatically incorporated.

This calculation was originally performed
by considering a string propagating in a non-trivial gauge-field background
that obeys the classical string equations of motion and thereby preserves
worldsheet conformal invariance \cite{Kaplunovsky};  an
alternative approach uses the background-field method
in which the gauge field is treated as a classical background in
which the string propagates, and is quantized accordingly \cite{DL}.
Both calculations yield similar results, however, and it is
found that $\tilde \Delta_G$ splits into two pieces of the form
\beq
    \tilde \Delta_G ~=~ k_G\, Y ~+~\Delta_G ~.
\label{threshgenform}
\eeq
Here $k_G$
is the affine level of the gauge group $G$;  $Y$ is a certain
quantity\footnote{This quantity $Y$ bears no relation to the hypercharge.}
    \cite{Yterm,Ytermsize,Kiritsis,PR}
which is {\it independent}\/ of the gauge group in question;
and the remaining term $\Delta_G$ has a highly non-trivial
dependence on the gauge group.
In general, $Y$ receives the contributions arising
from gravitational back-reactions
of background gauge fields and universal oscillator excitations
(and is expected to be relatively small \cite{Ytermsize}),
while $\Delta_G$ receives contributions from massive compactification modes.
Indeed, as we shall see, $\Delta_G$ simply tallies the contributions of
all of the massive string states as they propagate
around the one-loop torus diagram,
and thus represents the generalization of the field-theoretic diagram
as illustrated in Fig.~\ref{torusdiagram}.
The importance of such contributions to the running of the
effective low-energy couplings was pointed out early on (for example,
in Ref.~\cite{choi}), even before the formalism for calculating
these string corrections was developed.

The decomposition in Eq.~(\ref{threshgenform}) is particularly useful
because the term $k_G Y$ can simply be absorbed into
a redefinition of the string coupling at unification:
\beq
         {1\over g_{\rm string}^2} ~\to~
         {1\over g_{\rm string}^2} + {Y\over 16\pi^2}~.
\label{gstringrenormalized}
\eeq
This is convenient, for
a proper calculation of $Y$ is fairly complicated and depends
on having a suitable prescription for handling infrared
divergences in string theory.
We shall discuss this issue in Sect.~6.2.
Moreover, for phenomenological purposes,
we are often interested in only the {\it relative}\/ running
of the gauge couplings between different gauge group factors.
If these different gauge groups are realized with equal affine levels,
then the relative thresholds $\tilde \Delta_G-\tilde \Delta_{G'}$ will be
independent of $Y$.
Recall from Eq.~(\ref{preferredvalues}), for example, that equal affine
levels are precisely what are required for the non-abelian
group factors of the MSSM in order to achieve unification.
We shall therefore defer a discussion of $Y$ to Sect.~6.2, and
instead focus on the remaining term $\Delta_G$
in Eq.~(\ref{threshgenform}), which is found
(in the $\overline{\rm DR}$ renormalization group scheme)
to be given\footnote{
   Note that the expression in Eq.~(\ref{deltadef})
   actually contains an additional group-independent contribution $Y'$.
   Thus, one could perform a further decomposition of the form
   $\Delta_G= k_G Y' + \Delta'_G $, so that the full decomposition
   in Eq.~(\ref{threshgenform}) would take the final form
   $\tilde \Delta_G = k_G(Y+Y') + \Delta'_G$.
   However, because of the simple form of the intermediate
   expression (\ref{deltadef}) and its interpretation as the
   straightforward generalization of the field-theoretic result, this further
   decomposition is not traditionally used.}
by \cite{Kaplunovsky}:
\beq
 \Delta_G ~\equiv ~   \int_{\cal F} {d^2\tau\over ({\rm Im}\,\tau)^2}\,
     \biggl\lbrack B_G(\tau) ~-~ ({\rm Im}\,\tau)\,b_G \biggr\rbrack~.
\label{deltadef}
\eeq
Here $b_G$ is the one-loop beta-function coefficient
for the group $G$ (as determined in the effective theory by considering
only the massless modes of the string), and $\tau$ is the toroidal complex
modular
parameter whose integration over the modular group
fundamental domain ${\cal F}$ represents the summation over all
conformally-inequivalent toroidal geometries.
$B_G(\tau)$ is the so-called {\it modified partition function}.

This modified partition function $B_G(\tau)$ is defined as follows.
Recall that the ordinary string one-loop
partition function $Z(\tau)$ is defined
as
\beq
 Z(\tau) ~\equiv~ \sum_{\alpha}
        \,(-1)^F \, {\rm Tr}_\alpha \,
    \overline{q}^{\overline{H}_\alpha}\,q^{H_\alpha}~.
\label{Zdef}
\eeq
Here $\alpha$ labels the different sectors of the theory;
$F$ is the spacetime fermion number;
${\rm Tr}_\alpha$ indicates a trace over all of the states in
the Fock space of each sector $\alpha$;
$q\equiv \exp(2\pi i\tau)$;
and $(H_\alpha,\overline{H}_\alpha)$
respectively
represent the left- and right-moving worldsheet Hamiltonians in sector
$\alpha$.
Thus, the partition function $Z(\tau)$ essentially counts the numbers
of states (weighted by $+1$ for bosons, $-1$ for fermions)
at all mass levels in the theory.
Of course, for consistent string theories, the partition function
$Z(\tau)$ must be modular invariant.
This means that $Z(\tau)$ satisfies the equations
\beq
          Z(\tau) ~=~ Z(\tau+1)~=~Z(-1/\tau)~.
\eeq

The modified partitions $B_G(\tau)$ are defined in a similar manner,
and differ from the ordinary partition function $Z(\tau)$ only through
the insertion of a special operator into the trace:
\beq
        B_G(\tau) ~\equiv~ \sum_{\alpha}
        (-1)^F \, {\rm Tr}_\alpha ~{\cal Q}_G~
         \overline{q}^{\overline{H}_\alpha}\,q^{H_\alpha}~
\label{Bdef}
\eeq
where the inserted operator is
\beq
     {\cal Q}_G~\equiv~ ({\rm Im}\,\tau)^2
       ~\left({1\over 12}-\overline{Q_H}^2\right)\, {Q_G}^2~.
\label{Qinserteddef}
\eeq
Here $\overline{Q_H}$  is the (right-moving) spacetime helicity operator,
and $Q_G$ is the so-called ``gauge charge operator'' for the group $G$
(defined as the sum of the squares of the charges in the Cartan subalgebra
of $G$).
Thus, whereas the ordinary partition function $Z(\tau)$
tallies the {\it numbers}\/ of states,
the modified partition functions $B_G(\tau)$ tally their
spin states and {\it charges}\/ under the appropriate gauge groups $G$.
This is precisely what is required in order to determine the correction
to the running of the gauge couplings as in Eq.~(\ref{deltadef}).
In this context, recall that in field theory,
the one-loop $\beta$-function
coefficient can be written as
$b_G= {\rm Tr}\, \lbrack (-1)^F (1/12-h^2) Q_G^2\rbrack$
where $h$ is the helicity operator (with $h=0$ for scalars,
$\pm 1/2$ for fermions, and $\pm 1$ for vectors).
Likewise, in Eq.~(\ref{Qinserteddef}), $\overline{Q_H}$ is the analogous
helicity operator for string states of arbitrarily high spin.
Furthermore,
note that because the traces in Eq.~(\ref{Bdef}) include not only the massive
states
but also the massless states, the contribution of the latter must be
explicitly subtracted in order to determine the threshold corrections.
This is the origin of the term $-({\rm Im}\,\tau)b_G$ in the integrand
of Eq.~(\ref{deltadef}).
Finally, we remark that in string theories
with spacetime supersymmetry, we can omit the
factor of $1/12$ in Eq.~(\ref{Qinserteddef})
since its contribution is proportional to the
supertrace ${\rm Tr}\, (-1)^F =0$.

Unfortunately, one consequence of the insertion of the
charge operator ${\cal Q}_G$ into the trace is to destroy the
modular invariance of the original expression.
This is most easily seen as follows.  If $Q_I$
is the individual charge operator that corresponds to the
left-moving
elementary $U(1)$ current $\hat J_I$ ($I=1,...,22$),
then the gauge charge operator ${Q_G}^2$ in Eq.~(\ref{Qinserteddef})
will typically be a linear combination of products of the form
$Q_I Q_J$.
Let us now consider the contribution to the trace
in Eq.~(\ref{Zdef}) that comes from a particular lattice direction $I$
 {\it before}\/ these charge operators are introduced.
If $H_I\equiv Q_I^2/2$
is the Hamiltonian corresponding to the $I^{\rm th}$
lattice direction,
then this contribution to the total
trace will take the simple form
\beq
      f_I(\tau)~\equiv~ {\rm Tr} \, (-1)^F \, q^{H_I}~.
\eeq
Of course, such traces $f_I(\tau)$ are modular functions
(such as theta-functions), and transform
covariantly under the modular group.
However, if we now consider the effect of inserting the gauge
charge operators into the trace, we easily find that
\beq
      {\rm Tr} \, (-1)^F \,Q_I^2\, q^{H_I}
      ~=~ {1\over i\pi}\, {d\over d\tau}~f_I(\tau)~.
\label{insertedtraces}
\eeq
In other words,
the square of each individual gauge charge operator $Q_I$,
when inserted into a trace,
has a natural representation as a $\tau$-derivative
on that trace:
\beq
    Q_I^2 ~\Longleftrightarrow~
      {1\over i\pi}\, {d\over d\tau}~.
\eeq
The same result, complex-conjugated, also holds for the
right-moving helicity charge operator $\overline{Q_H}^2$.

Unfortunately,
such $\tau$-derivatives are not covariant with respect
to the modular group.   Indeed,
for modular functions of modular weight $k$,
the full covariant $\tau$-derivative would be
\beq
      D~\equiv~ {d\over d\tau} ~-~ {ik\over 2\,({\rm Im}\,\tau)}~.
\label{covderiv}
\eeq
Thus, the
charge-inserted traces in Eq.~(\ref{insertedtraces}) are not modular covariant,
and therefore the quantities $\Delta_G$ as defined in Eq.~(\ref{deltadef})
are also not modular invariant.  This indicates, {\it a priori}\/,
that the definition of the threshold corrections $\Delta_G$ is
inconsistent or at best incomplete.

\subsection{Regulating the infrared}

One possible solution to this problem, of course, is
that modular invariance might be restored if one considers
not just the partial threshold correction $\Delta_G$, but
rather the full result $\tilde \Delta_G$ which
also includes the universal contribution $k_G Y$.
Indeed, strictly speaking, it is only the full string threshold
correction $\tilde \Delta_G$ which should be modular invariant.
However, as we mentioned above, a proper calculation of $Y$
is a fairly complicated matter due to the presence of
infrared divergences, and historically this was not part
of the original calculation \cite{Kaplunovsky}  that evaluated $\Delta_G$.
In this section, we shall therefore briefly discuss some recent
developments concerning the calculation of $Y$ and the
regulation of infrared divergences.

In order to understand the problem of infrared divergences in string theory,
let us first recall that since string
theory currently exists in only a first-quantized formulation,
string calculations must be performed on-shell.
In particular,
in order to calculate quantities such as heavy string threshold corrections,
we require an on-shell method of computing the
coefficients of the gauge field-strength kinetic terms in the bare and one-loop
string effective Lagrangians.
As we discussed at the beginning of Sect.~6.1,
such a calculation is typically performed via the background field
method.
However, unlike the case in field theory, the analogous calculation
in string theory
is plagued by infrared divergences.  Such infrared divergences
arise due to the contributions of on-shell massless particles.

Therefore, in order to rigorously perform such a calculation in string theory,
we require a method of {\it regulating}\/ this infrared behavior.
In particular, one would like a regulator
with a number of special properties \cite{Kiritsis}.
First, it should be applicable to {\it any}\/ four-dimensional string
model with arbitrary particle content.  Second,
it should effectively introduce a mass gap into the string spectrum,
so as to consistently lift the massless string states.
Third, we wish to be able to remove this regulator smoothly,
and fourth, we would like this regulator to preserve modular
invariance.  Finally, there are also more technical requirements
we may wish to impose on a suitable regulator.  For example,
we would {\it a priori}\/ like our regulator to break as few spacetime
supersymmetries as possible,
and moreover to ensure that the vertex operators for spacetime fields
are well-defined on the worldsheet.
We would also like such a string regulator to permit
an unambiguous identification between the results of the
string theory calculation and those in the effective field theory.

Remarkably, there exist several regulators satisfying all
of these properties \cite{kounnas}.
Such regulators consist of replacing the conformal field
theory (CFT) of the flat four-dimensional spacetime with that of
an $N=4$ superconformal field theory.
This essentially corresponds to turning on the background fields
that are associated with the universal four-dimensional gravitational
background fields --- the metric $g_{\mu\nu}$, the anti-symmetric tensor
$B_{\mu\nu}$, and the dilaton $\phi$.
In general, several choices for this $N=4$ CFT are possible,
and one choice in particular, namely
$U(1)_Q\times SU(2)_{k}$,
has been investigated in detail \cite{Kiritsis}.
Here $Q$ corresponds to an introduced background charge,
and is related to the level $k\in 2\IZ$ of the $SU(2)$ factor via
$Q\equiv \sqrt{2/(k+2)}$.
This particular replacement therefore corresponds
to replacing the flat four-dimensional spacetime with a
four-dimensional manifold of the form $\IR\times S^3$
in which the three coordinates of the sphere are described
by an $SU(2)_k$ Wess-Zumino-Witten model, and
the background charge describes its overall scale factor.
Such a choice then not only introduces a mass gap
$\mu^2=(k+2)^{-1} M^2_{\rm string}$
into the string spectrum, but also preserves, as required, the
spacetime supersymmetry of $N=1$ string models.
Moreover, this choice also permits {\it exact}\/ solutions
to be obtained (to all orders in $\alpha'$) for background
gauge and gravitational fields.

Given this regulator, the calculation of the
threshold corrections then proceeds as before:  one turns
on the required background gauge fields, and calculates the
complete threshold corrections to the gauge couplings by
computing the bare and one-loop regulated effective Lagrangians.
Indeed, in the presence of the regulator, such a calculation can be
performed completely without dropping any universal pieces.
Thus, such a calculation in some sense goes beyond the
``field-theoretic'' computations originally
performed in Ref.~\cite{Kaplunovsky}.

The details of this calculation, as well as the final
expressions for the full threshold corrections $\tilde \Delta_G$,
can be found in Ref.~\cite{Kiritsis}.
As expected, these results generalize the simple expression
(\ref{deltadef}) in a fairly complicated way.
But certain features of the new expressions are noteworthy.
First, because the calculation throughout has been performed
in the presence of a regulator,
the results that are obtained
intrinsically include the gauge-independent term $k_G Y$
in Eq.~(\ref{threshgenform}).  In so doing, they also include
contributions  from additional gravitational
interactions such as, {\it e.g.}, dilaton tadpoles.
Such additional contributions were missed in the calculations
of Ref.~\cite{Kaplunovsky} which focussed on only the non-universal
contributions $\Delta_G$, but are generically present in the
full expressions and must be included.
Second, as required, it has been explicitly shown that
the final results are indeed independent of the specific regulator
chosen.
Most importantly for our purposes, however, is the observation
that the inclusion of these extra interactions
({\it i.e.}, the inclusion of the back-reaction from the
curved gravitational background) renders the resulting
expressions manifestly {\it modular invariant}.
Indeed, these extra back-reactions encoded within the universal
term $k_GY$ have
the net effect of restoring the second term in
the covariant derivative in Eq.~(\ref{covderiv}).
Thus, we see that the use of a carefully
chosen spacetime regulator enables modular-invariant
results to be obtained.

To date, there have been several threshold-correction calculations
that have been performed using this regulator.  In Ref.~\cite{PR},
for example,
the universal moduli-dependent string threshold corrections
$Y$ in Eq.~(\ref{threshgenform})
were evaluated for a certain class of $N=1$ orbifold models.
Moreover, in Refs.~\cite{DKLkiritsistwo,DKLkiritsis},
the full moduli-dependent threshold corrections were
calculated for various classes of
four-dimensional string models
with $N=2$ and $N=1$ spacetime supersymmetry.
In cases where we are only interested
in {\it relative}\/ threshold corrections, however,
it is sufficient for practical purposes
to continue to use the partial expressions (\ref{deltadef}).

\subsection{Required sizes of threshold corrections}

In general, it is hoped that the threshold corrections $\Delta_G$ will
be large enough to account for the discrepancy between the predicted
string unification scale and the extrapolated MSSM unification scale.
It is straightforward to determine how large the threshold corrections
must be in order to accomplish this.
Since the threshold corrections $\Delta_G$ contribute to
the running of the gauge couplings as in Eq.~(\ref{onelooprunning}),
their ultimate contributions to the low-energy couplings $\alpha_{\rm strong}$
and $\sin^2\theta_W$ are given in Eq.~(\ref{sinalphacorrections}) where now
$(\Delta_Y,\Delta_2,\Delta_3)$ represent the
corrections for the MSSM gauge factors that come from the heavy
string thresholds.
If we choose a ``proper'' normalization for the hypercharge generator
$Y$ via
\beq
             \hat Y ~\equiv ~ {1\over \sqrt{k_Y}}\, Y~,
\eeq
then $\Delta_{Y}=k_Y\Delta_{\hat Y}$ and consequently
for string models with equal non-abelian levels $k_2=k_3=1$,
the corrections (\ref{sinalphacorrections}) take the form
\beqn
     \Delta^{\rm (sin)} &=& -\,{k_Y\over 1+k_Y}\,
     {a\over 4\pi}\,(\Delta_{\hat Y}-\Delta_2)\nonumber\\
     \Delta^{(\alpha)} &=& -\,{1\over 1+k_Y}\,
     {1\over 4\pi}\,\left\lbrack k_Y(\Delta_{\hat Y}-\Delta_3)+
        (\Delta_2-\Delta_3)\right\rbrack~.
\label{newsinalphacorrections}
\eeqn
Thus, as predicted, only the {\it differences}\/ of threshold
corrections are necessary in order to predict the low-energy couplings.
It is for this reason that a computation of the gauge-independent term $Y$
in Eq.~(\ref{threshgenform}) is unnecessary for this
    analysis.\footnote{Of course, knowledge of $Y$ is still
    necessary in order to calculate the gravitational coupling $G_N$, or to
    calculate the three low-energy gauge couplings $\alpha_i$ individually
    [{\it i.e.}, without taking $\alpha_{\rm e.m.}(M_Z)$ as a known
    input parameter].
    Equivalently, taking all of these low-energy couplings as inputs,
    knowledge of $Y$ is necessary for determining the
    value of the string coupling $g_{\rm string}$;  in this context
    recall Eq.~(\ref{gstringrenormalized}).}

Given these corrections, it is simplest to discuss their effects
in terms of an {\it effective}\/ unification scale.
We have already seen in Figs.~\ref{couplings_vs_Ma} and \ref{couplings_vs_Mb}
that changes in the values of the low-energy couplings can be attributed
to changes in the effective unification scale $M_{\rm string}$.  Thus
the heavy string threshold corrections $\Delta^{(\sin)}$ and
$\Delta^{(\alpha)}$,
which shift the values of the low-energy couplings, equivalently
imply a change in the unification scale
from its original value $M_{\rm string}$
to some effective value
$M_{\rm string}^{\rm (eff)}$.
Given the one-loop RGE's (\ref{RGEs}),
we can easily determine the values of $M_{\rm string}^{\rm (eff)}$ for
our two low-energy couplings, obtaining
\beqn
   {\rm for}~ \sin^2\theta_W:&~~~&
   M_{\rm string}^{\rm (eff)}~=~
   M_{\rm string}\,\exp\left\lbrack
    {1\over 2}~ {(\Delta_{\hat Y} - \Delta_2) \over
     (b_{\hat Y} - b_2)}
     \right\rbrack\nonumber\\
   {\rm for}~ \alpha_{\rm strong}:&~~~&
   M_{\rm string}^{\rm (eff)}~=~
   M_{\rm string}\,\exp\left\lbrack
   {1\over 2}~{ k_Y(\Delta_{\hat Y}-\Delta_3) + (\Delta_2-\Delta_3)\over
    k_Y(b_{\hat Y}-b_3) + (b_2-b_3)}
     \right\rbrack~~~~~~~~
\label{effectivescales}
\eeqn
where of course $b_{\hat Y}\equiv b_Y/k_Y$.
Thus, we see that the effective scales $M_{\rm string}^{\rm (eff)}$
are decreased only if the terms within the exponentials are negative.
Moreover, we see that {\it unification}\/ of the couplings is
preserved if the two effective scales are equal to each other,
which occurs if and only if
\beq
          \Delta_i - \Delta_j ~{\buildrel ? \over =}~
                    \Delta\,(b_i - b_j)~
\label{unifpreserved}
\eeq
for all gauge group factors $G_i$ and $G_j$, with some constant
$\Delta$.
Thus, if Eq.~(\ref{unifpreserved}) is found to hold,
the gauge couplings will all meet at $M_{\rm string}^{\rm (eff)}$
rather than at $M_{\rm string}$.

Since we know that an extrapolation of the experimentally observed
low-energy couplings causes them to meet at the scale $M_{\rm MSSM}$,
we see that the relation (\ref{unifpreserved}) is our first condition
for successful gauge coupling unification via threshold corrections.
Indeed, this condition is also easy to obtain graphically via
the sketch in Fig.~\ref{paths}(c).
The required value of the proportionality constant $\Delta$
in Eq.~(\ref{unifpreserved}) is then determined by adjusting
$M_{\rm string}^{\rm (eff)}$ to coincide with
$M_{\rm MSSM}$.  Numerically, this yields
\beq
    \Delta^{\rm (required)} ~=~
        2\,\ln { M_{\rm MSSM} \over M_{\rm string} }
      ~\approx~ -6.0\pm 0.6~
\eeq
where the uncertainties reflect our ignorance concerning not only
the precise value of $M_{\rm MSSM}$ (which depends
on the value of the low-energy coupling $\alpha_{\rm strong}$),
but also the precise value of the string coupling $g_{\rm string}$
at unification (which in turn determines $M_{\rm string}$).
Nevertheless, using the central value $\Delta\approx -6$ and
given the one-loop beta-function coefficients
$(b_{\hat Y},b_2,b_3)=(33/5,1,-3)$, we find that we therefore require
the approximate relative corrections
\beq
          \Delta_{\hat Y}-\Delta_2 ~\approx~ -34    ~,~~~~~~~
          \Delta_{\hat Y}-\Delta_3 ~\approx~ -58    ~,~~~~~~~
          \Delta_{2}-\Delta_3 ~\approx~ -24    ~.
\label{diffsreqd}
\eeq
If we also include the effects of
the two-loop, Yukawa, and scheme-conversion corrections
(which arise even if
we assume our string theory to contain
only the standard MSSM gauge and matter structure
with the standard affine levels),
we find, using Eq.~(\ref{totalcorrections}), that the above
required relative threshold corrections (\ref{diffsreqd})
are modified to
\beq
          \Delta_{\hat Y}-\Delta_2 ~\approx~ -28    ~,~~~~~~~
          \Delta_{\hat Y}-\Delta_3 ~\approx~ -58    ~,~~~~~~~
          \Delta_{2}-\Delta_3 ~\approx~ -30    ~.
\label{finaldiffsreqd}
\eeq

It is clear, given these results, that the required sizes for the
threshold corrections (or at least for their {\it differences})
are quite large, typically of order $\Delta_G\sim {\cal O}(10^1-10^2)$.
The question then arises as to whether such large threshold corrections
can actually be obtained in string theory,
particularly in realistic four-dimensional string models.

\subsection{Sizes of threshold corrections:  Moduli dependence}

One model-independent way of estimating the sizes
of the threshold corrections $\Delta_G$ is to determine
their dependence on the various moduli which parametrize
the space of possible string vacua.
While at tree level the string gauge couplings
depend \cite{Witten} on only one modulus, namely the dilaton,
the quantum threshold corrections at higher orders in the
perturbation expansion depend on the precise properties of
not only the massless modes
but also the entire infinite towers of heavy states,
and these in turn generically depend on all of the moduli.
Thus, the threshold corrections are generally complicated functions
of all of the moduli fields, and consequently the running of the
gauge couplings of the four-dimensional group factors
in the effective low-energy theory also depends on
the expectation values of all of these moduli fields.
This important dependence has been studied by several groups
\cite{DKL,moduli,ILR,chemtob},
and is also crucial for understanding how
an effective potential might be generated which selects a preferred
string vacuum (thereby lifting the degeneracy of string vacua corresponding
to different values of the moduli VEV's).  This generation of an effective
potential can also
provide a means of dynamical supersymmetry breaking ({\it e.g.}, through
the formation of gaugino condensates).
Furthermore, understanding the dependence of the threshold corrections on
the moduli fields can yield important insights into the structure and
symmetries of the
string effective low-energy supergravity Lagrangian \cite{moduli}.
For our purposes, however, it will be sufficient to focus
our attention on only those general results which suggest how
threshold corrections might become large.

One of the first explicit calculations of the
moduli dependence of the threshold corrections was
performed in Ref.~\cite{DKL} for a restricted class of
$N=1$ string compactifications.
The essential idea behind this analysis is as follows.
First, it can be shown directly from Eqs.~(\ref{deltadef})
and (\ref{Bdef}) that
any string sectors with a full $N=4$
spacetime supersymmetry ({\it i.e.}, any sectors
giving rise to states in complete $N=4$ supermultiplets,
such as the completely untwisted Neveu-Schwarz sector)
make no contributions to the threshold corrections $\Delta_i$.
Likewise, while $N=1$ sectors may give contributions to the corrections
$\Delta_i$, these contributions are independent of the moduli.  Therefore, in
order
to calculate the moduli-dependence of the threshold corrections in any
supersymmetric string model, it is
sufficient to analyze only the contributions from those sectors of
the theory that preserve exactly $N=2$ spacetime supersymmetry.
In particular,
the moduli-dependence of the threshold corrections should be
the same for any two string models which share the same $N=2$ sectors.

Given this fact, one may first consider a special set of
four-dimensional string models with $N=2$ spacetime supersymmetry:
these are the string models which can be realized via toroidal
compactifications of six-dimensional string models with $N=1$ spacetime
supersymmetry.  For such models, the dependence of the threshold
corrections on the untwisted moduli can be directly calculated,
since the masses of all of the string states can be written as
analytic functions of such moduli (note that $N=1$ spacetime supersymmetry
guarantees that the {\it charges}\/ of such states are independent
of all moduli).
By directly computing the modified partition function $B_i(\tau)$
as a function of the moduli, and by analytically performing the
subsequent modular integrations, one then obtains\footnote{
  Note that, strictly speaking, the result in Eq.~(\ref{DKLresultone}) gives
  the threshold correction $\Delta_i'$ with {\it all}\/ group-independent
  contributions removed.  (See the second footnote in Sect.~6.1.)
  This is clear from the fact that this result is proportional
  to $b_i$.  }
\cite{DKL}:
\beq
    \Delta_i(T_\ell) ~=~ -b_i \, \sum_{\ell=1}^2 \, \ln \biggl\lbrack
            ({\rm Im}\, T_\ell)\,|\eta(T_\ell)|^4 \biggr\rbrack~+~ b_i\, X
\label{DKLresultone}
\eeq
where $X$ is a numerical constant.
Here $T_1$ and $T_2$ are respectively the two complex moduli
that parametrize the two-dimensional compactification from six dimensions:
in terms of the background metric $G_{IJ}$ and antisymmetric tensor
$B_{IJ}\equiv b\epsilon_{IJ}$ of the compactification two-torus $T^2$,
these moduli are $T_1\equiv 2(b+i\sqrt{\det\,G})$ and $T_2\equiv
(G_{12}+i\sqrt{\det\,G})/G_{11}$.
The Dedekind $\eta$ function appearing in Eq.~(\ref{DKLresultone}) is
defined as
\beq
          \eta(T) ~\equiv~ e^{\pi i T/12}\,\prod_{n=1}^{\infty}\,
              \biggl( 1- e^{2\pi i n T} \biggr)~.
\eeq
Because of the modular symmetries of the $\eta$ function,
the result in Eq.~(\ref{DKLresultone}) is invariant
under independent $SL(2,\IZ)$ transformations of the complex moduli
$T_1$ and  $T_2$, as required by the invariance of the string vacuum
under such transformations.  Remarkably, the result in Eq.~(\ref{DKLresultone})
also shows that the unification condition (\ref{unifpreserved}) is
always preserved for such $N=2$ string vacua.
Thus, in these theories,
the threshold corrections
can always be absorbed into a
redefinition of the unification scale
(provided, of course, that gauge coupling unification has already
occurred in the {\it absence}\/ of threshold corrections, a non-trivial
assumption given that these theories have $N=2$, rather than $N=1$,
supersymmetry).

The result in Eq.~(\ref{DKLresultone}) applies to only those
four-dimensional $N=2$ string
models which can be realized as toroidal compactifications of six-dimensional
$N=1$ string models.  However, by the arguments
discussed above, this result can easily be extended to any
four-dimensional $N=1$ orbifold string model that contains the same
$N=2$ sectors as such an $N=2$ model.
This wider class of $N=1$ string models consists of those whose
compactification six-tori $T^6$ can be decomposed as
$T^6=T^4\otimes T^2$ where $T^2$ is the compactification
two-torus of the corresponding $N=2$ string model.
Indeed, for such $N=1$ models, the result in Eq.~(\ref{DKLresultone})
generalizes to \cite{DKL}
\beq
    \Delta_i(T^a_\ell) ~=~ \sum_a \,- b^a_i \,c_a \, \sum_{\ell} \, \ln
\biggl\lbrack
            ({\rm Im}\, T^a_\ell)\,|\eta(T^a_\ell)|^4 \biggr\rbrack~+~ {\rm
constant}~.
\label{DKLresulttwo}
\eeq
Here the constant now also includes the moduli-independent
contributions of $N=1$ sectors,
and the sum over $a$ reflects the
fact that in such an $N=1$ model,
the complete set of $N=2$ sectors may be realizable
only as a direct sum of those from {\it different}\/ $N=2$ theories.
The sum over $a$ is thus a sum over all
of the different $N=2$ supersymmetry-preserving
twists in the theory, and the constants $c_a$ which
appear in Eq.~(\ref{DKLresulttwo})
are the ratios of the point group orders,
$c_a\equiv |G_a|/|G|$, where $G$ and $G_a$ are
respectively the orbifold groups of the original $N=1$ model
and its $N=2$ counterparts.
The complete sets of moduli $T^a_\ell$ which contribute
to Eq.~(\ref{DKLresulttwo})
for different orbifold groups depend
on the details of the relevant $N=2$ twists, and
are tabulated in Ref.~\cite{DKL}.
Note that unlike the simpler result in Eq.~(\ref{DKLresultone}), the
result in Eq.~(\ref{DKLresulttwo}) does {\it not}\/ generally satisfy the
unification
condition (\ref{unifpreserved}), since the beta-function
coefficients $b^a_i$ which appear in this expression are not those of the
original
$N=1$ string model, but rather those of the
corresponding $N=2$ string models.

Given the moduli-dependence indicated in Eq.~(\ref{DKLresulttwo}), we can
now determine for which points in moduli space
the threshold corrections $\Delta_i$ can be expected
to become large.
In particular, in the limit that an individual complex modulus $T$ becomes
large ({\it i.e.}, as $T\to i\infty$), we have
\beq
    \ln \biggl\lbrack ({\rm Im}\, T)\,|\eta(T)|^4 \biggr\rbrack~
       ~\sim~  -{\pi\over 3}\, {\rm Im}\,T~.
\eeq
Thus, to lowest order, we obtain \cite{DKL,INN}
\beq
       \Delta_i(T) ~\sim~ {\pi\over 3}\,\sum_{a,\ell}\,
             b^a_i \,c_a ~ {\rm Im}\,T_\ell^a~.
\label{limitresult}
\eeq
Note that this lowest-order result (\ref{limitresult}) is
consistent with worldsheet Peccei-Quinn symmetries, under which only
the real parts of $T_\ell^a$ are shifted.
For our purposes, however, the lesson from this result
is that large threshold corrections can be achieved for
 {\it large moduli}\/ ({\it i.e.}, for ${\rm Im}\,T\to\infty$),
as it is only in this limit that
the $\Delta_i$ can be made to grow without bound.

While the specific result in Eq.~(\ref{limitresult}) is derived only for
those four-dimensional $N=1$ supersymmetric models
which are $T^4\otimes T^2$ compactifications,
similar results have also been obtained for more general
four-dimensional $N=1$ models.
They have also been obtained via moduli-dependent
threshold calculations \cite{PR,DKLkiritsistwo,DKLkiritsis} based on
the full infrared-regulated expressions of Ref.~\cite{Kiritsis}.
Indeed, this behavior is fairly generic to string theory,
and can be avoided only in special cases of string models
without $N=2$ sectors \cite{antonlargeradius}
or models \cite{DKLkiritsistwo} in which taking the
limit ${\rm Im}\,T\to\infty$ increases the spacetime
supersymmetry to $N=4$ (for which the threshold
corrections are known to be suppressed).
Of course, the generic behavior in Eq.~(\ref{limitresult}) can also be
understood on the basis of simple physical intuition:
as the size of a modulus ({\it e.g.}, a radius of
compactification) is increased,
various momentum states become lighter and lighter.
Their contributions to the threshold corrections
therefore become more substantial, ultimately leading
to a decompactification of the theory.

Unfortunately, in realistic string models,
the moduli are not expected to be large.
Indeed, they are expected to settle at or near the self-dual
point, for which $T \sim {\cal O}(1)$  \cite{nolargemoduli}.
The question then arises as to whether there might be other
ways of obtaining large threshold corrections without having
large untwisted moduli.
Furthermore, realistic string models are typically
not as simple as the restricted class of $N=1$ string models
considered above, and usually contain not only untwisted moduli,
but also twisted moduli and moduli from Wilson line
deformations.  For this reason, it is useful to have
an argument for the sizes of the string threshold corrections
which is independent of a particular string construction or
class of compactifications.
Indeed, it would be preferable to have an
argument which depends on only the modular properties
of the threshold correction expressions in Eq.~(\ref{deltadef}).

Such an argument has recently been constructed \cite{unif2},
and the essence of this argument is as follows.
In order to meaningfully discuss the {\it size}\/ of the threshold
corrections, we need to have some expectation of the sizes of
typical one-loop amplitudes.
For this purpose, let us consider possibly the simplest
and most similar one-loop amplitude, the (dimensionless)
string one-loop cosmological constant, defined as
\beq
 \Lambda ~\equiv ~
   \int_{\cal F} {d^2\tau\over ({\rm Im}\,\tau)^2}\,
                Z(\tau)~.
\label{Lambdadef}
\eeq
Of course, this quantity is non-zero only in non-supersymmetric
string models, so we shall restrict ourselves to consideration
of non-supersymmetric string models in what follows.
This is sufficient for discussing the approximate sizes
of the string threshold corrections $\Delta_G$ even in
$N=1$ supersymmetric theories,  for it
has been demonstrated \cite{unif2} that the presence or absence of $N=1$
spacetime supersymmetry in a given string model does not significantly
affect the value of $\Delta_G$.
Now, for generic non-supersymmetric string models,
it has been shown \cite{KRDlambda} that $\Lambda \sim{\cal O}(10^1-10^2)$.
By contrast, the typical sizes of the threshold corrections
that are found in $N=0$ and $N=1$ string models are much smaller, typically
$\Delta_G\sim {\cal O}(1)$.
It is this suppression that we would like to explain,
especially since the expected unsuppressed values
$\Delta_G\sim{\cal O}(10^1-10^2)$ are precisely
what we would have required in order to achieve string-scale unification.
Clearly, comparing Eqs.~(\ref{deltadef}) and (\ref{Lambdadef}),
we see that this suppression must
arise due to the different integrands, $Z(\tau)$ versus
${\cal B}_G(\tau)\equiv \lbrace B_G(\tau)-({\rm Im}\,\tau)b_G\rbrace$.
Let us therefore expand these integrands as double power series in
$q\equiv e^{2\pi i\tau}$ and $\overline{q}$:
\beqn
    Z(\tau) & = & ({\rm Im}\,\tau)^{-1}\,
         \sum_{m,n} \, a_{mn}~ \overline{q}^m\,q^n \nonumber\\
    {\cal B}_G(\tau) & = & ({\rm Im}\,\tau)^{+1}\,
         \sum_{m,n} \, b^{(G)}_{mn}~ \overline{q}^m\,q^n~.
\label{ZBqexpansions}
\eeqn
In these expansions, the summations are over all (physical
and unphysical) states in the string spectrum,
and the coefficients $a_{mn}$ and $b^{(G)}_{mn}$ respectively
represent the total numbers and $G$-charges, weighted by $(-1)^F$,
of those states in the spectrum with right- and left-moving spacetime
(mass)$^2$ contributions $(m,n)\equiv (\alpha'M_R^2,\alpha' M_L^2)$.
Recall that string states are unphysical
if $m\not=n$ and tachyonic if $m+n<0$;  nevertheless all
string states potentially contribute to the one-loop
amplitudes $\Lambda$ and $\Delta_G$.

Given these expansions, we can then immediately
identify a variety of sources for the suppression of $\Delta_G$
relative to $\Lambda$.
Of course, we immediately observe that the two
expansions in Eq.~(\ref{ZBqexpansions})
contain different powers of ${\rm Im}\,\tau$ in the prefactor.
However, it turns out that \cite{unif2} that
this is not a major numerical effect, even after $\tau$-integration.
Thus, we can ignore this difference, and concentrate
on the coefficients $a_{mn}$ and $b_{mn}^{(G)}$.
It is here that we can identify three fundamental differences.

In general, the contributions to one-loop amplitudes
from string states with mass configurations
$(m,n)$ depend on the value of the sum $m+n$,
and decrease exponentially with increasing $m+n$.
Thus, the largest contributions to these amplitudes come from
tachyonic states.
Note that even though we exclude theories containing
 {\it physical}\/ tachyons ({\it i.e.}\/, states with $m=n<0$),
we nevertheless must consider the effects of {\it unphysical}\/
tachyons (states with $m\not=n$, $m+n<0$).
Indeed, such states are typically present in such string models.
Now, while the ordinary partition function $Z(\tau)$ counts
such states weighted only by their spacetime fermion numbers,
the {\it modified}\/ partition functions
$B_G(\tau)$ count such states multiplied by their gauge
charges.  Because these states are tachyonic,
however, they are typically gauge-neutral, for
any excitations that could possibly give a gauge charge to
such states would also increase their mass.
Thus, even though we have $a_{mn}\not=0$ for such states,
we nevertheless find that $b^{(G)}_{mn}=0$.
Consequently the unphysical tachyonic states, which typically make
large contributions to $\Lambda$,
make no contributions to the
threshold corrections $\Delta_G$.

The second source of suppression has to do with the
next-largest contributors, namely the physical massless states
with $m=n=0$.
While $\Lambda$ clearly receives contributions from massless states,
it is evident such states make no contributions
to $\Delta_G$, for their contributions have been
explicitly subtracted away in the definition of ${\cal B}_G(\tau)$.
Thus, even though $a_{00}\not =0$, we nevertheless always
have $b_{00}=0$.
Consequently the contributions of the {\it massless}\/ states
are also completely suppressed.

Finally, we turn to the contributions from the infinite towers
of {\it massive}\/ physical string states.
Remarkably, however, it is straightforward to show \cite{unif2}
that even these contributions are suppressed, for
while the {\it numbers}\/ of string states typically
increase {\it exponentially}\/ as a function of mass,
with the well-known Hagedorn behavior $a_{nn}\sim e^{C \sqrt{n}}$
for some constant $C>0$,
it turns out \cite{missusy} that the {\it charges}\/ of such string
states increase only {\it polynomially},
with $b^{(G)}_{nn}\sim n^{C'}$ for some $C'>0$.
This suppression essentially arises because the
insertion of the charge operator ${\cal Q}_G$
into the trace [as indicated in Eqs.~(\ref{Bdef}) and (\ref{Qinserteddef})]
has the effect of increasing the modular weight $k$ of the
integrand from its ordinary {\it negative}\/ value $k = -1$
[for $Z(\tau)$]
to the {\it positive}\/ value $k =+1$ [for $B_G(\tau)$].
Further details can be found in Ref.~\cite{unif2}.
Thus, even the contributions of the massive string states
are suppressed.

These are clearly simple arguments which are model-independent.
Indeed, given this last observation,
we now see that the only way to evade such arguments
are to somehow start with a smaller ``effective'' modular weight,
or equivalently a larger effective spacetime dimension.
However, must this necessarily imply a large internal modulus?
In other words, is a large internal modulus the only way
to produce a larger effective spacetime dimension
or smaller effective modular weight?

We have seen above that for the special compactifications
studied in Ref.~\cite{DKL}, this is indeed the case:  a larger
effective modular weight is achieved only via a large internal
modulus.
These are, however, very special compactifications.
Specifically, they are $(2,2)$ compactifications, which means that
they preserve $N=2$ worldsheet supersymmetry for both
the right-moving and left-moving
compactification degrees of freedom.
(This is in addition to the extra left-moving degrees
of freedom of the heterotic string, which are of course
non-supersymmetric in any case.)
However, for more general $(2,0)$ compactifications (in which there
are many more compactification moduli available,
including moduli from Wilson lines),
the situation becomes more complicated.
The moduli-dependence of the threshold corrections in certain
$(2,0)$ compactifications
has been recently calculated in Ref.~\cite{modulitwozero}, and it is found
that the dependence on the moduli
takes a form similar to that of Eq.~(\ref{DKLresulttwo}), except
that the $\eta$ functions are replaced by other more general modular
functions which mix the various compactification and Wilson-line moduli.
This replacement then turns out to imply that
for such $(2,0)$ string compactifications,
it is effectively a {\it product}\/ of moduli that
determines the size of threshold corrections.
Consequently, barring other cancellations, suitably chosen ${\cal O}(1)$
values for individual moduli have the potential to yield larger
threshold corrections \cite{NS}.
Unfortunately, this interesting mechanism has not yet been realized
within the context of realistic three-generation string models.

\subsection{Sizes of threshold corrections:
    Explicit calculations in realistic string models}

Another approach towards determining the sizes of
the threshold corrections, of course, is to directly
calculate them for actual, realistic string models.
Whereas the above expectations based on
studying the moduli dependence of the threshold corrections
can give general insights
about how these quantities {\it might}\/  become large,
it is necessary to study what actually happens within
the tight constraints of realistic string models.
For example, the mechanisms that give rise to large
threshold corrections may not be realizable or mutually
consistent with the mechanisms that give rise to other
desirable phenomenological features such as the appearance of
$N=1$ spacetime supersymmetry or three generations.
Many previous calculations
of the threshold corrections
have been performed for various types of orbifold
or free-fermionic models
\cite{Kaplunovsky,thresholdcalcs,DL,MNS,chemtob},
but in most cases
such models were typically not phenomenologically
realistic.
Indeed, the increased complexity of the realistic
string models
may substantially alter previous expectations.
It is therefore necessary to perform a detailed
calculation of the threshold corrections within the context
of actual realistic string models.
Such a calculation has recently been done \cite{unif1,unif2},
however, and in the rest of this section we shall discuss
the results.
Since these calculations are performed at specific points in the
string moduli space, the results that are obtained
essentially combine the moduli-dependent and moduli-independent
contributions into a single number.

\subsubsection{The models}

For such a detailed calculation, one requires
``realistic'' string models.
In particular, one requires four-dimensional string
models which have at least the following properties.
First, they should possess $N=1$ spacetime supersymmetry.
Second, they should give rise to an appropriate gauge group
--- {\it e.g.}, the MSSM gauge group $SU(3)\times SU(2)\times U(1)$,
the Pati-Salam gauge group $SO(6)\times SO(4)$,
or flipped $SU(5)\times U(1)$.
Third, the models should give rise to the proper
massless (observable) spectrum --- this includes not only
three complete chiral MSSM generations with the correct quantum
numbers and hypercharges, but also the correct MSSM Higgs scalar
representations.
Fourth, the models should be anomaly-free, so that
all $U(1)_Y$ anomalies should be cancelled across the
entire massless spectrum.
Finally, we may also even wish to demand certain
other ``realistic'' features
such as a semi-stable proton, an appropriate fermion mass
hierarchy, a heavy top quark, and so forth.

Remarkably, several level-one string models
exist in the literature
which meet all of these requirements.
In particular, these models
include the flipped $SU(5)$ model of Ref.~\cite{flipped} (which
realizes the flipped $SU(5)$ unification
scenario of Ref.~\cite{flippedsu5fieldtheory});
the $SO(6)\times SO(4)$ model of Ref.~\cite{alrmodel} (which
realizes the Pati-Salam unification scenario of Ref.~\cite{patisalam});
and various string models \cite{278model,274model}
in which the Standard-Model gauge group
$SU(3)\times SU(2)\times U(1)$
is realized directly at the Planck scale.
Similar models can also be found, {\it e.g.}, in Ref.~\cite{FNY}.
As required, all of these models have $N=1$ spacetime
supersymmetry and contain exactly three generations in
their massless spectra.
Moreover, these models also naturally incorporate the fermion
mass hierarchy with a heavy top-quark mass \cite{fermionmasses}.
Furthermore, unlike various string GUT models,
they also provide a natural mechanism that can yield
a semi-stable proton \cite{stableproton}.
These models can all be realized in the free-fermionic
construction \cite{freefermions,realfreefermions}, and their phenomenological
properties are ultimately a consequence of their common
underlying $\IZ_2\times \IZ_2$ orbifold structure \cite{orbifold}.  In the
free-fermionic construction, this structure is realized
through the so-called ``NAHE set'' \cite{NAHE}
of boundary-condition basis vectors.
These models then differ from each other through the addition
of Wilson lines (in the orbifold construction), or equivalently
through the imposition of additional boundary-condition basis vectors
(in the fermionic approach).
More complete descriptions of these models can be found
in the appropriate references.

\subsubsection{Results}

Within such realistic string models, the calculation of the
threshold corrections (\ref{deltadef}) is highly non-trivial.
First, there are a number of analytical subtleties:  these models
typically contain tens of thousands of sectors, all
of whose contributions must be systematically included;
one must carefully calculate the appropriate GSO projection
phases in each sector in order to verify that the
desired spectrum is produced;
these models often exhibit so-called ``enhanced gauge symmetries''
which lead to complicated gauge-group embeddings;
and one must find a method of {\it analytically}\/ removing
the logarithmic divergence due to the massless states
prior to integration.  Fortunately, there are a number of
self-consistency checks \cite{unif2} that can be performed
at various stages of the calculation.
Finally, there are also various numerical subtleties,
not the least of which concerns the performance of the modular
integrations in Eq.~(\ref{deltadef}).
Details concerning this calculation can
be found in Ref.~\cite{unif2}.

The results that are found, however, bear out the general
expectations based on the moduli-dependence
of the threshold corrections.
Indeed, not only are the threshold corrections found
to be unexpectedly
small in these models, but they are also found
to have the {\it wrong sign}\/!
Thus, in all of these realistic models,
the heavy string threshold
corrections effectively {\it increase}\/ the
string unification scale slightly, and
thereby exacerbate the disagreement with experiment.
For example, in the
$SU(3)\times SU(2)\times U(1)$ model of
Ref.~\cite{278model},
one finds the relative threshold
corrections \cite{unif1,unif2}
\beq
     \Delta_{\hat Y}-\Delta_{2}~\approx~  1.61~,~~~~~~~
     \Delta_{\hat Y}-\Delta_{3}~\approx~  5.05~.
\label{exampleresults}
\eeq
Since this string model has the standard $SO(10)$ embedding,
the levels $(k_Y,k_2,k_3)$ have their standard MSSM values $(5/3,1,1)$.  Thus,
by comparing against the desired results in Eq.~(\ref{finaldiffsreqd}),
we see that the magnitudes of these threshold differences are far too small,
and that, more importantly, their signs do not agree.
Each of the other realistic string models also exhibits
this behavior.
Moreover, similar results have also been observed in previous calculations
based on other less-realistic string
models \cite{Kaplunovsky,thresholdcalcs,DL}.
A recent proposal \cite{halyo} for fixing the sign
of the threshold corrections involves redistributing
the three generations
between the different $N=2$ sectors that arise
in such models;
however, this scenario has not yet been realized
in any consistent string model, and in any case still
requires either large moduli or small hypercharge
normalizations $k_Y<5/3$ in order to achieve unification.

One objection that might be raised against such
threshold calculations is that
they ignore the fact that in such $(2,0)$ string models,
one combination of the $U(1)$ factors is ``pseudo-anomalous''.
This $U(1)$ gauge symmetry is ``pseudo-anomalous'' in the sense that
the trace of the corresponding $U(1)$ charge
is non-vanishing when evaluated over the massless string states.
As we shall discuss in more detail in Sect.~8.2,
this non-vanishing trace gives rise to
a Fayet-Iliopoulos $D$-term which breaks supersymmetry and
destabilizes
the vacuum, and this in turn implies that the models must choose
non-zero VEV's for some of the scalar fields (twisted moduli)
so as to cancel the anomalous $U(1)$ $D$-term and restore
spacetime supersymmetry \cite{DSWshift}.
Since this corresponds to a shift in the string vacuum,
one might ask whether this shift
can modify the above calculations of the heavy string thresholds.
However, this is not the case,
for the contributions to the threshold corrections come from string
states weighing $\geq g_{\rm string}M_{\rm Planck}/\sqrt{8\pi}$,
whereas any extra masses acquired from shifting
the string vacuum are of higher order in $g_{\rm string}$.
Hence, these extra masses affect the
values of the thresholds only at higher
order, and can be ignored in such
analyses \cite{thresholdcalcs}.

Thus, we conclude that within the realistic string models examined
thus far,
string threshold corrections do not seem able
to resolve
the discrepancy between the MSSM unification scale and the string
unification scale.
Indeed, to date, despite various
attractive theoretical mechanisms as discussed above,
there do not exist any realistic string models
in which such mechanisms are realized and in which the threshold
corrections are sufficiently large.
Nevertheless, the attractiveness of these proposals
and the importance of understanding the moduli dependence
of threshold corrections
show that this ``path to unification'' merits further
examination.
Work in this area is continuing.


\setcounter{footnote}{0}
\section{Path \#4:  Light SUSY Thresholds, Intermediate-Scale Gauge Structure}

Heavy string threshold corrections are not the only effects that
involve adding ``correction terms'' to the renormalization
group equations of the MSSM.
As we discussed at the beginning of Sect.~6, two other possible
effects which may also  modify the running of the gauge
couplings are the effects from light SUSY thresholds and
from possible intermediate-scale gauge structure.
The effects from light SUSY thresholds are those
that arise due to the breaking of spacetime supersymmetry
at some intermediate scale.  Likewise, the effects from
possible intermediate-scale gauge structure arise in situations
in which there exists a larger gauge symmetry that breaks
at an intermediate scale to the gauge symmetry of the MSSM.
Such intermediate-scale gauge symmetry breaking
occurs, for example, in the flipped $SU(5)\times U(1)$ string model
or in the Pati-Salam $SO(6)\times SO(4)$ string model.
Such effects may also arise
even in $SU(3)\times SU(2)\times U(1)$ string models due to the
breaking of extra custodial symmetries \cite{custodial}.  In both cases, such
intermediate-scale gauge structure introduces non-trivial
corrections to the running of the gauge couplings.

In this section we shall briefly summarize some recent
calculations \cite{unif2} of these effects as they arise
within realistic string models.
As we discussed at the beginning of Sect.~6.5, it is important
to analyze these effects within actual, realistic string models.
In isolation,
each of these effects can be made to assume any potential size
due to their many adjustable parameters,
and hence
studying these effects in such abstract settings is not particularly
meaningful.  Within the tight constraints of realistic string
models, however, and with two-loop, Yukawa-coupling, and scheme-conversion
effects included, we shall see that
the range of possibilities is significantly narrowed
and in some cases altogether eliminated.

Unlike the heavy string thresholds, the light SUSY thresholds
and intermediate-scale gauge thresholds are calculated and analyzed
directly in the low-energy effective field theory derived
from a particular string model.  Thus, the correction terms
$\Delta_i$ that
they introduce into the renormalization group
equations (\ref{onelooprunning})
have the same form as they would take in an
ordinary field-theoretic analysis, and our goal is to
determine if, within realistic string models,
they can have the sizes
that are necessary in order to predict the
experimentally observed values of the low-energy couplings.
As before, these required sizes are indicated in Eq.~(\ref{finaldiffsreqd}).

\subsection{Light SUSY thresholds}

It has been assumed thus far that the thresholds
for supersymmetric particles are located only at or near $M_Z$;
this was, for example, the assumption underlying
Fig.~\ref{introfigb}.  By contrast, the light SUSY threshold
corrections are those corrections that arise
from considering the more general situation in which
supersymmetric particles have other (intermediate-scale)
masses.
In the one-loop MSSM renormalization group
equations (\ref{onelooprunning}), the correction
terms $\Delta_i$ arising
from such light SUSY thresholds are
given by
\beq
          \Delta_i ~=~ -\,\sum_{\rm sp} \, b_i^{\rm (sp)} \,
         \ln\, {M_{\rm sp}^2\over M_Z^2} ~,
\label{lightsusythresholds}
\eeq
where the summation is over the MSSM sparticle states, with
corresponding masses and one-loop beta function coefficients
$M_{\rm sp}$ and $b_i^{\rm (sp)}$ respectively.
This term, when added to the one-loop RGE's (\ref{onelooprunning}),
essentially
subtracts the contributions from the sparticles between $M_{\rm sp}$
and $M_Z$ so that below each $M_{\rm sp}$
only the contributions from the non-supersymmetric particles of
the Standard Model remain.
Retracing the algebraic steps leading from Eq.~(\ref{onelooprunning}) to
Eq.~(\ref{sinalphacorrections}), we then find that in string models
with $k_2=k_3=1$, the light SUSY
thresholds (\ref{lightsusythresholds})
lead to the following corrections to the
low-energy couplings:
\beqn
    \Delta^{(\sin)} &=&
            {k_Y\over 1+k_Y}\,
          {a\over 2\pi}\,
       \sum_{\rm sp}\,
       \biggl\lbrack
        b_{\hat Y}^{\rm (sp)} -b_{2}^{\rm (sp)}
       \biggr\rbrack
      \,\ln\, {M_{\rm sp}\over M_Z}\nonumber\\
    \Delta^{(\alpha)} &=&
          {1\over 2\pi}\,
       \sum_{\rm sp}\,
     \left\lbrack \,
      \left( k_Y\over 1 +k_Y\right)\,
        b_{\hat Y}^{\rm (sp)} ~+~
      \left( 1\over 1 +k_Y\right)\,
        b_{2}^{\rm (sp)} ~-~ b_3^{\rm (sp)} \right\rbrack\,
      \ln\, {M_{\rm sp}\over M_Z} ~.~~~~
\label{delsinalphalightsusy}
\eeqn

In order to evaluate these corrections, the next step
is to assign the sparticle masses $M_{\rm sp}$
according to some parametrization.
The conventional choice is to assume that the
SUSY-breaking
occurs via soft  SUSY-breaking terms in the MSSM Lagrangian;
these masses $M_{\rm sp}$ are then given in terms of the
soft SUSY-breaking parameters $m_0$, $m_{1/2}$, and $\mu$.
At tree level, $m_0$ is the mass of all of the scalar
superpartners, while $m_{1/2}$ is the mass of the
supersymmetric fermions and $\mu$ is the mass of the higgsino.
At the one-loop level, however, these
masses evolve as follows.
Neglecting the contributions
from Yukawa couplings and electroweak VEV's (which in turn
implies we are neglecting any non-diagonal supersymmetric scalar mass matrix
elements as well as any $D$-term contributions to the sparticle masses),
we find
that for the supersymmetric fermions ({\it i.e.}, for the
gluino $\tilde g$, wino $\tilde W$, and higgsino $\tilde h$),
the corresponding masses are
\beq
  M_{\tilde g}={{\alpha_3(M_{\tilde g})}
         \over{\alpha_{\rm string}(M_{\rm string})}}\,m_{1/2}~,~~~~~
  M_{\tilde W}={{\alpha_2(M_{\tilde W})}
         \over{\alpha_{\rm string}(M_{\rm string})}}\,m_{1/2}~,~~~~~
  M_{\tilde h}=\mu~
\label{gwhmasses}
\eeq
where we may approximate
$\alpha_3(M_{\tilde g})\approx \alpha_3(M_Z)$
and
$\alpha_2(M_{\tilde W})\approx \alpha_2(M_Z)$
for simplicity.
Likewise, for the supersymmetric bosons ({\it i.e.}, for the
scalar superpartners of the Standard Model fermions),
the masses are given as
\beq
      M_{\tilde p}^2~=~ m_0^2~+~f_{\tilde p}\, m_{1/2}^2
\label{mptilde}
\eeq
where the coefficients $f_{\tilde p}$ for the different sparticles
are given in terms of their hypercharges $Y_{\tilde p}$ by
\beq
       f_{\tilde p}~=~
     \epsilon_3\, c_3(M_{\tilde p})~+~ \epsilon_2\, c_2(M_{\tilde p})~+~
       Y^2_{\tilde p}\,c_Y(M_{\tilde p})~
\label{cpcoef1}
\eeq
with
\beqn
        c_3(m_{\tilde p})&=&-{8\over 9}\,\left\lbrack 1-
                      (1- 3X)^{-2} \right\rbrack \nonumber\\
        c_2(m_{\tilde p})&=&{3\over 2}\,\left\lbrack 1-
                      (1+ X)^{-2} \right\rbrack \nonumber\\
        c_Y(m_{\tilde p})&=&{2k_1\over 11}\,\left\lbrack 1-
                      (1+ 11X/k_1)^{-2} \right\rbrack
\label{cpcoef2}
\eeqn
and
\beq
     X~\equiv ~{1\over 2\pi} \,\alpha_{\rm string}\,
         \ln{M_{\rm string}\over{M_{\tilde p}}}~.
\label{Xdef}
\eeq
In Eq.~(\ref{cpcoef1}), $\epsilon_3=1$ for color triplets and $=0$
for color singlets;   likewise,
$\epsilon_2=1$ for electroweak doublets and $=0$ for electroweak singlets.
As with the gauginos,
one may again approximate $M_Z$ for $M_{\tilde p}$ in Eq.~(\ref{Xdef}) if
desired.
Thus, while the gaugino and supersymmetric fermion masses continue
to be equal to $m_{1/2}$ and $m_{0}$
respectively
at the string scale $M_{\rm string}$, we see from Eqs.~(\ref{gwhmasses}) and
(\ref{mptilde}) that they evolve quite differently below this scale.
The beta-function coefficients $b^{\rm (sp)}_i$ that govern this
running for each of
the sparticle representations are standard,
and are tabulated for convenience in Ref.~\cite{unif2}.

\begin{figure}[ht]
\centerline{\epsfxsize 3.5 truein \epsfbox {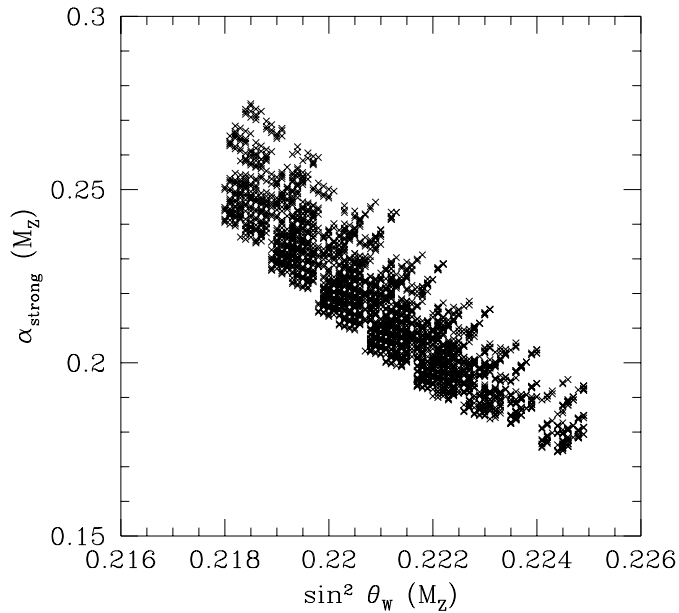}}
\caption{Scatter plot for $\lbrace\sin^2\theta_W(M_Z),
 \alpha_{3}(M_Z)\rbrace$ when the SUSY-breaking
   parameters $\lbrace m_0,m_{1/2},M_h,M_{\tilde h},M_{\rm string}\rbrace$
   are varied as in Eq.~(\protect\ref{sps}),
   assuming only the MSSM spectrum between the $Z$ scale and
   the string scale.}
\label{lightsusyfig}
\end{figure}

Assuming universal boundary conditions for the soft SUSY-breaking
terms ({\it i.e.}, assuming that $m_0$ and $m_{1/2}$ are the same
for all sparticles), one can then analyze the possible light
SUSY threshold contributions for a wide range of points in the
parameter space $\lbrace m_0, m_{1/2}, M_h, M_{\tilde h}\rbrace$,
where $M_h$ is the mass of the pseudoscalar Higgs.
Such an analysis was recently performed in Ref.~\cite{unif2},
with the parameters allowed to vary within the following ranges
and with the following interval sizes:
\begin{equation}
\begin{tabular}{c|ccc}
Parameter $X$ & $X_{\rm min}$  & $X_{\rm max}$  &   $\Delta X$  \\
\hline
$m_0$~(GeV) & 0 &  600 & 200 \\
$m_{1/2}$~(GeV) & 50 &  600 & 150 \\
$M_h$~(GeV) & 100 &  500 & 200  \\
$M_{\tilde h}$~(GeV)     & 100 &  500 & 200 \\
$M_{\rm string}$~(GeV) & ~~$3\times10^{17}$~~ &
     ~~$7\times10^{17}$~~ &  ~~$5\times10^{16}$~~   \\
\end{tabular}
\label{sps}
\end{equation}
In this analysis, the top-quark mass was taken to be $M_t=175$ GeV (with the
other quark and lepton masses set to zero).
The results for the low-energy couplings $\alpha_{\rm strong}(M_Z)$
and $\sin^2\theta_W(M_Z)$ are shown as a scatter plot
in Fig.~\ref{lightsusyfig}.

It is apparent from Fig.~\ref{lightsusyfig} that
such light SUSY thresholds are generally insufficient to resolve
the discrepancy between $M_{\rm string}$ and $M_{\rm MSSM}$,
and that the predicted low-energy couplings continue to disagree with
their experimentally measured values.
It turns out \cite{unif2} that this conclusion remains valid even
if {\it non-universal}\/ boundary conditions are allowed
for the soft SUSY-breaking terms (as is expected in string
models \cite{nonuniversal}).
Indeed, even if one neglects the
sparticles whose contributions to the low-energy couplings
push them in the wrong
direction, one finds that successful predictions for the
low-energy couplings require that the remaining sparticles have masses
$\sim 100$ TeV \cite{unif2}.  Such large masses are generally
considered to be unnatural, and require fine-tuning.
Hence light SUSY thresholds appear incapable of
providing sufficiently large correction terms to the one-loop
RGE's of the MSSM.

\subsection{Intermediate-scale gauge structure}

Another similar possibility is that intermediate-scale gauge structure
might provide the corrections that are necessary in order to reconcile the
low-energy couplings with string-scale unification.
As we stated at the beginning of this section,
such intermediate
gauge structure can arise in realistic string models
either through extended Planck-scale gauge groups
which break to that of the MSSM at some intermediate
scale  (such as occurs in the flipped $SU(5)\times U(1)$ or
$SO(6)\times SO(4)$ string models \cite{flipped,alrmodel}), or through the
breaking
of certain custodial symmetries (as can occur \cite{custodial} in particular
$SU(3)\times SU(2)\times U(1)$ string models \cite{274model}).

Like those of light SUSY thresholds, the effects of
intermediate-scale gauge structure are analyzed purely in the
low-energy effective field theory,
and are parametrized  by $M_I$, the intermediate mass scale of the
gauge symmetry breaking.
Of course, any calculation of these effects must also be done
in conjunction with the effects from two-loop contributions,
Yukawa couplings, and scheme conversion.
Unlike the effects of light SUSY thresholds, however, the
effects from intermediate-scale gauge structure cannot be analyzed in a
model-independent fashion, since
they ultimately depend on the intricate details of the particular
symmetry breaking scenario.  They therefore must be subjected to a
detailed model-by-model analysis.

Such an analysis has recently been performed \cite{unif1,unif2}
for the class of realistic models discussed in Sect.~6.5.1,
and the results are rather surprising.
It is found that despite the
appearance of an extra adjustable degree of freedom (namely $M_I$,
the scale of the breaking), such effects cannot resolve
the discrepancy.  Moreover, in each of the relevant
string models examined, it turns out that taking $M_I<M_{\rm string}$
only {\it increases}\/ the disagreement with the low-energy data!
Details concerning this result can be found in Ref.~\cite{unif2}.
Thus, at least for this class of models, it does not appear
that intermediate-scale gauge structure
can be used to reconcile the string predictions with the
low-energy couplings.

It is important to note that intermediate-scale gauge structure can
nevertheless play a useful role in bringing about string-scale
gauge coupling unification if it occurs in conjunction with some
of the other features we have discussed.  Indeed, such a
successful combination of effects was implicitly assumed for
the sketch in Fig.~\ref{paths}(d).
Moreover, as we shall now discuss,
one natural candidate which may serve this purpose (and which
generically arises along with intermediate-scale gauge structure)
is extra matter beyond the MSSM.


\setcounter{footnote}{0}
\section{Path \#5:  Extra Matter Beyond the MSSM}

The final possibility for reconciling the predicted string and MSSM
unification scales via renormalization-group ``correction terms''
involves the appearance of extra matter beyond the MSSM.
While many of the above ``paths to unification''
assumed only the MSSM spectrum between the $Z$ scale and the
string scale, string models often contain
additional states beyond those of the MSSM.
Indeed, in such cases these states are {\it required}\/
for internal consistency, for the string
spectrum is tightly constrained by many symmetries --- among them
modular invariance --- and the arbitrary removal of such states
would result in worldsheet anomalies.  Thus, such ``unwanted''
states are typically present at the massless level
in realistic string spectra, and their effects must be included.
Such matter has been considered, for example, in
Refs.~\cite{Gaillard,flipped,274model,unif1,unif2,allanach}.

\subsection{Extra matter in realistic string models}

The detailed properties of such extra non-MSSM matter
are of course highly model-dependent, but within the class
of realistic string models discussed in Sect.~6.5.1, such matter
has certain generic properties.  First, such matter appears
in a majority of these models.  Second, it typically
appears in vector-like representations;  thus such states
are non-chiral.  Third, because such states are non-chiral,
they can be given mass and  become superheavy.
Finally, in the realistic string models,
such matter typically appears in the form of extra color triplets
or electroweak doublets with special hypercharge assignments.
These representations include
$(\rep{3},\rep{2})_{1/6}$, $(\rep{\overline{3}},\rep{1})_{1/3}$,
$(\rep{1},\rep{2})_{1/2}$,
$(\rep{\overline{3}},\rep{1})_{1/6}$, and $(\rep{1},\rep{2})_{0}$,
and have the following beta-function coefficients:
\beqn
   ({\bf 3},{\bf 2})_{1/6} \,:&&~~~~~~~
(b_3,b_2,b_1)=(1,\,3/2,\,1/10)\nonumber\\
   (\overline{{\bf 3}},{\bf 1})_{1/3} \,:&&~~~~~~~
(b_3,b_2,b_1)=(1/2,\,0,\,1/5)\nonumber\\
   ({\bf 1},{\bf 2})_{1/2} \,:&&~~~~~~~
(b_3,b_2,b_1)=(0,\,1/2,\,3/10)\nonumber\\
   (\overline{{\bf 3}},{\bf 1})_{1/6} \,:&&~~~~~~~
(b_3,b_2,b_1)=(1/2,\,0,\,1/20)\nonumber\\
   ({\bf 1},{\bf 2})_{0} \,:&&~~~~~~~ (b_3,b_2,b_1)=(0,\,1/2,\,0)~.
\label{typreps}
\eeqn

While the first three of these representions can clearly be
embedded into standard $SU(5)$ or $SO(10)$ representations,
it is clear from their hypercharges that the latter two
representations cannot be, and are therefore
truly exotic.  From the point of view of gauge coupling
unification, however, these types of extra matter are precisely what
are needed, for the representations listed in Eq.~(\ref{typreps}) have
small values of $b_1$ relative to their values of ($b_2$,$b_3$),
and consequently they have the potential to substantially modify
the running of the $SU(2)$ and $SU(3)$
gauge couplings {\it without}\/ seriously affecting the $U(1)$ coupling.
It is evident from the sketch in Fig.~\ref{paths}(e)
that this is precisely what must happen if the
scale of gauge coupling unification is to be increased.
Thus, while such representations are entirely unexpected from the
field-theory point of view,
they are precisely what would be needed in order
to reconcile the discrepancy between
the string and MSSM unification scales.

This is an encouraging observation, but
the question then arises:
in the realistic string models, do such matter states actually
appear with the
correct multiplicities and in the appropriate combinations
to achieve successful string-scale
gauge coupling unification?  In other words, do the realistic
string models contain the proper combinations of such
representations that can reconcile the discrepancy between
the string scale and the MSSM unification scale?

To answer such a question is a straightforward exercise,
and can be handled in the low-energy effective field theory.
The one-loop ``correction terms'' introduced by such matter are given
as
\beq
          \Delta_i ~=~ +\,\sum_{I} \, b_i^{(I)} \,
         \ln\, {M_{I}^2\over M_Z^2}
\label{extramatterthresholds}
\eeq
where the summation is over the extra intermediate-scale
matter states with masses $M_I$.
Thus, comparing this expression
with Eq.~(\ref{lightsusythresholds}),
we find the effects of such matter states on the
values of the low-energy couplings
are given by Eq.~(\ref{delsinalphalightsusy})
with the `sp' subscript replaced by `$I$\/'
and with an overall change of sign.
Of course, as before, for a rigorous calculation
one must also include the effects of
two-loop corrections, Yukawa couplings, and scheme conversion.
One must also include
the effects of light SUSY thresholds as well as
the calculated model-dependent
heavy string thresholds
corrections (see Sects.~6 and 7).
One then seeks to determine, given the particular combinations
of non-MSSM representations that
appear in a particular string model, whether there exists a
window in the parameter space
$\lbrace M_I \rbrace$ of non-MSSM masses
for which a successful string-scale unification of
the gauge couplings is achieved.

For the realistic free-fermion models of Sect.~6.5.1,
the results of such an analysis are as follows \cite{unif1,unif2}.
For {\it some}\/ of these models ({\it e.g.}, the model of
Ref.~\cite{278model}),
it is found that the required combinations of non-MSSM matter
do {\it not}\/ appear.
This is remarkable, for it indicates that despite the
presence of the new
intermediate scales $\lbrace M_I\rbrace$,
successful low-energy predictions are still difficult
to obtain.
For example, as a general feature,
successful gauge
coupling unification requires
both extra color triplets and electroweak doublets to
appear {\it simultaneously}, and such representations
often do not appear together in these particular
models \cite{unif1,unif2}.

\begin{figure}[htb]
\centerline{\epsfxsize 3.5 truein \epsfbox {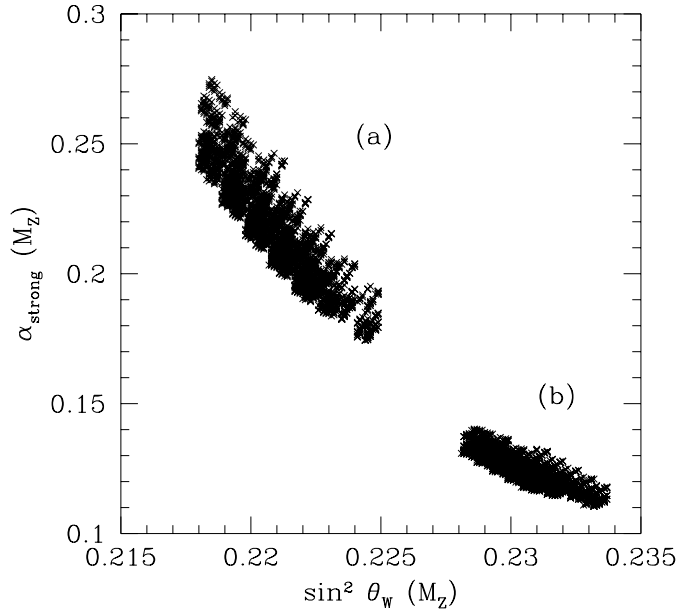}}
\caption{~~Scatter plot of $\lbrace\sin^2\theta_W(M_Z),
       \alpha_{\rm strong}(M_Z)\rbrace$
        when the SUSY-breaking parameters
     $\lbrace m_0,m_{1/2},M_h,M_{\tilde h},M_{\rm string}\rbrace$
      are varied as in Eq.~(\protect\ref{sps}).
     Region (a) plots the same points as in Fig.~\protect\ref{lightsusyfig},
    which assumes only the MSSM spectrum between the $Z$ scale and
    the string scale, while region (b)
       also includes the effects of the string-predicted extra matter
      discussed in the text.}
\label{finalfig}
\end{figure}

By contrast, for other realistic string models ({\it e.g.}, that
in Ref.~\cite{274model}),
the required matter appears in precisely
the combinations that can do the job.  Thus,
these models potentially allow successful
gauge coupling unification at the string scale ---
 {\it i.e.}, the correct low-energy
couplings can be achieved.
In the string model of Ref.~\cite{274model}, for example,
there appear two pairs of
$(\rep{\overline{3}},\rep{1})_{1/3}$
color triplets, one pair of
$(\rep{\overline{3}},\rep{1})_{1/6}$
color triplets,
and three pairs of
$(\rep{1},\rep{2})_{0}$ electroweak doublets.
This particular combination of representations and hypercharge
assignments opens up a sizable window
in which the low-energy data and string unification can
be reconciled:  if the triplets
all have equal masses in the approximate range
$ 2\times 10^{11} \leq M_3 \leq 7\times 10^{13}$ GeV,
with the doublet masses in the corresponding range
$ 9\times 10^{13} \leq M_2 \leq 7\times 10^{14}$ GeV,
then the discrepancy is removed.
This situation is illustrated in Fig.~\ref{finalfig},
which shows how the predicted low-energy couplings
in Fig.~\ref{lightsusyfig} are modified
when this extra matter is included in the analysis.
Details and other scenarios for each of the other
realistic string models can be found in Ref.~\cite{unif2}.

It is of course no surprise that extra non-MSSM matter at intermediate
scales can have such a profound effect on the running of the gauge
couplings, or equivalently on the values of low-energy observables.
What is perhaps more remarkable, however, is that
string theory, which predicts an unexpectedly high unification
scale, often also simultaneously predicts precisely
the extra exotic matter that is necessary in order to reconcile
this higher scale with low-energy data.  As we have already discussed,
the appearance of such matter is not at all arbitrary, and is
required for the internal consistency of the theory.
It is therefore encouraging that the states which appear
are often exactly the ones that can also reconcile
the predicted string scale of unification
with the observed low-energy couplings.
Indeed, it recently been proposed \cite{AEFdarkmatter} that
such matter may even serve as a potential dark-matter candidate.

Extra matter at intermediate mass scales
can also play a useful role in string models
for which the Planck-scale gauge group is {\it not}\/ that
of the MSSM.
In such cases, intermediate-scale matter {\it and}\/ intermediate-scale
gauge structure can together bring about gauge coupling unification at
the string scale.
For example, as we have already discussed in Sect.~4.4,
certain $G\times G$ string models have the feature that
they naturally give rise to extra moduli states transforming
in the adjoint of the gauge group.    If these states
have the appropriate mass scales, then they can effectively
raise the unification scale to the string scale and thereby
render unnecessary the use of a grand-unified group in the
first place \cite{bachas,FGM}.
Another recent idea employing
extra matter states has been proposed \cite{LN}
for the flipped $SU(5)$ string model.
Although the intermediate-scale gauge structure of this
model does not permit successful gauge coupling
unification by itself (as discussed in Sect.~7),
extra non-MSSM matter
in the {\bf 10} and {\bf 4} representations of the hidden
gauge groups $SO(10)$ and $SU(4)$ of this model
can nevertheless
potentially alter the running of the gauge couplings
in a manner which enables one to identify the
scale of $SU(5)\to SU(3)\times SU(2)$ breaking
with $M_{\rm MSSM}$, while identifying the scale of
the subsequent unification of $SU(5)$ and $U(1)$
gauge couplings with $M_{\rm string}$.
As with the other extra-matter proposals, however,
this proposal also requires
that the intermediate mass scales take certain
special values which are subject to a host of
dynamical and (in some cases) non-perturbative effects.
Thus, the viability of such schemes
awaits a detailed model-by-model analysis.

Finally, we remark that extra states beyond the MSSM
may also take the form of gauge bosons from extra $U(1)$
gauge symmetries.  Indeed, like the extra matter discussed above,
such extra $U(1)$ gauge symmetries
are also rather generic in realistic string models, and may
also lead to observable experimental consequences (including
a change in the unification scale at the two-loop order through
kinetic-mixing effects \cite{kinetic}).
Various phenomenological implications of such extra $U(1)$
gauge symmetries have recently been discussed in Ref.~\cite{zprime}.

\subsection{Determining the mass scale of extra matter}

Although extra exotic states beyond the MSSM naturally appear
in the context of various realistic string models,
it is a completely different question to verify
that this extra non-MSSM matter actually obtains the required
masses in these models.
As we have already remarked, such states typically appear
in vector-like representations, and thus can become superheavy.
The masses that such states will obtain can be determined
by analyzing the cubic and higher-order terms in
the superpotential.  For this reason,
the precise values of these masses depends on many factors.
These include the particular
vacuum shift which is required in order to break
the ``pseudo-anomalous'' $U(1)$ gauge
symmetry and restore spacetime supersymmetry \cite{DSWshift},
as well as the particular subsequent low-energy SUSY-breaking scheme
employed.

Determining the masses of such exotic states therefore
depends on a detailed analysis of the superpotential,
and can be done only model-by-model.
In one model, however, it has been estimated
\cite{Alonmassestimate} that the
masses of certain extra triplet states are approximately
$M_3\approx 10^{11}$ GeV.
Similar intermediate mass scales have also been
anticipated in, {\it e.g.}, Ref.~\cite{DKMNS}.
As we have seen,
this sort of intermediate scale is just what is required
for successful gauge coupling unification at the string scale.

It may seem surprising that string theory can generate
such intermediate mass scales
that are substantially below the Planck scale.
However, such intermediate mass scales can arise in a variety of ways.
One way is through the effects that arise
when a hidden-sector gauge group becomes strong
at some intermediate scale.
Another way that has been extensively exploited in the literature
is through the vacuum shift which generally arises in such $(2,0)$ models.
It is therefore important to understand why this vacuum
shift arises, and how it can be used to generate intermediate mass
scales.

This vacuum shift arises as follows.
As we mentioned above, most $(2,0)$ models have,
in addition to their observable- and hidden-sector gauge groups, a
``pseudo-anomalous'' $U(1)$ gauge symmetry [typically denoted $U(1)_X$,
with generator $Q_X$].  Here the adjective ``pseudo-anomalous'' simply
means that ${\rm Tr}\, Q_X\not=0$ where the trace is evaluated over the full
string spectrum (or equivalently over only the massless string
spectrum, since massive states come in vector-like
pairs with opposite values of $Q_X$).
In field theory, of course, the non-vanishing of this trace
implies the existence of a triangle anomaly.
In string theory, however,
all gauge and gravitational anomalies are automatically cancelled
by the Green-Schwarz mechanism \cite{GSmech},
and the anomalies associated with having ${\rm Tr}\, Q_X\not=0$
are no exception.  Indeed, in this case the Green-Schwarz
mechanism works by ensuring that the anomalous variation
of the field-theoretic $U(1)_X$ triangle diagram
can be cancelled via a corresponding
non-trivial $U(1)_X$ transformation of the
string axion field.
This axion field arises generically in string theory
as the pseudo-scalar partner of the dilaton, and couples universally
to all gauge groups.
However, the existence of such a mechanism implies that
anomaly cancellation in string theory does not require cancellation of
${\rm Tr}\,Q_X$ by itself, and consequently
a given string model can remain non-anomalous even while having
${\rm Tr}\, Q_X\not=0$.
Indeed, this is the generic case for most realistic $(2,0)$ models.

The danger posed by this situation arises
from the fact that such a non-vanishing trace leads to
the breaking of spacetime supersymmetry at one-loop order
through the appearance of a
one-loop Fayet-Iliopoulos $D$-term of the form \cite{DSWshift}
\beq
         {g_{\rm string}^2 \, {\rm Tr}\, Q_X \over 192 \,\pi^2 }\,
          M_{\rm Planck}^2
\label{Dterm}
\eeq
in the low-energy superpotential.
This in turn destabilizes the string ground state by generating a
dilaton tadpole at the two-loop level,
and signals that the original model cannot be consistent.
The solution to this problem, however, is
to give non-vanishing vacuum expectation values (VEV's) to
certain scalar fields (twisted moduli) $\phi$ in the string
model in such a way that the offending $D$-term in Eq.~(\ref{Dterm}) is
cancelled and spacetime supersymmetry is restored.
In string moduli space,
this procedure is equivalent to moving
to a nearby point at which the string ground
state is stable, and consequently this procedure is
referred to as {\it vacuum shifting}\/.

In general, the specific VEV's that parametrize this vacuum shift
can be determined by solving the various $F$- and $D$-term
flatness constraints.
In a given realistic string model, however,
these constraints are typically quite complicated,
and it is not always clear that a simultaneous solution
to all of the appropriate constraints exists.
Of course, the existence of such a solution is a
necessary prerequisite for the construction of a
 ``realistic'' string model, and such solutions
do exist for all of the models we have discussed
thus far.
In fact, one finds that the required VEV's are
typically small in these models, of the order
$\langle \phi \rangle/M_{\rm Planck} \sim {\cal O}(1/10)$.
Thus all of these models have
stable vacua which are relatively ``near'' those
points at which they were originally constructed.
This is fortunate, for it guarantees that many of
the phenomenological properties of such models will
not depend too crucially on whether they
are evaluated in the original vacuum (for which
a convenient conformal-field-theory description may exist,
simplifying the calculation),
or in the shifted, stable vacuum (at which the true
model is presumed to sit).
For example, heavy string threshold corrections
are relatively insensitive to such vacuum shifts,
chiefly because vacuum shifting
does not significantly affect the masses of those states
which were already massive before the shift \cite{thresholdcalcs}.
This was discussed in Sect.~6.5.2.

Vacuum shifting does have important consequences
for the massless states, however.
For example, vacuum shifting
clearly requires that those scalar fields
receiving VEV's be charged under $U(1)_X$.
Thus, the act of vacuum shifting breaks $U(1)_X$, with the
$U(1)_X$ gauge boson ``eating'' the axion to become massive.
In fact, since the scalars $\phi$ which are charged under $U(1)_X$
are also often charged under other gauge symmetries as well,
giving VEV's to these scalars typically causes further gauge
symmetry breaking.
For our purposes, however, the most important effect of vacuum
shifting is that it can generate effective superpotential mass
terms for vector-like states $\Psi$ that would otherwise be massless.
Indeed, upon replacing the scalar fields $\phi$ by their VEV's in
the low-energy superpotential,
one finds that higher-order non-renormalizable couplings
can become lower-order effective mass terms:
\beq
       {1\over M_{\rm Planck}^{n-1}}\,
        \phi^n \,\overline{\Psi}\Psi ~\to~
       {1\over M_{\rm Planck}^{n-1}}\,
        \langle\phi\rangle^n \,\overline{\Psi}\Psi ~.
\label{effmassterm}
\eeq

Given this observation, it is then straightforward
to estimate the typical sizes that such mass terms will have.
As we stated above,
the $D$- and $F$-flatness constraints typically
yield VEV's of the order
$\langle \phi \rangle/M_{\rm Planck} \sim {\cal O}(1/10)$.
This ratio is ultimately the origin of the intermediate mass scales.
Due to various selection rules stemming from
hidden string gauge symmetries,
effective mass terms for non-MSSM states
typically appear in the superpotential
only at a high order, {\it e.g.}, $n\sim 5$ in Eq.~(\ref{effmassterm}).
Consequently, the effective mass terms that are generated
after the vacuum shift are schematically
of the order $\langle \phi\rangle^n / M_{\rm Planck}^{n-1}\sim
   (1/10)^n M_{\rm Planck}$.
This then generates the desired intermediate mass scales.

Thus, we see
that the intermediate mass scales
required for string-scale
gauge coupling unification
can be generated quite naturally --- the only ingredients
are a vacuum shift that is relatively small (but still of order one),
coupled with hidden-sector symmetries to
enforce selection rules at relatively low orders in the
superpotential.
As we have seen, both of these features arise
quite naturally in realistic string models.
We nevertheless emphasize that a rigorous calculation of
these intermediate mass scales can be quite involved, and must
ultimately be done on a model-by-model basis.

\subsection{Semi-perturbative unification and dilaton runaway}

As we have discussed, the appearance of extra matter beyond the MSSM can
have the effect of raising the gauge coupling unification scale.
Indeed, we have seen that in the realistic string models, this extra matter
typically
comes in {\it incomplete}\/ $SU(5)$ or $SO(10)$
multiplets [{\it i.e.}, such matter cannot be
assembled into only {\bf 5} or {\bf 10} representations of
$SU(5)$], and it is precisely this feature which enables
the introduction of such matter to raise the one-loop unification
scale.   By contrast, the introduction of {\it complete}\/ GUT multiplets
cannot raise the one-loop unification scale, and instead only
raises the value of the coupling $g_{\rm string}$ at unification.

Despite this fact, it turns out that extra matter in
complete multiplets may nevertheless be extremely useful
in string theory.
Until now, we have had little to say about the value of $g_{\rm string}$,
and have simply remarked that its value is fixed by the expectation value
of a certain modulus field, the dilaton $\phi$, via a relation
of the form $g_{\rm string}\sim e^{-\langle \phi \rangle}$ \cite{Witten}.
It is natural to wonder, therefore, how the vacuum expectation
value of the dilaton is fixed.
Unfortunately,
this question is particularly difficult to answer because
of some general results \cite{dilatonstabilize}
which assert that in supersymmetric string theories,
the dilaton has a potential which is classically flat and which
remains flat to all orders in perturbation theory.
Of course, a dilaton potential can be generated
if small non-perturbative effects are included
(the common example being gaugino condensation due to
unknown hidden-sector dynamics),
but it has been shown \cite{dilatonstabilize}
that all such potentials must nevertheless
vanish as $\langle \phi\rangle \to \infty$.
Thus, if the string coupling $g_{\rm string}$ is presumed to be very weak
(so that the corresponding string theory is perturbative),
then the shape of the dilaton potential
forces the dilaton vacuum expectation value to increase without bound.
This in turn implies that weakly-coupled string theories are not stable,
and that they eventually become free theories.
This is the so-called {\it dilaton runaway problem}\/.

How then might one stabilize the dilaton in order to avoid this problem?
This issue has been discussed in Ref.~\cite{banksdine}.
One possibility might be that the true value of $g_{\rm string}$ is
actually quite large, so that the above
predictions at weak coupling might be avoided
by unknown strong-coupling effects.
Unfortunately, this would mean that one can no longer analyze
the string model through perturbation theory, which
has been the basis of our analysis of gauge coupling unification.
Furthermore, recent results in string duality \cite{duality} suggest that
various
strings at strong coupling are equivalent to other theories at weak
coupling, and therefore the original dilaton runaway problem may not be avoided
after
all, but may instead simply re-emerge in the dual theory.\footnote{
      Note, however, that such a ``dual'' theory may have different
      unification properties than the original theory.
      This is the basis of a strong-coupling unification scenario that
      will be discussed in Sect.~10.  We also remark that
      there exist values of coupling and compactification volume
      of heterotic strings for which all possible
      dual theories are also strongly coupled.  This is discussed
      in Ref.~\cite{Shirman}. }

Thus, the best hope might be that somehow the value of the string coupling
$g_{\rm string}$ lies at some {\it intermediate}\/ value.  This option
has been called {\it semi-perturbative unification} \cite{semipert}.
Indeed, not only might this scenario avoid
the dilaton runaway problem, but it might still permit a perturbative
analysis.  It may at first seem inconsistent that
unknown {\it non-perturbative}\/ string
effects are being invoked to stabilize
the dilaton while perturbation theory is assumed to be valid
for analyzing the low-energy effective field theory.  However, it has
been shown  \cite{shenker} that the strength of non-perturbative effects
in string theory typically grows as $\sim \exp (-a/g_{\rm string})$
where $a$ is a constant of order one,
whereas in field theory such effects typically grow more slowly,
as $\sim \exp(-8\pi^2/g_{\rm string}^2)$.
Thus, precisely for such intermediate values of $g_{\rm string}$,
it is possible that the dilaton can be stabilized without sacrificing
a perturbative treatment of the low-energy effective field theory.
By ``intermediate'' values we refer to couplings which may be as
high as $\alpha_{\rm string}\sim 0.3$ or $0.4$.
While the usual MSSM value for the unified coupling $g_{\rm string}\sim 0.7$
(or $\alpha_{\rm string}\sim 1/25$)
might also be sufficiently strong to enable
such string-theoretic effects to stabilize the dilaton in this manner,
the above ``intermediate'' values
may be able to do this more effectively.

This, then, provides a string-theoretic motivation for intermediate
values of $g_{\rm string}$.  Moreover, as we have discussed above,
it turns out that such intermediate values of
$g_{\rm string}$ can be achieved --- without ruining agreement
with the measured values of the low-energy couplings ---
through the introduction of extra non-MSSM matter
in complete GUT multiplets.
However, there is also one further benefit to having extra matter of this form.
We have already remarked that such matter is not capable of increasing
the one-loop scale of unification.  However, since the
gauge coupling unification in this scenario is only {\it semi}\/-perturbative,
we have less reason than before to trust that a one-loop analysis is
sufficient.
Indeed, it is possible that higher-loop effects can raise the unification
scale.  A detailed analysis of this question has recently been
performed \cite{semipert}, and indeed one finds that it is possible to add
particular numbers of complete multiplets [{\it e.g.}, $\rep{5}$ or $\rep{10}$
representations of $SU(5)$, or $\rep{16}$ representations of $SO(10)$]
so that the new unification scale approaches the predicted one-loop string
scale
without destroying the validity of perturbation theory.
Such a scenario was sketched in Fig.~\ref{paths}(f).
This route to unification is particularly noteworthy,
given that the predicted one-loop string scale
itself is enhanced due to the increased value of $g_{\rm string}$.
Unfortunately, no calculations of the corresponding higher-loop corrections
to the predicted string unification scale currently exist.

Thus, semi-perturbative unification achieved through extra matter in complete
GUT multiplets is an intriguing field-theoretic scenario whereby the
extrapolated unification scale can be raised and the string dilaton runaway
problem may be avoided.
There are, however, three points that must be stressed.
First, it is important to realize that this scenario does not
 {\it solve}\/ the dilaton stabilization problem;  it merely
evades the dilaton {\it runaway}\/ problem.
Second, as we have stated above, even
the usual MSSM unified coupling $\alpha_{\rm string}\sim 1/25$ is in
some sense ``intermediate'',
since it too permits sufficiently large string-theoretic non-perturbative
corrections to potentially stabilize the dilaton.
However, such values $\alpha_{\rm string}\sim 1/25$ fail, by themselves,
to lead to a raising of the unification scale.
Finally, as we have seen at the beginning of this section,
even the introduction of extra matter in {\it incomplete}\/
multiplets can have the effect of changing the unification coupling
while preserving unification and increasing the unification scale;
this is precisely what occurs, for example, in the realistic string model
underlying Fig.~\ref{finalfig}.
It will nevertheless be interesting to construct realistic string
models with extra matter in complete GUT multiplets,
in order to test whether the semi-perturbative gauge coupling unification
mechanism with $\alpha_{\rm string}\sim 0.3-0.4$ can actually be realized
in string theory.


\setcounter{footnote}{0}
\section{Path \#6:  Strings without Supersymmetry}

Another possible scenario for gauge coupling unification
--- one which is much more unconventional and speculative
than any considered thus far ---
involves string models {\it without}\/ spacetime supersymmetry.
This remarkable possibility rests upon the little-exploited
observation (see Fig.~\ref{introfiga}) that within the
 {\it non}\/-supersymmetric Standard Model,
the $SU(2)$ and $SU(3)$ gauge couplings
already unify at the string scale!
Indeed, only the hypercharge coupling fails to unify
at the same point.
However, the running of the hypercharge coupling depends
on the normalization $k_Y$ of the hypercharge generator,
and while this is taken to be $k_Y=5/3$ in the MSSM
[regardless of whether any $SU(5)$  GUT theory is envisaged],
we have seen in Sect.~5 that in string theory,
the normalization $k_Y$ has completely different
origins, and the choice $k_Y=5/3$ is not required or preferred.
Thus, simply by adjusting $k_Y$,
we can achieve a simultaneous
unification of all of the gauge couplings, at the string
scale, and {\it without supersymmetry}.

\begin{figure}[thb]
\centerline{\epsfxsize 3.5 truein \epsfbox {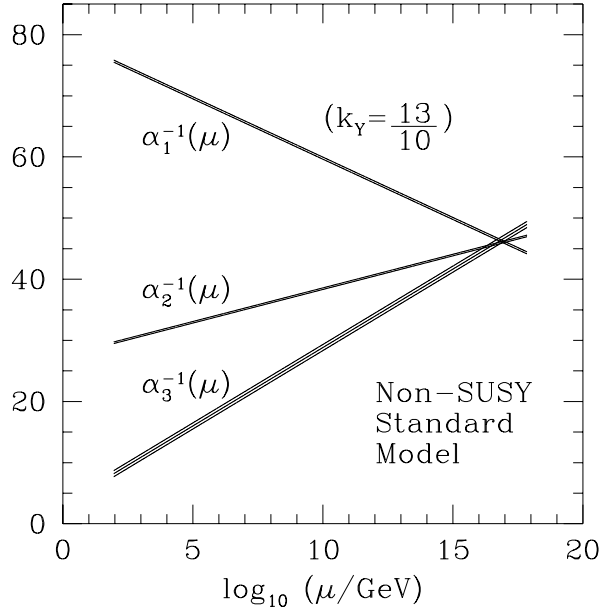}}
\caption{ Gauge coupling unification in the Standard Model
  can be achieved {\it without supersymmetry}
  for a hypercharge normalization $k_Y\approx 13/10$.
  In this scenario, the unification scale is naturally
  close to the string scale $\sim {\cal O}(10^{17}~ {\rm GeV})$.}
\label{nonsusystringfig}
\end{figure}

It is straightforward to calculate the value of $k_Y$ that is
required in this scenario,
and we find that we require $k_Y\approx 13/10$.
It is important that this value is
well within the constraints imposed by string theory
(and most importantly, satisfies $k_Y\geq 1$,
as required for a consistent hypercharge assignment for the right-handed
electron).
Therefore, instead of the unification relation given in
Eq.~(\ref{MSSMunification}),
we find that we now have the {\it non}\/-supersymmetric unification relation
\beq
           {13\over 10}\,\alpha_Y (M_{\rm SM}) ~=~
           \alpha_2 (M_{\rm SM}) ~=~
           \alpha_3 (M_{\rm SM}) ~\approx~ {1\over 45} ~
\label{SMunification}
\eeq
which holds at the new, higher, ``Standard Model unification scale''
\beq
           M_{\rm SM} ~\approx~ 10^{17}\,{\rm GeV}~.
\label{MSM}
\eeq
The unification of gauge couplings in this scenario is
shown in Fig.~\ref{nonsusystringfig}, which should be compared
against Fig.~\ref{introfiga}.
Thus, we conclude that spacetime supersymmetry is not required
on the basis of gauge coupling unification, provided we
normalize the hypercharge generator accordingly.

Note that although this scenario is similar in spirit to that
discussed in Sect.~5, the key difference here is
that {\it no supersymmetry is required}\/.  Thus,
the running of the gauge couplings is governed by
the beta functions of the Standard Model rather than those
of the MSSM, and neither includes the effects of, nor requires
the existence of, any MSSM superpartners.
Furthermore, the resulting unification
automatically occurs at a higher scale $\approx 10^{17}$ GeV
than is comfortably achieved in the supersymmetric case
with $k_Y\approx 1.5$, and may thus be more desirable from
a string-theory perspective.

Given that no experimental evidence for spacetime supersymmetry
currently exists, this possibility  (though highly unorthodox)
merits examination.
In particular, three groups of serious questions naturally arise:
\begin{itemize}
\item Can suitable non-supersymmetric
    string models even be constructed?  In other words,
    is spacetime supersymmetry required for the self-consistency
    of the string?
\item  If we dispense with spacetime supersymmetry, how
     can we reproduce many of the other phenomenological
      benefits that supersymmetry provides, such as controlling
      divergences and providing a
     technical solution to the gauge hierarchy problem?
\item  How can we explain the vanishing of the cosmological
      constant?
\end{itemize}
To date, there are no satisfactory answers to these questions
(indeed, the third question has no resolution even within the MSSM,
given the fact that SUSY must be broken somewhere at or above
the electroweak scale \cite{Lambdareview}).
There are, however, certain tantalizing hints that if string theory
can solve any one of these problems, it may solve them all
simultaneously.
Furthermore, if this does happen, such a solution might take a form
that is completely
unexpected from a field-theory point of view.

To see how such a scenario might work,
let us begin by discussing the first issue:
the construction of suitable non-supersymmetric string models.
Most of our experience building string models has focused
on string models possessing $N=1$ spacetime supersymmetry.
Not only is $N=1$ supersymmetry desired on a phenomenological
basis (as a result of gauge coupling unification within
the $N=1$ MSSM),
but the presence of supersymmetry naturally eliminates certain
problems that could otherwise arise in string theory, such as the appearance
of tachyons in the string spectrum.  One might think, therefore,
that it is impossible to construct non-supersymmetric
string models which are free of tachyons.  Fortunately, it turns
out that this is not the case:  just as there are millions
of $N=1$ supersymmetric string models, there are
millions of non-supersymmetric tachyon-free models.  Indeed,
one must simply choose the GSO projections
in such a way as to project such tachyonic states out of the spectrum.
Perhaps the most famous example of such a self-consistent string model is
the ten-dimensional tachyon-free $SO(16)\times SO(16)$ heterotic string
model \cite{so16so16string}.
There are likewise a plethora of such models in four dimensions
\cite{KRDlambda}.
While some of these models can be understood as resulting from
supersymmetric models via a form of Scherk-Schwarz
SUSY-breaking \cite{scherkschwarz,antonlargeradius,DKLkiritsistwo},
others bear no relation to supersymmetric models at all.

To build a suitable non-supersymmetric string model,
then, we would simply seek to construct a string model whose low-energy
spectrum reproduces the Standard Model, rather than the MSSM.
Presumably this could be achieved by starting with some of
the realistic MSSM-reproducing string models, and then projecting
out the superpartners of the massless states.
In principle, this could be done either by
removing the superpartner sectors from the theory altogether,
or by carefully adjusting the GSO projections in such a way
that the superpartners of massless states are removed but no
tachyons are introduced.
The string-predicted tree-level unification of gauge couplings, as
given in Eq.~(\ref{unification}), remains valid even without spacetime
supersymmetry.

Thus, at tree level, there does not appear to be any barrier
against constructing the desired four-dimensional,
non-supersymmetric, realistic, tachyon-free string models.
Beyond tree level, however, certain subtle issues arise.
In particular, while non-supersymmetric string models are perfectly
valid solutions (or classical vacua) of string theory
at tree level, they suffer from various
instabilities at one loop.  Perhaps the most serious of these
concerns the dilaton:  it turns out that for generic non-supersymmetric
string models, the dilaton develops a non-vanishing one-point function.
This implies that the dilaton effective potential contains a linear
term, which in turn indicates that the non-supersymmetric
ground state is not stable.
This ground state is then presumed to ``flow'' to some other point
in the string moduli space at which stability is restored.

Is supersymmetry therefore required in order to cancel this
dilaton one-point function and provide a stable ground state?
The answer to this question is not known.
At one-loop order, the dilaton one-point function is
proportional \cite{stringreview} to the dimensionless
string cosmological constant $\Lambda$ defined
in Eq.~(\ref{Lambdadef}).
Thus, at one-loop order, the dangerous dilaton one-point function
does not arise in string models with $\Lambda=0$.
It is interesting that in string theory, the problem of
the dilaton is so intimately connected with the problem of
the cosmological constant.\footnote{
    An interesting additional connection between
    these two problems has recently been
    proposed in Ref.~\cite{Wittencos}.
   }
Of course, models with spacetime supersymmetry have $\Lambda=0$
(to all orders in perturbation theory \cite{allorders}).
However, despite numerous attempts and various
proposals \cite{lambdarefs,KRDlambda},
it is not known whether there exist four-dimensional
 {\it non-supersymmetric}\/ string models which also
have vanishing cosmological constant.
Thus, it is not yet known whether there exist any
non-supersymmetric string models which are stable beyond tree level.

Let us now turn to the remaining question:
without supersymmetry, how might the gauge hierarchy problem be solved?
As we shall see, this issue is also tied to the above questions.
Let us first recall the situation in field theory.
In a supersymmetric field theory, a tree-level
gauge hierarchy is stabilized at higher loops because the contributions
of bosonic states cancel those of fermionic states level-by-level,
at all masses.
Thus, the divergences which would otherwise destabilize the
gauge hierarchy are absent.
This cancellation is often encoded in the
cancellation of the mass supertraces
\beq
    \Str M^{2\beta} ~\equiv ~ \sum_{\rm states}\,(-1)^F \,M^{2\beta}
\eeq
for various $\beta$.   For example, $\Str M^4$ controls the logarithmic
divergence in the vacuum energy density, while $\Str M^2$
and $\Str M^0\equiv \Str \bone$ respectively control the
quadratic and quartic divergences.
Of course, at some scale it is necessary to break the supersymmetry,
and in field theory this is usually done either softly or spontaneously.
One then finds that $\Str\bone $
continues to vanish, while $\Str M^2$ becomes non-zero.
Thus, the quartic divergences are still absent,
and the value of $\Str M^2$
is then constrained so as to preserve
the technical solution to the gauge hierarchy problem.
In practice,
this means that the numerical size of
$\Str M^2$ is bounded by two phenomenological
observations:  no superpartners have yet been observed
(which sets a minimum for $\Str M^2$), and
the Higgs mass should not be too large (which sets
a maximum).
The important point here, however, is that one achieves
a technical resolution of the gauge hierarchy problem
in field theory
by introducing supersymmetry at some high scale, and
then breaking it softly at a much lower scale.

For a non-supersymmetric string model --- which by definition
has no supersymmetry at {\it any}\/ scale --- it might seem
that such an achievement might be beyond reach.
However, it has recently been shown \cite{supertraces}
that this is not the case.
In particular, if we evaluate the supertraces
(as appropriate for string theory) by summing
over the contributions of all string states, both massive
and massive, in the presence of a suitable regulator,
\beq
    \Str M^{2\beta} ~\equiv ~
        \lim_{y\to 0} \,  \sum_{\rm states}\,(-1)^F \,M^{2\beta}\,
               e^{-y M^2}~,
\eeq
then it turns out that
 all four-dimensional tachyon-free non-supersymmetric string models
 {\it automatically}\/ satisfy the supertrace relations \cite{supertraces}
\beq
       \Str \bone ~=~0~~~~~~{\rm and}~~~~~~
       \Str M^2 ~=~ -{3\over 4\pi^2\,\alpha'}\,\Lambda~.
\label{supertracerelations}
\eeq
Thus, we see that the required sorts of
supertrace relations are automatically satisfied in string theory,
even without supersymmetry appearing at any scale!
Moreover, it has been found \cite{missusy,supertraces}
these stringy supertrace relations (\ref{supertracerelations})
do {\it not}\/ hold multiplet-by-multiplet
(as they do in field theory),
but instead hold as the result of delicate cancellations
at all mass levels
throughout the entire string spectrum.
Indeed, the spectrum of a typical non-supersymmetric string model
need not exhibit any (broken) multiplet structure at all,
and in particular need not contain (at any mass level)
the superpartners of low-energy states.
Nevertheless, the numbers of bosonic and fermionic states at
all string mass levels always miraculously adjust
themselves in such a way that a given surplus of
bosonic states at any one level is delicately balanced by
surpluses of fermionic states at other levels and
the supertrace relations (\ref{supertracerelations}) are maintained.

\begin{figure}[thb]
\centerline{\epsfxsize 3.5 truein \epsfbox {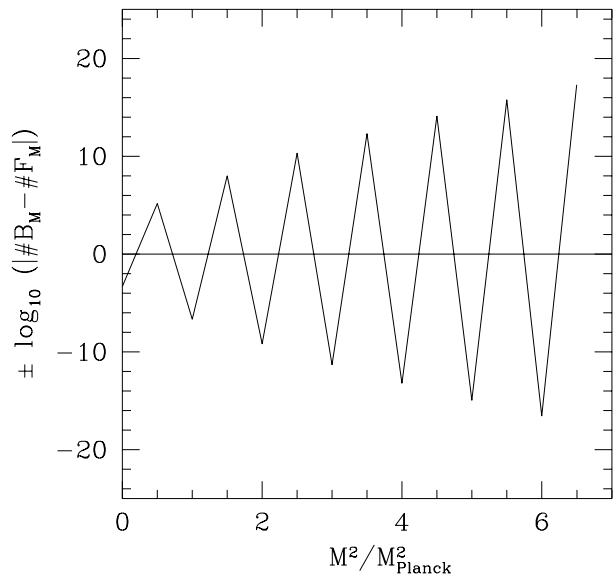}}
\caption{Alternating boson/fermion surpluses
    in the $D=10$ non-supersymmetric tachyon-free
    $SO(16)\times SO(16)$ heterotic string.
   For each mass level in this model (either integer or half-integer
   in Planck-mass units), we plot
  $\pm \log_{10}(|B_M-F_M|)$ where $B_M$ and $F_M$ are respectively
  the corresponding numbers of spacetime bosonic and fermionic states.
  The overall sign is chosen positive if $B_M>F_M$, and negative otherwise.
   The points are connected in order to stress the alternating, oscillatory
   behavior of the boson and fermion surpluses throughout the string
   spectrum.
   These oscillations
   insure that $ \Str M^0= \Str M^2= \Str M^4= \Str M^6=0$
   in this model, even though there is no spacetime supersymmetry.  }
\label{missusyfig}
\end{figure}

Such a delicate balancing of boson/fermion surpluses has been
called a ``misaligned supersymmetry'' \cite{missusy}, and is illustrated
in Fig.~\ref{missusyfig} for the case of the non-supersymmetric
ten-dimensional $SO(16)\times SO(16)$ string \cite{so16so16string}.
In this particular string,
one has a surplus of 2112 fermionic states at the massless level,
and these are balanced by a surplus of 147,456 bosonic states at the
first excited level, which in turn are balanced a surplus
of 4,713,984 fermionic states
at the next excited level, and so forth.
Indeed, as a general property, one finds that
such bosonic and fermionic surpluses typically
alternate as the mass is increased \cite{missusy}.
Moreover, for non-supersymmetric strings in ten dimensions,
one also finds \cite{supertraces} that the four-dimensional
results (\ref{supertracerelations}) are generalized in such
a way that not only does $\Str \bone$ vanish, but in fact
$\Str M^2$, $\Str M^4$, and $\Str M^6$ {\it all vanish as well!}~
It is indeed remarkable that such conditions can be satisfied
in string theory, especially given the infinite towers of states
whose degeneracies increase exponentially as a function of mass.

This, then, may be an alternative solution
to the gauge hierarchy problem, a novel way
in which string theory manages to maintain finite amplitudes
and constrain the corresponding mass supertraces,
even without supersymmetry or TeV-scale superpartners.
Indeed, string theory appears to ``dress up'' a non-supersymmetric
massless spectrum with a ``cloud'' of infinitely many Planck-scale
states in just the right way to enforce finiteness and cancel the
supertraces.
Of course, for four-dimensional
string models, we see from Eq.~(\ref{supertracerelations})
that the phenomenologically preferred values for
$\Str M^2$ require a one-loop cosmological constant
which is extremely small in Planck-scale units.
Thus, once again, such a solution to the gauge hierarchy
problem is closely tied to the question of vacuum stability,
and to the problem of the cosmological constant as a whole.

These intriguing results indicate that
non-supersymmetric string theories have some remarkable
properties that mimic the properties of supersymmetric
string theories, though in unexpected ways.
Moreover, these extra finiteness properties  --- coupled
with our ability to adjust $k_Y$ in string theory ---
may imply that spacetime supersymmetry is not as critical in
string theory as it is in ordinary field theory, at least as
far as gauge coupling unification or the gauge hierarchy problem
are concerned.
Non-supersymmetric strings therefore remain
an interesting, though largely unexplored, alternative
avenue towards gauge coupling unification.


\setcounter{footnote}{0}
\section{Path \#7:  Strings at Strong Coupling}

Another possible resolution of the gauge coupling problem  ---
one which has only recently been proposed ---
grows out of exciting developments involving
newly discovered string dualities \cite{duality}.
Such dualities have the potential for describing the strong-coupling
limits of string theories in various dimensions (including
the phenomenologically appealing four-dimensional heterotic string
theories we have been discussing here), and may ultimately shed
light on complex issues such as supersymmetry breaking
and the selection of the string vacuum.
Indeed, recent developments seem to indicate that perhaps
the string itself is not fundamental to ``string'' theory,
but that string theory is only an effective theory arising
as a certain weak-coupling limit of a more general theory
(an eleven-dimensional theory recently called ``M-theory'')
which appears to have deep connections to membrane theory
and eleven-dimensional supergravity.
If so, these developments have the potential to substantially
alter our view not only of string theory
itself, but also its phenomenological implications.

For the purposes of this article, however,
the phenomenological implication that most concerns us
is a possible {\it strong-coupling}\/ solution to
the gauge coupling unification problem.
Until now, all of the different ``paths to unification'' that we have
discussed are essentially {\it perturbative}, and rely on
the assumption that the four-dimensional string coupling,
as well as the coupling of the ten-dimensional theory
from which it is derived via compactification, are weak.\footnote{
    Note that in this context, the semi-perturbative unification scenario
     discussed in Sect.~8.3 still may be considered to be
     a ``weak-coupling'' scenario, for like the other scenarios
    we have considered, it too permits a perturbative treatment
     within the framework of the heterotic string.}
Indeed, it is only in such a limit that the tree-level unification
relations (\ref{unification})
are expected to hold.  It is precisely these relations, however,
which are the source of the discrepancy
between the string and MSSM unification scales, for by unifying
the Newtonian gravitational coupling $G_N$ with the gauge couplings
$g_i$ through $g_{\rm string}$, this relation ultimately
connects the gauge coupling unification
scale to the gravitationally-determined string scale (or Planck
scale).  Indeed, from Eq.~(\ref{unification}), we find
\beq
      G_N ~=~ {\textstyle{1\over 2}}\, \alpha'\, \alpha_{\rm string}~.
\label{weakcouplingGN}
\eeq

Given that Eq.~(\ref{weakcouplingGN}) is the prediction of
weakly coupled heterotic string theory, the
question then arises as to whether this relation can somehow
be altered or loosened at strong coupling.
General arguments which explain why such a loosening might occur
at strong coupling can be found in Ref.~\cite{banksdine},
and a simple heuristic argument (based on Ref.~\cite{newwitten}) is as follows.
It is a straightforward matter to relate the four-dimensional
string coupling $\alpha_{\rm string}$ to the ten-dimensional
string coupling $\alpha_{10}$ and the volume $V$ of the corresponding
six-dimensional compactification manifold:
\beq
   \alpha_{\rm string}~\sim~ {(\alpha')^3 \, \alpha_{10} \over V}~.
\label{couplingvolume}
\eeq
If we now identify $V\sim M_{\rm string}^{-6}$,
we find from Eq.~(\ref{weakcouplingGN}) that
\beq
         G_N ~\sim~ {\alpha_{\rm string}^{4/3} \over
               M_{\rm string}^2 \, \alpha_{10}^{1/3} }~.
\eeq
Thus, if we imagine strictly identifying $\alpha_{\rm string}$ and $M_{\rm
string}$
with the values $\alpha_{\rm MSSM}$ and $M_{\rm MSSM}$ at unification
(thereby taking gauge coupling unification within the MSSM as
an experimental {\it input}\/ to fix the scale of string theory),
we find that we must make $\alpha_{10}$ very large
in order to reduce $G_N$ to its experimentally observed value.
In other words, increasing $\alpha_{10}$ has the effect of increasing
the gravitational scale $M_{\rm string}$ above $M_{\rm MSSM}$, as desired.
Of course, for $\alpha_{10} \gg 1$ (as would be necessary), the perturbative
underpinning of these calculations breaks down, and one must deal directly
with the strongly-coupled strings in ten dimensions.

These abstract observations are, in principle, not new.  However, thanks
to the recently proposed string dualities, such a calculation
can now be done \cite{newwitten}.
Since the strongly coupled ten-dimensional $SO(32)$ string is apparently
described by a weakly coupled ten-dimensional Type~I string \cite{type1dual},
the relations (\ref{unification}) that were derived for heterotic strings
can, in this case, simply be replaced by their Type~I counterparts.
Likewise, for the ten-dimensional $E_8\times E_8$ string, the conjectured
dual \cite{e8dual} (a certain eleven-dimensional theory compactified on
$S^1/\IZ_2$)
allows one to obtain an appropriate corresponding
unification relation generalizing Eq.~(\ref{unification}) for this case as
well.
Remarkably, in each case, it then turns out \cite{newwitten} that the
new unification relations thus obtained are modified relative to
Eq.~(\ref{unification})
in such a way that, after compactification to four dimensions,
the experimental discrepancies may be resolved.

To be more specific, let us first consider the
case of the strongly coupled $SO(32)$ string, or equivalently
the weakly coupled Type~I string.
What are the unification relations for the Type~I string?
Recall that in the heterotic string,
the gauge and gravitational interactions arise in the same (closed-string)
sector;  their respective couplings $g_{\rm gauge}$ and $\kappa\sim
\sqrt{G_N/\alpha'}$
therefore have the same dilaton-dependence $e^{-\phi}$, and indeed one finds
$\kappa\sim g_{\rm gauge}$.
This is the origin \cite{Ginsparg} of the heterotic
unification relations (\ref{unification}).
In Type~I strings, by contrast,
the gravitational interactions continue to arise in the closed-string sector,
but the gauge interactions now arise in the open-string sector.
Their couplings are therefore related as $\kappa\sim g^2$, and their
dilaton-dependence does not cancel.
Indeed, one finds that the weak-coupling heterotic unification relations
(\ref{unification}) are replaced by the Type~I relations \cite{newwitten}:
\beq
    8\pi \,e^{\phi}\, {G_N\over \alpha' } ~=~ g_{\rm gauge}^2~
\label{unificationtype1}
\eeq
where $\phi$ is the ten-dimensional Type~I dilaton
and $g_{\rm gauge}$ is the unified gauge coupling constant.
This result, without modification,
holds in the ten-dimensional theory and in any lower-dimensional
theory obtained from such a string via compactification.
However, while
this new relation does not disturb the unification of the {\it gauge}\/
couplings, the final relation to the Newton gravitational coupling $G_N$
is changed relative to Eq.~(\ref{unification}):
instead of Eq.~(\ref{weakcouplingGN}), we now find
\beq
     G_N ~=~ {\textstyle{1\over 2}}\, e^{-\phi}\, \alpha'\,
         \alpha_{\rm string}~
\label{newGN}
\eeq
where $\alpha_{\rm string}\equiv g_{\rm gauge}^2 /(4\pi)$.
Thus, by taking the ten-dimensional Type~I string to be sufficiently
weakly coupled ({\it i.e.}, by taking $e^{-\phi}$ to be sufficiently small),
the experimental value of $G_N$ can be reconciled with
the MSSM unification scale.
Moreover, as stressed in Ref.~\cite{newwitten},
the results in Eqs.~(\ref{unificationtype1}) and (\ref{newGN})
hold directly for the weakly coupled Type~I string independently of
its relation to the strongly coupled $SO(32)$ heterotic string,
and hence could form the basis for a successful gauge coupling
unification scenario directly within the framework of Type~I
phenomenology.  Indeed, along these lines,
several four-dimensional Type~I string models with $N=1$ supersymmetry
have recently been constructed \cite{TypeImodels}
using various orientifold techniques \cite{TypeItechniques};
moreover, heavy string threshold corrections for certain types of
Type~I string models
have also recently been calculated \cite{bachasfabre}.
Of course, the presence of the {\it a priori}\/ unfixed parameter $e^{-\phi}$
in these relations implies that the Type~I string is not as predictive
as the heterotic.

The situation for the ten-dimensional $E_8\times E_8$ heterotic string is
similar,
though somewhat more complicated due to the fact that this string is dual
not to another string, but rather to an eleven-dimensional ``M-theory''
compactified on $S^1/\IZ_2$.
The resulting gauge and gravitational coupling
unification relations therefore depend not only on the
geometrical details of this compactification --- such as the
length of the line segment $S^1/\IZ_2$ ---
but also on the precise relations between the ten-dimensional
gauge couplings and the eleven-dimensional
gravitational couplings, on various special topological
constraints, and on the intricacies of certain eleven-dimensional
strong-coupling expansions.
These issues are all discussed in Ref.~\cite{newwitten}.
One finds, however, that in this case the
unification relations (\ref{unification}) now take the form \cite{newwitten}
\beq
    16 \pi^3\, \left( {4\pi \over \kappa}\right)^{2/3} \rho ~G_N ~=~
      g_i^2\,k_i
\label{unificationM}
\eeq
where $\kappa$ is the underlying eleven-dimensional gravitational coupling,
and where $\rho$ is the length of $S^1/\IZ_2$.
As before, this result holds for the ten-dimensional theory
as well as for any lower-dimensional theory obtained via compactification.
While there are certain self-consistency constraints on the ranges
of values that $\kappa$ and $\rho$ may have in order for this analysis
to be valid \cite{newwitten},
experimentally acceptable values of $G_N$ can
once again apparently be accommodated:  one requires, in particular,
that $\rho$ be substantially larger than the
eleven-dimensional Planck length, which suggests that at intermediate
energy scales there should be an effective {\it fifth}\/ dimension
that becomes accessible before ultimately realizing
the full eleven-dimensional theory.
Note that various phenomenological investigations regarding possible
large internal dimensions in connection with supersymmetry breaking
can be found in Refs.~\cite{scherkschwarz,antonlargeradius,DKLkiritsistwo}
and more recently in Ref.~\cite{newanton};
the latter paper also proposes an alternative unification scenario which
differs from the one discussed here.
It also turns out that there are limits on the size of
the internal six-dimensional compactification volume \cite{newkap};
recall that this issue becomes especially relevant in light of the fact that
the (presumably strong) ten-dimensional coupling $\alpha_{10}$ and the
(presumably weak) four-dimensional coupling $\alpha_{\rm string}$ are
related to each other through a (presumably large) six-dimensional
compactification
volume, as indicated in Eq.~(\ref{couplingvolume}).
Several recent attempts at understanding the general
phenomenological implications of such strong-coupling and large-volume
compactification scenarios can be
found in Ref.~\cite{banksdinehorava}.

Thus, from this analysis, we see that the solution to the gauge
coupling problem in string theory may involve not only perturbative
effects, but also intrinsically non-perturbative physics.
It will be interesting to see whether realistic
models ({\it i.e.}, Type~I models or M-theory compactifications) can be
constructed which exploit these observations.


\setcounter{footnote}{0}
\section{Conclusions}

In this article, we have reviewed the status of the various approaches
that have been taken in recent years towards understanding the
unification of gauge couplings within string theory.
As we have seen, the chief issue in all of these approaches is
the need to reconcile the unification
scale predicted within the framework of the MSSM with the unification
scale expected within the framework of heterotic string theory --- or
equivalently, the need to reconcile the values of the low-energy
gauge couplings as measured by experiment
with the values predicted by string theory.
The approaches we have reviewed in this article
include the possibilities of building string GUT models;
models with non-standard affine levels and hypercharge normalizations;
and models with large heavy string threshold corrections, light SUSY
thresholds,
intermediate-scale gauge structure, or extra
matter beyond the MSSM.
We have also discussed two more speculative proposals:
strings without spacetime supersymmetry, and strings at strong coupling.
It is clear that there has been considerable progress in each of these
areas.

At present, it would appear that some of these approaches have
met with more success than others.  For example,
although heavy string threshold corrections have the theoretical potential
to be sufficiently large, and although there exist various attractive
mechanisms
by which they might actually become large, no phenomenologically
realistic string models have been constructed which realize these
mechanisms and actually have large threshold corrections.  Rather, all of the
calculations that have been performed within realistic string
models seem to indicate that these heavy string threshold
corrections are unexpectedly small, and moreover have signs
that enlarge, rather than diminish, the discrepancy between
the two unification scales.
Likewise, we have seen in a model-independent way that light SUSY thresholds
are also typically too small to resolve the low-energy discrepancies,
and we similarly found that in string models realizing
a variety of realistic symmetry-breaking scenarios
[such as, {\it e.g.}\/, the flipped $SU(5)$ or
$SO(6)\times SO(4)$ scenarios], the presence of such
intermediate-scale gauge structure also does not
enable a reconciliation of the two scales.

As far as realistic string models are concerned, the
best route to unification seems to be the appearance
of extra matter beyond the MSSM.  Although the introduction
of such matter might seem {\it ad hoc}\/ from a field-theory
perspective, we have seen that such extra matter appears
as a generic feature in most realistic string models
and has the potential to alter the running of the gauge couplings
significantly.  While some realistic string models do not
contain such extra non-MSSM matter in
the combinations that are required in order to achieve successful string-scale
unification of the gauge couplings, we have seen that
other string models apparently do contain such matter in the
right combinations.
Thus, these models succeed in resolving the discrepancy
between the string and MSSM unification scales, and thereby predict
the correct values of the low-energy couplings --- provided, of course,
that this extra matter sits at the appropriate intermediate
mass scales.  Unfortunately, these mass scales are typically set by a
variety of stringy mechanisms, and require detailed model-dependent
analyses before concrete predictions can be made.
It is nevertheless a significant fact that such string models
even manage to provide a {\it window}\/ in the parameter space of
intermediate mass scales for which such a reconciliation is possible.
Indeed, this non-trivial possibility arises because such string models
give rise to just the right combinations of extra non-MSSM matter, in just
the right exotic representations and with just the right non-standard
hypercharge assignments, in order to achieve the correct
low-energy predictions for the gauge couplings.
Within such models, therefore, the appearance of such matter
essentially becomes a prediction of string-scale unification.

The status of the other two approaches --- namely,
those based on constructing string GUT models and
models with non-standard levels and hypercharge normalizations ---
is perhaps less clear.
As we have discussed, these approaches tend to run into technical
subtleties which have, for some time, impeded progress.
In the case of the string GUT models,  the primary historical
difficulty has been the construction
of three-generation (or more generally, odd-generation) models;
likewise, in the case of string models with non-standard hypercharge
normalizations $k_Y$, the primary difficulty has been the construction
of models with $k_Y<5/3$.  However, we have seen that both
of these problems now appear to be solved.
Thus, although no {\it realistic}\/ models have yet
been constructed in either class,
these recent developments suggest that the problems of
finding realistic models
in either class are only technical (rather than somehow fundamental).
Once realistic models of these sorts are eventually constructed
(and assuming that their related phenomenological problems
of the proton lifetime and fractionally charged states can be solved),
a detailed model-dependent analysis of their unification
properties can then take place.

As a result of their different approaches and assumptions,
each of these ``paths to unification''
has different virtues.  Perhaps the strongest virtue of the
string GUT approach is that
it is relatively simple and elegant, and that it does not
regard the unification of the experimental gauge
couplings within the MSSM as an accident.
While gauge coupling unification at the MSSM scale is also a natural
feature of the approaches based on
strings with extra complete GUT multiplets
or strings at strong coupling,
this feature is not shared by the approaches that rely on
light SUSY thresholds or extra non-MSSM matter in incomplete multiplets.
Likewise, whether this feature is shared by the heavy string
threshold approach depends entirely on whether these heavy
string thresholds $\Delta_i$ ultimately satisfy the
unification relations (\ref{unifpreserved}).
In some string models, {\it e.g.}, the $N=2$
string models relevant for Eq.~(\ref{DKLresultone}),
such relations are automatically satisfied.
Thus, gauge coupling unification is
guaranteed in such models (assuming it was somehow already
present {\it before}\/ the corrections are included),
and it only remains to determine
the scale of unification.  In the more realistic $N=1$ models,
by contrast, the heavy string thresholds $\Delta_i$ do not
 {\it a priori}\/ satisfy Eq.~(\ref{unifpreserved}).  Hence
such models do not necessarily incorporate a unification of the
gauge couplings at any scale, and
it would seem unlikely to find a realistic string model
in this class
whose heavy threshold corrections magically manage to preserve
gauge coupling unification at the MSSM scale.
Of course, the viability of this approach ultimately depends on
finding precisely such a model.

Although it might seem attractive to retain gauge
coupling unification at the MSSM scale within string theory,
it must be emphasized that
what is most important is {\it not}\/ that a particular
approach be consistent with gauge coupling unification at some
intermediate scale, but rather that the predictions of string theory
be directly reconciled with the experimentally observed values
of the low-energy couplings.
Indeed, all of the approaches that we have discussed in this review
are designed with this goal in mind.  Furthermore, as we have seen,
extra matter beyond the MSSM appears to be a generic prediction
of string theory {\it regardless}\/ of the particular type of string model
under consideration.  It is therefore unjustified to ignore such states
when constructing, {\it e.g.}, a string GUT model, and the presence
of such extra matter can destroy what would otherwise have been a
successful unification of gauge couplings.
Thus string theory, it seems, would appear to favor a more complicated
route to unification than any of these
paths in isolation would suggest, and we can imagine that
features borrowed from many or all of these paths
may ultimately play a role.
Indeed, string theory is likely to be far richer in its
possibilities than simple field-theoretic extrapolations
would suggest.

There are, therefore, many open avenues for future investigation.
As far as string GUT models are concerned, it would clearly be desirable
to have an efficient method of surveying the possible
classes of three-generation models, and refining the techniques
for their construction.  Work along these lines is continuing.
Similarly, in order to build models with $k_Y<5/3$, it is necessary
to further analyze the MSSM embeddings which yield the appropriate
values of the hypercharge normalization, and to
attempt to construct string models in which any fractionally
charged states can be confined or made heavy.
Work here too is continuing.
Another obvious line of inquiry that should be undertaken
is to carefully study the role
that the extra non-MSSM states play in the realistic
string models.  In particular, it is clearly
necessary to actually {\it calculate}\/ the masses
that such states have in these models, and to
carefully evaluate their effects on low-energy physics.
Similarly, for the non-supersymmetric string models,
it is an important problem
not only to construct realistic models which reproduce
the Standard Model at low energies and which
achieve gauge coupling unification
in the manner described in Sect.~9,
but also to solve the accompanying
vacuum stability, technical gauge hierarchy, and cosmological
constant problems.
As we have discussed, these problems are undoubtedly related
in some deep (although as yet mysterious) manner.
The cosmological constant problem, of course, is one that
must be faced in {\it any}\/ string model, given that
spacetime supersymmetry must ultimately be broken at
some scale at or above the electroweak scale.
Finally, while the approach based on string non-perturbative effects
is perhaps the most daring, it is still far too early
to imagine the phenomenological implications and problems
that this approach may entail.

There are also other issues which we have not discussed in this
review, but which are equally important for
understanding the phenomenology of realistic string models.
For example, in this review we have focused exclusively
on the unification of the gauge couplings, but
in string theory we also expect special relations
between the {\it Yukawa couplings}\/ at the unification scale.
This is a whole subject unto itself which is of critical
importance, and which involves issues and scenarios
similar to those we have considered here.
Indeed, by calculating the Yukawa couplings
and studying their evolution below the string scale,
it is possible to approach the question of determining
the low-energy masses of the fermions and the textures of the
low-energy mass matrices.
This would also be an important test for string
phenomenology.

Finally,
there are also other more general issues which,
through crucial for eventual string phenomenology,
are not amenable to
study through the phenomenological
sorts of investigations we have been discussing here.
These include, for example, the important questions
of the selection of the string vacuum, and of supersymmetry breaking.
Such issues must of course be resolved before string
theory can make any
absolute phenomenological predictions.
Unfortunately,
it is not yet clear what shape the answers to these
questions will take.
One possibility, however, is that
progress will come from
exciting developments
concerning the implications
of the recently discovered string dualities.
Indeed, these dualities formed the basis of the strong-coupling
proposal for gauge coupling unification that was discussed in Sect.~10.

In conclusion, then, we have seen that
although string theory has the potential for
accurately describing
many properties of the observed low-energy world,
the scenarios and
mechanisms by which this can be accomplished appear to be
varied and numerous.
This is both a blessing and a curse.
On the one hand, it is important to study the extent to
which all of the different approaches can be realized in
a self-consistent manner, for it is by testing the bounds
of realistic model-building that we can truly determine
the limits and opportunities that string phenomenology provides.
On the other hand,  assuming that one unique vacuum solution
of string theory ultimately describes nature,
it will be necessary to develop methods
of determining which vacuum solution this might be, and
what its properties are.  Thus, we expect that some combination
of both approaches will ultimately prove most useful in
determining the
extent to which string theory is capable of describing
the physical world.

\vfill\eject
\leftline{\large\bf Acknowledgments}
\medskip

It is a pleasure to thank numerous people for informative discussions
on the work described herein.  These include
 M.~Alford,
 I.~Antoniadis, K.S.~Babu, C.~Bachas, S.~Barr, S.~Chaudhuri,
 G.~Cleaver, J.~Distler, L.~Dolan, J.~Erler, A.~Faraggi,
 D.~Finnell, A.~Font, L.~Ib\'a\~nez, Z.~Kakushadze, V.~Kaplunovsky,
 E.~Kiritsis, C.~Kolda, C.~Kounnas, P.~Langacker, J.~Louis, J.~Lykken,
 J.~March-Russell, M.~Moshe, R.~Myers, P.~Nath, H.P.~Nilles,
 J.~Pati, M.~Peskin, M.~Petropoulos, F.~Quevedo, J.~Rizos,
 A.~Sagnotti, S.~Stieberger, S.-H.H.~Tye, F.~Wilczek, and E.~Witten.
I also wish to thank
 T.~Damour, A.~Faraggi, E.~Kiritsis, J.~March-Russell,
 R.~Myers, Y.~Nir, A.~Sagnotti, and S.-H.H.~Tye
for careful preliminary readings of this article, and for numerous comments
and suggestions.
This article originally grew out of an invited review talk given
at the Conference {\it ``Unification:  From the Weak Scale to the Planck
Scale''}\/ which was held at the Institute for Theoretical Physics in Santa
Barbara,
California, in October 1995, and I benefitted greatly from conversations
with many participants both at this conference and at the workshop that
preceded it.
My greatest debt, however, is to my collaborators
from whom I have learned much of this subject, and
who were co-authors on some of the work described herein.
These include
Alon Faraggi (with whom I collaborated on portions of the material
presented in Sects.~5 through 8);
John March-Russell (portions of Sects.~4 and 5);  and
Moshe Moshe and Robert Myers (portions of Sect.~9).
Indeed, many of the ideas in this
review have been drawn from the papers
that resulted from these collaborations, and I thank
my co-authors for their
extraordinary indulgence in this regard.
This work was supported in part by DOE Grant No.\ DE-FG-0290ER40542.


\vfill\eject

\bigskip
\medskip

\bibliographystyle{unsrt}

\begin{thebibliography}{999}

\bibitem{unifreview}
   In this reference we gather together some of the ``classic'' papers
   pertaining to the unification of gauge groups and gauge couplings
   in field theory.
   The first papers to suggest a possible grand unification of the
   Standard Model gauge groups were: \\
   J.C. Pati and A. Salam, \PRD{8}{73}{1240};
            \PRL{31}{73}{661}; \PRD{10}{74}{275};\\
   H. Georgi and S.L. Glashow, \PRL{32}{74}{438}.\\

   This idea in turn requires a unification of gauge couplings,
   and was discussed in:\\
   H. Georgi, H.R. Quinn, and S. Weinberg, \PRL{33}{74}{451}.\\

   Note that the above paper makes use of prior results concerning the
   running of gauge couplings in non-abelian gauge theories which can
   be found in:\\
   D.J. Gross and F. Wilczek, \PRL{30}{73}{1343};\\
   H.D. Politzer, \PRL{30}{73}{1346}.\\

   The effects of incorporating supersymmetry into the unification
   scenario were discussed in:\\
   S. Dimopoulos, S. Raby, and F. Wilczek, \PRD{24}{81}{1681};\\
   S. Dimopoulos and H. Georgi, \NPB{193}{81}{150};\\
   N. Sakai, {\it Zeit.\ Phys.}\/ {\bf C11} (1981) 153;\\
   L.E. Ib\'a\~nez and G.G. Ross, \PLB{105}{81}{439}.\\

   The first two-loop calculations of gauge coupling unification
   within the MSSM can be found in:\\
   M.B. Einhorn and D.R.T. Jones, \NPB{196}{82}{475};\\
   W.J. Marciano and G. Senjanovic, \PRD{25}{82}{3092}.\\

   Subsequent calculations of gauge coupling unification and
   precision fits can be found in, {\it e.g.}\/:\\
      U. Amaldi \etal, \PRD{36}{87}{1385};\\
    J. Ellis, S. Kelley, and D.V. Nanopoulos, \PLB{249}{90}{441};
                   \PLB{260}{91}{131};  \NPB{373}{92}{55};\\
     P. Langacker and M. Luo, \PRD{44}{91}{817}.\\

    For recent detailed analyses of gauge coupling unification,
     including the effects of low-energy thresholds, see {\it e.g.}\/:\\
        U. Amaldi, W. de Boer, and H. F{\"u}rstenau, \PLB{260}{91}{447};\\
          H. Arason \etal, \PRD{46}{92}{3945};\\
         F. Anselmo \etal, \NUVC{105}{1992}{1179};\\
   V. Barger, M.S. Berger, and P. Ohmann,
      \PRD{47}{93}{1093}, hep-ph/9209232;\\
   P. Langacker and N. Polonsky,
     \PRD{47}{93}{4028}, hep-ph/9210235;
          \PRD{52}{95}{3081}, hep-ph/9503214;\\
   M. Carena, S. Pokorski, and C.E.M. Wagner,
     \NPB{406}{93}{59}, hep-ph/9303202;\\
   A.E. Faraggi and B. Grinstein,
     \NPB{422}{94}{3}, hep-ph/9308329;\\
   P.H. Chankowski, Z. Pluciennik, and S. Pokorski,
     \NPB{439}{95}{23}, hep-ph/9411233;\\
   P. Langacker, hep-ph/9411247;\\
   R. Barbieri, P. Ciafaloni, and A. Strumia,
             \NPB{442}{95}{461}, hep-ph/9411255;\\
   J. Bagger, K. Matchev, and D. Pierce,
        \PLB{348}{95}{443}, hep-ph/9501277;\\
   J. Ellis, hep-ph/9512335;\\
   D. Pierce, J. Bagger, K. Matchev, and R. Zhang,
             hep-ph/9606211;\\
   W. de Boer \etal, hep-ph/9609209.  \\

   For general reviews of grand-unified theories, see, {\it e.g.}\/:\\
   P. Langacker, \PRT{72}{81}{185};\\
   L. Hall, in {\it 1984 TASI Lectures
     in Elementary Particle Physics}\/
      (TASI Publications, Ann Arbor, Michigan, 1984);\\
   G.G. Ross, {\it Grand Unified Theories}\/ (Benjamin/Cummings,
    Menlo Park, California, 1985);\\
   R. Mohapatra, {\it Unification and Supersymmetry}\/ (Springer-Verlag,
    New York, 1986);\\
   P. Frampton, {\it Gauge Field Theories}\/, Chap.~7 (Benjamin/Cummings,
    Menlo Park, California, 1986);\\
   and references therein.
\bibitem{stringreview}  For introductions and reviews, see:\\
         M.B. Green, J.H. Schwarz, and E. Witten,
       {\it Superstring Theory, Vols.\ 1 \& 2}\/
        (Cambridge University Press, Cambridge, 1987);\\
       M. Dine, ed., {\it String Theory in Four Dimensions}\/
       (North-Holland, Amsterdam, 1988);\\
       B. Schellekens, ed., {\it Superstring Construction}\/
       (North-Holland, Amsterdam, 1989);\\
       J. Polchinski, {\it What is String Theory?,
        1994 Les Houches Summer School Lectures},
         hep-th/9411028.
\bibitem{phenreview}  For recent general discussions on
      various aspects of string phenomenology and
     string model-building, see, {\it e.g.}\/:\\
      L.E. Ib\'a\~nez, hep-th/9112050; hep-th/9505098;\\
      I. Antoniadis, hep-th/9307002;\\
      M. Dine, hep-ph/9309319;\\
      J.L. Lopez, {\it Surveys H.E. Phys.}\/ {\bf 8} (1995) 135,
      hep-ph/9405278;\\
      A.E. Faraggi, hep-ph/9405357;\\
      J.L. Lopez and D.V. Nanopoulos, hep-ph/9511266;\\
      J.D. Lykken, hep-ph/9511456;\\
      Z. Kakushadze and S.H.-H. Tye, hep-th/9512155;\\
      F. Quevedo, hep-th/9603074; \\
       and references therein.
\bibitem{unif1}  K.R. Dienes and A.E. Faraggi, \PRL{75}{95}{2646},
                 hep-th/9505018.
\bibitem{unif2}  K.R. Dienes and A.E. Faraggi, \NPB{457}{95}{409},
                 hep-th/9505046.
\bibitem{unif3}  K.R. Dienes, A.E. Faraggi, and J. March-Russell,
                 \NPB{467}{96}{44}, hep-th/9510223.
\bibitem{unif4}  K.R. Dienes and J. March-Russell, \NPB{479}{96}{113},
                 hep-th/9604112.
\bibitem{SUSYreviews} For reviews, see:\\
       R. Arnowitt and P. Nath, {\it Applied N=1
            Supergravity}\/ (World Scientific, Singa\-pore, 1983);\\
       H.P. Nilles, \PRT{110}{84}{1};\\
      H.E. Haber and G. L. Kane, \PRT{117}{85}{75};\\
      	A.B. Lahanas and D.V. Nanopoulos, \PRT{145}{87}{1}.\\
      Recent lectures on the phenomenology of the MSSM can be found in:\\
        X. Tata, hep-ph/9510287;\\
        J.L. Lopez, hep-ph/9601208;\\
        J. Bagger, hep-ph/9604232;\\
         M. Drees, hep-ph/9611409;\\
         M. Dine, hep-ph/9612389.
\bibitem{exptcouplings}
   See, {\it e.g.}: \\
   N. Polonsky,  hep-ph/9411378;\\
   P. Langacker, hep-ph/9412361;\\
   and references therein.
\bibitem{framp}
     P. Frampton and S.L. Glashow, \PLB{131}{83}{340};\\
     D. Chang, R.N. Mohapatra, and M.K. Parida, \PRL{52}{84}{1072};\\
     D. Chang, R.N. Mohapatra, J. Gipson, R.E. Marshak, and M.K. Parida,
          \PRD{31}{85}{1718};\\
     U. Amaldi \etal, \PLB{281}{92}{374};\\
     L. Lavoura and L. Wolfenstein, \PRD{48}{93}{264}.
\bibitem{heterotic} D.J. Gross, J.A. Harvey, J.A. Martinec, and R. Rohm,
			\PRL{54}{85}{502}; \NPB{256}{86}{253}.
\bibitem{phensucc}
     P. Candelas, G.T. Horowitz, A. Strominger, and E. Witten,
                   \NPB{258}{85}{46};\\
    E. Witten, \NPB{258}{85}{75};\\
    J.D. Breit, B.A. Ovrut, and G.C. Segr\'e, \PLB{158}{85}{33};\\
      A. Sen, \PRL{55}{85}{33};\\
     A.E. Faraggi, \NPB{428}{94}{111}, hep-ph/9403312.
\bibitem{Ginsparg}  P. Ginsparg, \PLB{197}{87}{139}.
\bibitem{KMreview} For pedagogical introductions to affine Lie
   algebras, also known as \KM\ algebras, see, {\it e.g.}:\\
    P. Goddard and D. Olive, \IJMP{1}{86}{303};\\
    P. Ginsparg, {\it Applied Conformal Field Theory},
     in {\it Fields, Strings, and Critical Phenomena:
     Proceedings of Les Houches, Session XLIX, 1988}, eds.\
     E. Br\'ezin and J. Zinn-Justin (Elsevier, 1989);\\
      J. Fuchs, {\it Affine Lie Algebras and Quantum Groups}\/
   (Cambridge University Press, Cambridge, England, 1992).\\
    A history of these algebras can be found in the Appendix of:\\
     M.B. Halpern, E. Kiritsis, N.A. Obers, and K. Clubok,
     \PRT{265}{96}{1}, hep-th/9501144.
\bibitem{KacMoody}
     V.G. Ka\v{c}, {\it Funct.\ Anal.\ App.}\/ {\bf 1} (1967) 328;\\
     R.V. Moody, {\it Bull.\ Am.\ Math.\ Soc.}\/ {\bf 73} (1967) 217.
\bibitem{Halpern}
     K. Bardakci and M.B. Halpern, \PRD{3}{71}{2493}.
\bibitem{Witten}
      E. Witten, \PLB{155}{85}{151}.
\bibitem{scales}  M. Dine and N. Seiberg, \PRL{55}{85}{366};\\
          V.S. Kaplunovsky, \PRL{55}{85}{1036}.\\
   A subsequent analysis along these lines can also be found in:\\
   R. Petronzio and G. Veneziano, \MODA{2}{87}{707}.
\bibitem{Kaplunovsky}
        V.S. Kaplunovsky, \NPB{307}{88}{145}, hep-th/9205070;
          Erratum:  {\it ibid.}\/ {\bf B382} (1992) 436.
\bibitem{stringguts}
   D.C. Lewellen, \NPB{337}{90}{61};\\
   A. Font, L.E. Ib\'a\~nez, and F. Quevedo, \NPB{345}{90}{389};\\
   J.A. Schwartz, \PRD{42}{90}{1777};\\
   S. Chaudhuri, S.-W. Chung, G. Hockney, and J. Lykken,
      \NPB{456}{95}{89}, hep-ph/9501361;\\
   J. Erler, \NPB{475}{96}{597}, hep-th/9602032;\\
   G. Cleaver, hep-th/9604183.
\bibitem{aldaone}
   G. Aldazabal, A. Font, L.E. Ib\'a\~nez, and A.M. Uranga,
    \NPB{452}{95}{3},  hep-th/9410206.
\bibitem{aldatwo}
   G. Aldazabal, A. Font, L.E. Ib\'a\~nez, and A.M. Uranga,
     \NPB{465}{96}{34}, hep-th/9508033.
\bibitem{shynew}  S. Chaudhuri, G. Hockney, and J. Lykken,
     \NPB{469}{96}{357}, hep-th/9510241.
\bibitem{KTtwo}  Z. Kakushadze and S.-H.H. Tye,
    \PRL{77}{96}{2612}, hep-th/9605221;
    \PRD{54}{96}{7520}, hep-th/9607138;
    \PLB{392}{97}{335}, hep-th/9609027;
     hep-th/9701057.
\bibitem{KTthree}  Z. Kakushadze and S.-H.H. Tye,
     hep-th/9610106.
\bibitem{orbifoldconstr}
     L.J. Dixon, J. Harvey, C. Vafa, and E. Witten, \NPB{261}{85}{678};
       \NPB{274}{86}{285}.
\bibitem{asymmorbifolds}
     K. Narain, M. Sarmadi, and C. Vafa,
        \NPB{288}{87}{551};  \NPB{356}{91}{163};\\
    L.E. Ib\'a\~nez, J. Mas, H.P. Nilles, and F. Quevedo,
         \NPB{301}{88}{157}.
\bibitem{oldorbmodels}
   L.E. Ib\'a\~nez, H.P. Nilles, and F. Quevedo, \PLB{187}{87}{25};
\PLB{192}{87}{332};\\
   L.E. Ib\'a\~nez, J.E. Kim, H.P. Nilles, and F. Quevedo,
\PLB{191}{87}{282};\\
   J.A. Casas and C. Mu\~noz, \PLB{214}{88}{63}.
\bibitem{KT}  Z. Kakushadze and S.-H.H. Tye, hep-th/9512156;\\
      Z. Kakushadze, G. Shiu, and S.-H.H. Tye,
     \PRD{54}{96}{7545}, hep-th/9607137.
\bibitem{freefermions}  H. Kawai, D.C. Lewellen, and S.-H.H. Tye,
                {\it Nucl.\ Phys.}\/ {\bf B288} (1987) 1;\\
                I. Antoniadis, C. Bachas, and C. Kounnas,
                {\it Nucl.\ Phys.}\/ {\bf B289} (1987) 87.
\bibitem{realfreefermions}
          H. Kawai, D.C. Lewellen, J.A. Schwartz, and S.-H.H. Tye,
                 \NPB{299}{88}{431}.
\bibitem{Eeightmodel}  H. Kawai, D.C. Lewellen, and S.-H.H. Tye,
          \PRD{34}{86}{3794}.
\bibitem{ELN}
   J. Ellis, J.L. Lopez, and D.V. Nanopoulos, \PLB{245}{90}{375}.
\bibitem{KDso10paper}
         K.R. Dienes, \NPB{488}{97}{141}, hep-ph/9606467.
\bibitem{WZW}
    E. Witten, \CMP{92}{86}{455}.
\bibitem{GxGmodels}
   A.A. Maslikov, S.M. Sergeev, and G.G. Volkov, \PLB{328}{94}{319};
        \PRD{50}{94}{7440};  \IJMP{9}{94}{5369};\\
   A.A. Maslikov, I. Naumov, and G.G. Volkov,
           \IJMP{11}{96}{1117}, hep-ph/9512429.
\bibitem{Finnell} D. Finnell, \PRD{53}{96}{5781}, hep-th/9508073.
\bibitem{GxGfield}
   R. Barbieri, G. Dvali, and A. Strumia,
         \PLB{333}{94}{79}, hep-ph/9404278;\\
   R.N. Mohapatra, \PLB{379}{96}{115}, hep-ph/9601203.
\bibitem{bachas}  C. Bachas, C. Fabre, and T. Yanagida,
    \PLB{370}{96}{49}, hep-th/9510094.\\
   A subsequent two-loop analysis of the effects of such
   adjoint scalars can be found in:\\
   M. Bastero-Gil and B. Brahmachari, hep-ph/9610374.
\bibitem{FGM}  A.E. Faraggi, B. Grinstein, and S. Meshkov, \PRD{47}{93}{5018},
       hep-ph/9206238.
\bibitem{flippedsu5fieldtheory}
       S.M. Barr, \PLB{112}{82}{219}; \PRD{40}{89}{2457}; \\
       J.P. Derendinger, J.E. Kim, and D.V Nanopoulos, \PLB{139}{84}{170}.
\bibitem{patisalam} J.C. Pati and A. Salam, \PRD{8}{73}{1240};
            \PRL{31}{73}{661}; \PRD{10}{74}{275}.
\bibitem{flipped}
    I. Antoniadis, J. Ellis, J. Hagelin, and D.V. Nanopoulos,
                  \PLB{231}{89}{65}; \\
    J.L. Lopez, D.V. Nanopoulos, and K. Yuan, \NPB{399}{93}{654},
           hep-th/9203025.
\bibitem{alrmodel}  I. Antoniadis, G.K. Leontaris, and J. Rizos,
    {\it Phys.\ Lett.}\/ {\bf B245} (1990) {161}.\\
    Since its introduction, this model has been analyzed in
    several papers.  See, {\it e.g.}\/:\\
   G.K. Leontaris and N.D. Tracas, {\it Z.\ Phys.}\/ {\bf C56} (1992) 479;
                               \PLB{279}{92}{58};\\
   A. Murayama and A. Toon, \PLB{318}{93}{298};\\
   O. Korakianitis and N.D. Tracas, \PLB{319}{93}{145}, hep-ph/ 9306255;\\
    G.K. Leontaris and N.D. Tracas, \PLB{372}{96}{219}, hep-ph/9511280.
\bibitem{Wen}
      X.G. Wen and E. Witten, \NPB{261}{85}{651};\\
      G. Athanasiu, J. Atick, M. Dine, and W. Fischler, \PLB{214}{88}{55}.
\bibitem{Schellekens}
       A.N. Schellekens, \PLB{237}{90}{363}.
\bibitem{DLT}
     P. Dimopoulos, G.K. Leontaris, and N.D. Tracas, hep-ph/9604265.
\bibitem{ibanez} J.A. Casas and C. Mu\~noz, \PLB{214}{88}{543}; \\
                 L.E. Ib\'a\~nez, \PLB{318}{93}{73}, hep-ph/9308365.
\bibitem{prevattempts}
           See, {\it e.g.}, H. Kawabe, T. Kobayashi, and N.  Ohtsubo,
          \NPB{434}{95}{210}, hep-ph/9405420; and references therein.
\bibitem{allanach}
   B.C. Allanach and S.F. King, \NPB{473}{96}{3}, hep-ph/9601391.
\bibitem{FIQS}
   See, {\it e.g.},  Appendix C of:\\
   A. Font, L.E. Ib\'a\~nez, F. Quevedo, and A. Sierra, \NPB{331}{90}{421}.
\bibitem{exptinputs}
    Particle Data Group, L. Montanet \etal, \PRD{50}{94}{1173}.
\bibitem{NAHE}  The so-called ``NAHE'' set was first
     constructed in the context of the flipped $SU(5)$
     string model of Ref.~\protect\cite{flipped},
      and was named and discussed in a model-independent
      way in:\\
      A.E. Faraggi, \NPB{387}{92}{239}, hep-th/9208024.
\bibitem{274model}   A.E. Faraggi, \PLB{302}{93}{202}, hep-ph/9301268.
\bibitem{fractionalexpt}
     For a recent review, see:\\
     P.F. Smith, {\it Ann. Rev. Nucl. Part. Sci.}\/
            {\bf 39} (1989) 73.
\bibitem{schellyank}
      A.N. Schellekens and S. Yankielowicz, \NPB{327}{89}{673};
         \PLB{227}{89}{387}.
\bibitem{keni}   K. Intriligator, \NPB{332}{90}{541}.
\bibitem{prevschell}
   I. Antoniadis and K. Benakli, \PLB{295}{92}{219},
     hep-th/9209020.
\bibitem{DSWshift}
         M. Dine, N. Seiberg, and E. Witten, \NPB{289}{87}{589};\\
              J.J. Atick, L.J. Dixon, and A. Sen, \NPB{292}{87}{109};\\
         M. Dine, I. Ichinose, and N. Seiberg,  \NPB{293}{87}{253};\\
              S. Cecotti, S. Ferrara, and M. Villasante, \IJMP{2}{87}{1839}.
\bibitem{Huet}
   P. Huet, \NPB{350}{91}{375}.
\bibitem{Alonmassestimate}
        A.E. Faraggi, \PRD{46}{92}{3204}.
\bibitem{DL}
   L. Dolan and J.T. Liu, \NPB{387}{92}{86}, hep-th/9205094;\\
   D.M. Pierce, \PRD{50}{94}{6469}, hep-th/9508178.
\bibitem{Yterm}   I. Antoniadis, E. Gava, and K.S. Narain,
         \NPB{383}{92}{93}, hep-th/9204030;\\
        B. de Wit, V.S. Kaplunovsky, J. Louis, and D. L\"ust,
     \NPB{451}{95}{53}, hep-th/9504006;\\
       I. Antoniadis, S. Ferrara, E. Gava, K.S. Narain, and T.R. Taylor,
          \NPB{447}{95}{35}, hep-th/9504034.
\bibitem{Ytermsize}   S. Kalara, J.L. Lopez, and D.V. Nanopoulos,
        \PLB{269}{91}{84}.
\bibitem{Kiritsis}
          E. Kiritsis and C. Kounnas,
    {\it Nucl.\ Phys.\ Proc.\ Suppl.}\/ {\bf 41} (1995) 331, hep-th/9410212;
         \NPB{442}{95}{472}, hep-th/9501020;
          hep-th/9507051;
    {\it Nucl.\ Phys.\ Proc.\ Suppl.}\/ {\bf 45} (1996) 207, hep-th/9509017.
\bibitem{PR}
     P.M. Petropoulos and J. Rizos, \PLB{374}{96}{49}, hep-th/9601037.
\bibitem{choi}
     K. Choi, \PRD{37}{88}{1564}.
\bibitem{kounnas}
      C. Kounnas, M. Porrati, and B. Rostand, \PLB{258}{91}{61};\\
      E. Kiritsis, C. Kounnas, and D. L\"ust, \IJMP{9}{94}{1361}; \\
      C. Kounnas, \PLB{321}{94}{26}.
\bibitem{DKLkiritsistwo}
      E. Kiritsis, C. Kounnas, P.M. Petropoulos, and J. Rizos,
       \PLB{385}{96}{87}, hep-th/9606087.
\bibitem{DKLkiritsis}
      E. Kiritsis, C. Kounnas, P.M. Petropoulos, and J. Rizos, hep-th/9608034.
\bibitem{DKL}
         L.J. Dixon, V.S. Kaplunovsky, and J. Louis,
           \NPB{355}{91}{649};\\
            V.S. Kaplunovsky and J. Louis, \NPB{444}{95}{191}, hep-th/9502077.
\bibitem{moduli} See, {\it e.g.}, the following papers and
       references therein:\\
      I. Antoniadis, K.S. Narain, and T.R. Taylor, \PLB{267}{91}{37};\\
        J.P. Derendinger, S. Ferrara, C. Kounnas,
        and F. Zwirner, \PLB{271}{91}{307}; \NPB{372}{92}{145};\\
      I. Antoniadis, E. Gava, and K.S. Narain, \PLB{283}{92}{209},
          hep-th/9203071;\\
         G. Lopes Cardoso and B.A. Ovrut, \NPB{369}{92}{351};\\
         D. Bailin and A. Love, \PLB{292}{92}{315};\\
         P. Mayr and S. Stieberger, \NPB{407}{93}{725}, hep-th/9303017;
                                  \NPB{412}{94}{502}, hep-th/9304055;\\
         D. Bailin, A. Love, W.A. Sabra, and S. Thomas,
             \PLB{320}{94}{21}, hep-th/9309133;
              \MODA{9}{94}{67}, hep-th/9310008;
             \MODA{10}{95}{337}, hep-th/9407049;
             \PLB{378}{96}{113}, hep-th/9602132.
\bibitem{ILR}   L.E.~Ib\'a\~nez, D.~L\"ust, and G.G. Ross,
        \PLB{272}{91}{251}, hep-th/9109053.
\bibitem{chemtob}  M. Chemtob, \PRD{53}{96}{3920}, hep-th/9506178.
\bibitem{INN}
     L.E. Ib\'a\~nez and H.P. Nilles, \PLB{169}{86}{354};\\
    H.P. Nilles, \PLB{180}{86}{240}.
\bibitem{antonlargeradius}
     I. Antoniadis, \PLB{246}{90}{377};\\
     I. Antoniadis, C. Mu\~noz, and M. Quir\'os,
     {\it Nucl.\ Phys.}\/ {\bf B397} (1993) 515.
\bibitem{nolargemoduli}
      A. Font, L.E. Ib\'a\~nez, D. L\"ust, and F. Quevedo,
          \PLB{245}{90}{401};\\
     M. Cveti\v{c}, A. Font, L.E. Ib\'a\~nez,
      D. L\"ust, and F. Quevedo, \NPB{361}{91}{194};\\
    B. de Carlos, J.A. Casas, and C. Mu\~noz,
    \NPB{399}{93}{623},  hep-th/9204012.
\bibitem{KRDlambda}  K.R. Dienes, \PRL{65}{90}{1979};
		       Ph.D. dissertation (Cornell University, 1991).
\bibitem{missusy}  K.R. Dienes, \NPB{429}{94}{533}, hep-th/9402006;
        hep-th/9409114 (published in Proceedings of PASCOS '94);
       hep-th/9505194 (published in Proceedings of Strings '95).
\bibitem{modulitwozero}
  P. Mayr and S. Stieberger, hep-th/9412196;
   \PLB{355}{95}{107}, hep-th/9504129.
\bibitem{NS}     H.P. Nilles and S. Stieberger, \PLB{367}{96}{126},
         hep-th/9510009.
\bibitem{thresholdcalcs}
   I. Antoniadis, J. Ellis, R. Lacaze, and D.V. Nanopoulos,
                  \PLB{268}{91}{188}.
\bibitem{MNS}
   P. Mayr, H.P. Nilles, and S. Stieberger, \PLB{317}{93}{53},
                  hep-th/9307171.
\bibitem{278model}  A.E. Faraggi, \PLB{278}{92}{131}.
\bibitem{FNY}
    A.E. Faraggi, D.V. Nanopoulos, and K. Yuan, \NPB{335}{90}{347}.
\bibitem{fermionmasses}
   See, {\it e.g.}\/:\\
   J.L. Lopez and D.V. Nanopoulos, \NPB{338}{90}{73};
    \PLB{251}{90}{73};
    \PLB{256}{91}{150};
    \PLB{268}{91}{359};\\
   J. Rizos and K. Tamvakis, \PLB{251}{90}{369};\\
   A.E. Faraggi, \PLB{274}{92}{47};
      \PLB{377}{96}{43}, hep-ph/9506388;
      hep-ph/9601332.
\bibitem{stableproton}
   A.E. Faraggi, \NPB{428}{94}{111}, hep-ph/9403312.
\bibitem{orbifold}
     A.E. Faraggi,
    {\it Phys.\ Lett.}\/ {\bf B326}, {62} (1994);
    \NPB{407}{93}{57}, hep-ph/9210256.
\bibitem{halyo}   E. Halyo, hep-ph/9509323.
\bibitem{custodial}  A.E. Faraggi, \PLB{339}{94}{223}, hep-ph/9408333.
\bibitem{nonuniversal}
      L.E.~Ib\'a\~nez and D.~L\"ust, \NPB{382}{92}{305},
             hep-th/9202046;\\
                 V.S. Kaplunovsky and J. Louis, \PLB{306}{93}{269},
hep-th/9303040;\\
                 A. Brignole, L.E.~Ib\'a\~nez, and C. Mu\~noz,
\NPB{422}{94}{125},
             hep-ph/9308271.
\bibitem{Gaillard}
      I. Antoniadis, J. Ellis, S. Kelley, and D.V. Nanopoulos,
       \PLB{272}{91}{31};\\
           S. Kelley, J.L. Lopez, and D.V. Nanopoulos, \PLB{278}{92}{140};\\
     D. Bailin and A. Love, \PLB{280}{92}{26};  \MODA{7}{92}{1485};
      Erratum: {\it ibid.}\/ {\bf A7} (1992) 2963;\\
    M.K. Gaillard and R. Xiu, {\it Phys.\ Lett.}\/
      {\bf B296} (1992) 71, hep-ph/9206206;\\
      S.P. Martin and P. Ramond, {\it Phys.\ Rev.}\/
      {\bf D51} (1995) 6515, hep-ph/9501244.
\bibitem{AEFdarkmatter}
   S. Chang, C. Corian\`o, and A.E. Faraggi, hep-ph/9603272;
        \NPB{477}{96}{65}, hep-ph/9605325;\\
   A.E. Faraggi, hep-ph/9608420.
\bibitem{LN}
    J.L. Lopez and D.V. Nanopoulos,  \PRL{76}{96}{1566}, hep-ph/9511426.
\bibitem{kinetic}
     B. Holdom, \PLB{166}{86}{196};\\
     F. del Aguila, G. Coughlan, and M. Quir\'os, \NPB{307}{88}{633};\\
     C. Hattori, M. Matsuda, T. Matsuoka, and D. Mochinaga,
       {\it Prog.\ Theor.\ Phys.}\/ {\bf 90} (1993) 895, hep-ph/9307305;\\
     K.S. Babu, C. Kolda, and J. March-Russell,
           \PRD{54}{96}{4635},  hep-ph/9603212;\\
     K.R. Dienes, C. Kolda, and J. March-Russell, hep-ph/9610479 
     (to appear in {\it Nucl.\ Phys.}\/ {\bf B}\/).
\bibitem{zprime}
     M. Cveti\v{c} and P. Langacker,
     \PRD{54}{96}{3570},  hep-ph/9511378;
    \MODA{11}{96}{1247}, hep-ph/9602424;\\
     A.E. Faraggi and M. Masip, \PLB{388}{96}{524}, hep-ph/9604302;\\
     J.L. Lopez and D.V. Nanopoulos,
      \PRD{55}{97}{397}, hep-ph/9605359;\\
     J.D. Lykken, hep-ph/9610218.
\bibitem{DKMNS}
   M. Dine, V.S. Kaplunovsky, M. Mangano, C. Nappi,
      and N. Seiberg, \NPB{259}{85}{549}.
\bibitem{GSmech}
   M.B. Green and J. Schwarz, \PLB{149}{84}{117}.
\bibitem{dilatonstabilize}
    M. Dine and N. Seiberg, \PLB{162}{85}{299}.
\bibitem{banksdine}
   T. Banks and M. Dine, \PRD{50}{94}{7454}, hep-th/9406132;\\
   M. Dine, hep-th/9508085.
\bibitem{duality}
    The literature on string dualities is large and rapidly growing.
    For recent overviews, see:\\
    J. Polchinski, hep-th/9511157;
           {\it Rev.\ Mod.\ Phys.}\/ {\bf 68} (1996) 1245, hep-th/ 9607050;
           hep-th/9611050;\\
    J. Schwarz, hep-th/9607201;\\
    M. Duff, hep-th/9608117;\\
    M. Dine, hep-th/9609051;\\
    A. Sen, hep-th/9609176;\\
    M.R. Douglas, hep-th/9610041;\\
      P.K. Townsend, hep-th/9612121;\\
     S. F\"orste and J. Louis, hep-th/9612192.
\bibitem{Shirman}
   M. Dine and Y. Shirman, \PLB{377}{96}{36}, hep-th/9601175.
\bibitem{semipert}
   K.S. Babu and J.C. Pati, \PLB{384}{96}{140}, hep-ph/9606215;\\
   C. Kolda and J. March-Russell, hep-ph/9609480.\\
   An earlier analysis of the effects of extra complete multiplets
   in producing semi-perturbative unification can also be found in:\\
   R. Hempfling, \PLB{351}{95}{206}, hep-ph/9502201.
\bibitem{shenker}
    S. Shenker, in
    Proceedings of {\it Cargese 1990: Random surfaces and quantum gravity},
    eds.\ E. Brezin and S.R. Wadia, pp.\  191-200;\\
    J. Polchinski, \PRD{50}{94}{6041}, hep-th/9407031;\\
    E. Witten, \NPB{443}{95}{85}, hep-th/9503124.
\bibitem{Lambdareview}  For a review of the cosmological constant problem
          in field theory, see:\\
       S. Weinberg, {\it Rev.\ Mod.\ Phys.}\/ {\bf 61} (1989) 1;
astro-ph/9610044.
\bibitem{so16so16string}
            L.J. Dixon and J. Harvey,
             {\it Nucl.\ Phys.}\/ {\bf B274} (1986) 93;\\
     L. Alvarez-Gaum\'e, P. Ginsparg, G. Moore, and C. Vafa, {\it Phys.
     Lett.}\/ {\bf B171} (1986) 155.
\bibitem{scherkschwarz}
        J. Scherk and J.H. Schwarz, \PLB{82}{79}{60};\\
        R. Rohm, \NPB{237}{84}{553};\\
         C. Kounnas and M. Porrati, \NPB{310}{88}{355};\\
         S. Ferrara, C. Kounnas, M. Porrati, and F. Zwirner,
         \NPB{318}{89}{75};\\
       C. Bachas, hep-th/9503030; hep-th/9509067;\\
       J.G. Russo and A.A. Tseytlin, \NPB{461}{96}{131}, hep-th/9508068;\\
       A.A. Tseytlin, hep-th/9510041.\\
     For recent reviews of methods of supersymmetry-breaking in
     string theory, see:\\
        T.R. Taylor, hep-ph/9510281;\\
        F. Quevedo, hep-th/9511131.
\bibitem{Wittencos}
   E. Witten, \MODA{10}{95}{2153}, hep-th/9506101.
\bibitem{allorders} G. Moore, J. Harris, P. Nelson,
      and I. Singer, \PLB{178}{86}{167};
      Erratum: {\it ibid.}\/ {\bf B201} (1988) 579.
\bibitem{lambdarefs}
       See, {\it e.g.}:\\
      H. Itoyama and T.R. Taylor, \PLB{186}{87}{129};\\
      P. Ginsparg and C. Vafa, \NPB{289}{87}{414};\\
      G. Moore, \NPB{293}{87}{139};
        Erratum: {\it ibid.}\/ {\bf B299} (1988) 847;\\
      T.R. Taylor, \NPB{303}{88}{543};\\
      E. Alvarez and M.A.R. Osorio, {\it Z.\ Phys.}\/ {\bf C44} (1989) 89;\\
     J. Balog and M.P. Tuite, \NPB{319}{89}{387};\\
    K.R. Dienes, \PRD{42}{90}{2004};\\
     T. Gannon and C.S. Lam, \PRD{46}{92}{1710}, hep-th/9201028.
\bibitem{supertraces}
       K.R. Dienes, M. Moshe, and R.C. Myers,
     {\it Phys.\ Rev.\ Lett.}\/ {\bf 74} (1995) 4767, hep-th/9503055;
     hep-th/9506001 (published in Proceedings of Strings '95).
\bibitem{newwitten}
   E. Witten, \NPB{471}{96}{135}, hep-th/9602070;\\
   P. Ho\v{r}ava and E. Witten, \NPB{475}{96}{94}, hep-th/9603142.
\bibitem{type1dual}
   E. Witten, \NPB{443}{95}{85}, hep-th/9503124;\\
   A. Dabholkar, \PLB{357}{95}{307}, hep-th/9506160;\\
   C.M. Hull, \PLB{357}{95}{545},  hep-th/9506194;\\
   J. Polchinski and E. Witten, \NPB{460}{96}{525}, hep-th/9510169.
\bibitem{e8dual}
   P. Ho\v{r}ava and E. Witten, \NPB{460}{96}{506}, hep-th/9510209.
\bibitem{TypeImodels}
   For recent constructions of four-dimensional Type~I string
   models with $N=1$ spacetime supersymmetry, see, {\it e.g.}\/:\\
   M. Berkooz and R. Leigh, hep-th/9605049;\\
   C. Angelantonj, M. Bianchi, G. Pradisi, A. Sagnotti, and
      Y.S. Stanev, \PLB{385}{96}{96}, hep-th/9606169;\\
   R. Gopakumar and S. Mukhi, \NPB{479}{96}{260}, hep-th/9607057. \\
   In particular, the second paper presents the only {\it chiral}\/
   four-dimensional Type~I string model constructed to date.
\bibitem{TypeItechniques}
   See, {\it e.g.}\/:\\
   A. Sagnotti, in Proceedings of {\it Cargese 1987:  Non-Perturbative
   Quantum Field Theory}\/, eds.\ G. Mack \etal\  (Plenum, 1988), p.\  521;\\
   Z. Bern and D.C. Dunbar, \PLB{203}{88}{109};  \NPB{319}{89}{104};\\
   D.C. Dunbar, \NPB{319}{89}{72};\\
   P. Ho\v{r}ava, \NPB{327}{89}{461};  \PLB{231}{89}{251};\\
   J. Dai, R.G. Leigh, and J. Polchinski,
    {\it Mod.\ Phys.\ Lett.}\/ {\bf A4} (1989) 2073;\\
   G. Pradisi and A. Sagnotti, \PLB{216}{89}{59};\\
   M. Bianchi and A. Sagnotti, \PLB{247}{90}{517};  \NPB{361}{91}{519};\\
   E.G. Gimon and J. Polchinski, \PRD{54}{96}{1667}, hep-th/9601038;\\
   E.G. Gimon and C.V. Johnson,
    \NPB{477}{96}{715}, hep-th/9604129;
    \NPB{479}{96}{285}, hep-th/9606176;\\
   A. Dabholkar and J. Park, \NPB{477}{96}{701}, hep-th/9604178;\\
   K. Dasgupta and S. Mukhi, \PLB{385}{96}{125}, hep-th/9606044;\\
   J.D. Blum and A. Zaffaroni, \PLB{387}{96}{71}, hep-th/9607019;\\
   C. Angelantonj, M. Bianchi, G. Pradisi, A. Sagnotti,
       and Y.S. Stanev, \PLB{387}{96}{743}, hep-th/9607229;\\
   J.D. Blum, hep-th/9608053.
\bibitem{bachasfabre}
   C. Bachas and C. Fabre, \NPB{476}{96}{418}, hep-th/9605028.
\bibitem{newanton}
   I. Antoniadis and M. Quir\'os, hep-th/9609209.
\bibitem{newkap}
   E. Caceres, V.S. Kaplunovsky, and I.M. Mandelberg, hep-th/9606036.
\bibitem{banksdinehorava}
   T. Banks and M. Dine, \NPB{479}{96}{173}, hep-th/9605136;
     hep-th/9608197;  hep-th/9609046;\\
   P. Ho\v{r}ava, \PRD{54}{96}{7561}, hep-th/9608019.
\end{thebibliography}

\vfill\eject

\end{document}